\documentclass[a4paper,10pt]{article}
\usepackage[english]{babel}
\usepackage{jheppub}
\usepackage{caption}
\usepackage{subfig}
\usepackage{float}
\usepackage[T1]{fontenc}
\usepackage{graphicx}
\usepackage{epsfig}
\usepackage{amsmath}
\usepackage{amssymb}
\usepackage{float}
\usepackage{placeins}
\usepackage{braket}
\usepackage{slashed}
\usepackage{mathdots}
\usepackage{lipsum}
\usepackage{bigints}
\usepackage{cleveref}
\usepackage{url}

\allowdisplaybreaks[4]

\newcommand{\masterNew}[1]{m_{#1}}

\title{Two-loop master integrals for pseudo-scalar quarkonium and leptonium production and decay}

\preprint{{
\footnotesize 
\begin{tabular}{l} 
  BONN-TH-2022-14\\ CERN-TH-2022-093\\ TTP22-037
\end{tabular}
}}

\author[a,b]{Samuel Abreu,}
\author[c]{Matteo Becchetti,}
\author[d]{Claude Duhr,}
\author[e]{Melih~A.~Ozcelik}
\affiliation[a]{Theoretical Physics Department, CERN, Geneva 1211, Switzerland}
\affiliation[b]{Higgs Centre for Theoretical Physics, School of Physics and Astronomy,
The University of Edinburgh, Edinburgh EH9 3FD, Scotland, UK}
\affiliation[c]{Physics Department, Torino University and INFN Torino,
Via Pietro Giuria 1, I-10125 Torino, Italy}
\affiliation[d]{Bethe Center for Theoretical Physics, Universit\"{a}t Bonn, D-53115, Germany}
\affiliation[e]{Institute for Theoretical Particle Physics, KIT, 76128 Karlsruhe, Germany}
\emailAdd{samuel.abreu@cern.ch}
\emailAdd{matteo.becchetti@unito.it}
\emailAdd{cduhr@uni-bonn.de}
\emailAdd{melih.oezcelik@kit.edu}

\abstract{We compute the master integrals relevant for the two-loop 
corrections to pseudo-scalar quarkonium and leptonium production and 
decay. We present both analytic and high-precision numerical results. 
The analytic expressions are given in terms of multiple polylogarithms (MPLs), 
elliptic multiple polylogarithms (eMPLs) and iterated integrals of Eisenstein series. 
As an application of our results, we obtain for the first time an analytic expression for the two-loop amplitude for 
para-positronium decay to two photons at two loops.}

\newcommand{\beq}{\begin{equation}}

\newcommand{\eeq}{\end{equation}}
\newcommand{\nn}{\nonumber}
\newcommand{\bea}{\begin{eqnarray}}
\newcommand{\eea}{\end{eqnarray}}
\newcommand{\bfig}{\begin{figure}}
\newcommand{\efig}{\end{figure}}
\newcommand{\bc}{\begin{center}}
\newcommand{\ec}{\end{center}}

\DeclareMathOperator{\Cl}{Cl}
\DeclareMathOperator{\Li}{Li}
\DeclareMathOperator{\Ree}{Re}
\DeclareMathOperator{\Ime}{Im}
\DeclareMathOperator{\K}{K}
\DeclareMathOperator{\F}{F}

\date{}

\begin{document}
\maketitle
\flushbottom


\section{Introduction}
\label{sec:intro}

Current approaches to the calculation of higher-order corrections 
in perturbative Quantum Field Theory (QFT) are based on the decomposition of an observable
into an independent set of Feynman integrals, usually called master integrals. 
While the details of the decomposition strongly depend on the specific QFT, 
the master integrals only depend on the kinematics of the process and the loop order.
The computation of these Feynman integrals is thus an interesting problem in its
own right, providing one of the crucial ingredients for the calculation
of higher-order corrections to various physical observables.

Over the last decades, a lot of effort has been 
put into developing new techniques aimed at the efficient evaluation of Feynman 
integrals, both analytical and numerical.
This effort has been focused on developing a better understanding of the classes
of functions that these integrals evaluate to. It can be shown very generically
that Feynman integrals evaluate to iterated integrals \cite{ChenSymbol},
with a non-trivial branch-cut structure that is constrained by physical 
considerations \cite{Landau:1959fi}. These very generic arguments still
leave a large space of functions to explore, but it is by now well known
that many Feynman integrals evaluate to iterated integrals with a very
specific underlying geometry. The simplest kind of functions one finds
are \emph{multiple polylogarithms} (MPLs) \cite{GoncharovMixedTate},
which, roughly speaking, correspond to iterated integrals over rational
functions.
The mathematical properties of MPLs are well understood. In particular,
several computational tools have been developed to work with this class of functions, 
both for their analytic manipulation and their numerical evaluation.
It is however known that starting at the two-loop order a new class of iterated
integrals appears, where the integration kernels involve square roots of cubic or 
quartic polynomials which define an elliptic curve. If a single elliptic curve is
present, the iterated integrals can be written in terms of 
\emph{elliptic multiple polylogarithms} (eMPLs) \cite{brown2013multiple,Broedel:2014vla,Broedel:2017kkb}. 
Despite many recent developments, our understanding of the analytic structure of eMPLs is not nearly
as mature as in the case of MPLs, and their numerical evaluation is still very challenging.
Under certain conditions eMPLs can be expressed in terms of iterated integrals of Eisenstein 
series~\cite{ManinModular,Brown:mmv},
for which more advanced numerical evaluation strategies are known \cite{Duhr:2019rrs}.
Computing sets of master integrals involving elliptic curves is  at the forefront of
the problems that can currently be tackled.

In this paper we consider the complete set of master integrals appearing in 
the calculation of the two-loop amplitudes describing the production or decay
of a pseudo-scalar bound state of a pair
of massive fermions of the same flavour, be they quarkonium or leptonium bound 
states. For instance, the set of master integrals we consider is sufficient 
to compute the next-to-next-to-leading order (NNLO) QCD corrections to the 
hadro-production of a pseudo-scalar ($c\overline{c}$) bound state (usually 
called $\eta_c$) at the LHC, or the NNLO QED corrections to the decay of an 
($e^+e^-$) bound state (usually called para-positronium) into two photons. 
These processes are of great phenomenological interest. For instance,
quarkonium production offers interesting opportunities to study the interplay
between the perturbative and non-perturbative regimes of QCD 
\cite{Kramer:2001hh,Lansberg:2019adr,QuarkoniumWorkingGroup:2004kpm,
Lansberg:2006dh,Andronic:2015wma}.
The master integrals contributing to the NNLO corrections to such processes 
involve both MPLs and eMPLs and, while some of them
were already known in the literature \cite{Gehrmann:1999as,Bonciani:2003hc,Anastasiou:2006hc,
Beerli:2008zz,Bonciani:2009nb,Gehrmann:2010ue,Chen:2017xqd,vonManteuffel:2017hms,Chen:2017soz,
DiVita:2018nnh,Chen:2019zoy,Gerlach:2019kfo,Becchetti:2019tjy,Mandal:2022vju}, the complete
set was not yet available in analytic form.

The purpose of this paper is two-fold: we present both analytic expressions for 
a complete set of two-loop master integrals contributing to the processes discussed above,
and high-precision numerical evaluations of the integrals which can then be used for phenomenological
studies. Regarding the analytic calculation, we obtain analytic expressions for all integrals
by direct integration of their parametric representation. As already noted, we find that 
the integrals can be expressed in terms of MPLs, eMPLs and iterated integrals of Eisenstein series.
The set of elliptic Feynman integrals involves two different elliptic curves: 
one elliptic curve belongs to the same family as the sunrise integral \cite{Sabry,Broadhurst:1987ei,Bogner:2017vim,Adams:2015gva,Adams:2015ydq,Bogner:2019lfa,Broedel:2017siw,Campert:2020yur}, while the second is an elliptic curve that appears in certain master integrals for 
$t\overline{t}$ production at hadron colliders \cite{vonManteuffel:2017hms,Broedel:2019hyg}. 
Importantly, these  two elliptic curves appear in independent sets of master integrals.
Regarding the numerical evaluation of the master integrals, we present high-precision numerical results valid
to 1000 digits. These numbers are obtained by numerically solving
systems of differential equations within two slightly different approaches. More specifically,
the results valid to 1000 digits are obtained with the auxiliary mass flow 
method \cite{Liu:2017jxz,Liu:2021wks,Liu:2022chg} as implemented in $\tt{AMFlow}$ \cite{Liu:2022chg}, 
and they are validated with results obtained with the generalised power series expansion \cite{Moriello:2019yhu,Hidding:2020ytt} 
as implemented in $\tt{diffexp}$ \cite{Hidding:2020ytt}.
These high-precision
numerical results allow us to identify relations between the coefficients in the Laurent expansion
of the master integrals in the dimensional regulator $\epsilon$ using the \texttt{PSLQ} algorithm~\cite{pslq}. 
These relations are important to obtain more compact analytic results, and it would certainly be interesting 
to understand how they can be generated more systematically.
Our results can be found in a $\tt{Mathematica}$-readable format at 
ref.~\cite{masterIntegralsGit}.

The calculation of the various amplitudes which can be written in terms of the set of master integrals
we consider in this paper will be discussed in detail in a companion paper \cite{Abreu:2022cco}.
We nevertheless present here analytic results for the two-loop QED corrections to the decay of true para-positronium, 
which were first obtained in numerical form more than 20 years ago in refs.~\cite{Czarnecki:1999ci,Czarnecki:1999gv},
adding to the small but increasing number of physical quantities involving elliptic integrals which are known in 
analytic form \cite{Honemann:2018mrb,Abreu:2019fgk,Prausa:2020psw,Badger:2021owl,Caola:2022vea}.

The paper is structured as follows. 
In \cref{sec:setup} we discuss the set of master integrals we will consider in this paper. 
We present two types of relations beyond integration-by-parts identities which arise in degenerate 
kinematic configurations such as the ones corresponding to the amplitudes we consider.
In \cref{sec:analytic} we describe the analytic computation of the master integrals by 
direct integration, and we characterise the solutions in terms of MPLs, eMPLs and iterated 
integrals of Eisenstein series. 
In \cref{sec:numerics} we summarise the main steps we followed for the numerical evaluation of
the master integrals, along with the checks we did on our results. We also obtain relations
between the elliptic master integrals using the high-precision numerical evaluations.
Finally, in \cref{sec:parap} we discuss our results for the two-loop amplitude for the 
para-positronium decay to two photons, before we present our conclusion and outlook
in \cref{sec:conclusion}.


\section{Master integrals} 
\label{sec:setup}

\subsection{Kinematics and conventions}

We consider the master integrals required to compute the two-loop perturbative corrections to the
production or decay of a pseudo-scalar bound state of two massive fermions. This bound state can be
a quarkonium bound state, in which case we consider higher-order QCD corrections, or a leptonium bound
state, in which case we consider QED corrections.
For concreteness, the discussion of this section focuses on the production of such a bound state, 
but it is clear that it also holds for its decay.

The perturbative corrections to the production of a bound state of massive fermions
are systematically accounted for by considering the corrections to the short-distance process 
\begin{equation}\label{eq:proc}
a{\left(k_1\right)}b{\left(k_2\right)}\rightarrow Q{\left(p_1\right)}\overline{Q}{\left(p_2\right)}\,,
\end{equation}
where $Q$ and $\overline{Q}$ are fermions of mass $m_Q$ and the initial-state particles $ab$ can be two gluons $(gg)$, two photons $(\gamma \gamma)$ or
a photon and a gluon $(\gamma g)$.
We will consider the process in \cref{eq:proc} at leading order in an expansion
in the relative velocity $v$ of the $Q\overline{Q}$ pair in the bound-state rest frame.
This amounts to equating the heavy-fermion momenta $p_1$ and $p_2$, and
the kinematics effectively degenerate to those of a 
three-point process. More explicitly,
assuming $k_1$ and $k_2$ incoming and $p_1$ and $p_2$ outgoing, we have
\begin{equation}\label{eq:kin1}
  k_1^2=k_2^2=0, \qquad p^2=\frac{1}{2}k_1\cdot k_2=m_Q^2 \quad \text{ where }\quad p=p_1=p_2=\frac{1}{2}\left(k_1+k_2\right).
\end{equation}
The usual Mandelstam variables associated with four-point kinematics become
\begin{equation}\label{eq:kin2}
  s = \left(k_1+k_2\right)^2=4m_Q^2\,,
  \qquad
  t = \left(k_1-p_1\right)^2=-m_Q^2\,,
  \qquad
  u = \left(k_1-p_2\right)^2=-m_Q^2\,.
\end{equation}
Besides the perturbative corrections we consider here, there are also
corrections related to higher orders in an expansion in $v$ 
(see, e.g., ref.~\cite{Bodwin:1994jh} and references therein).
All these contributions must in general be taken into account 
for phenomenological predictions for the production or decay of quarkonium or leptonium bound
states, but they fall outside of the scope of this paper.

We will focus on the calculation of the master integrals that 
contribute to the two-loop amplitudes for the process in \cref{eq:proc}.
For quarkonium states, this will allow us to compute the two-loop amplitudes for the production and decay 
of the colour-singlet pseudo-scalar state $^1S_0^{[1]}$ in both the $gg$ and $\gamma \gamma$ channels, 
or the production and decay of the pseudo-scalar colour-octet state $^1S_0^{[8]}$ 
in the $gg$ and $\gamma g$ channels.
For leptonium states, our set of integrals is sufficient to compute the 
two-loop QED corrections to para-positronium decaying into two photons, 
for which a numerical computation has already been performed in 
refs.~\cite{Czarnecki:1999ci,Czarnecki:1999gv}, or the equivalent process for (true) muonium or tauonium.
We will return to the decay of para-positronium in section~\ref{sec:parap}. 
The calculation of the two-loop corrections to the production and decay of a pseudo-scalar bound state, 
both in colour-singlet and colour-octet, 
will be described in detail in a companion paper~\cite{Abreu:2022cco}. 

To determine the set of master integrals (MIs) that contribute to the process in \cref{eq:proc},
we use standard techniques for the calculation of scattering amplitudes (we refer the reader
to ref.~\cite{Abreu:2022cco} for more details). Let us nevertheless highlight here a consequence of the 
degenerate kinematics of \cref{eq:kin1,eq:kin2}. For generic four-point kinematics at two loops there are nine 
independent scalar products involving at least one loop momentum, but due to the degenerate kinematics of \cref{eq:kin1,eq:kin2} 
only seven of them are independent. This fact must be taken into account when mapping all
integrals into topologies, and we use the program \texttt{Apart} \cite{Feng:2012iq}
to implement partial-fraction relations systematically. Once all integrals have been sorted into topologies, we reduce them to MIs
using integration-by-parts (IBP) relations \cite{Chetyrkin:1981qh,Chetyrkin:1979bj}.
We perform the IBP reductions with publicly available codes such as 
$\tt{FIRE}$~\cite{Smirnov:2019qkx},  $\tt{LiteRed}$~\cite{Lee:2013mka} 
and $\tt{Kira}$~\cite{Klappert:2020nbg}. 

\begin{figure}
\captionsetup[subfloat]{labelformat=simple}
\renewcommand*\thesubfigure{$\masterNew{\arabic{subfigure}}$} 
\centering
\subfloat[]{\includegraphics[width =3 cm]{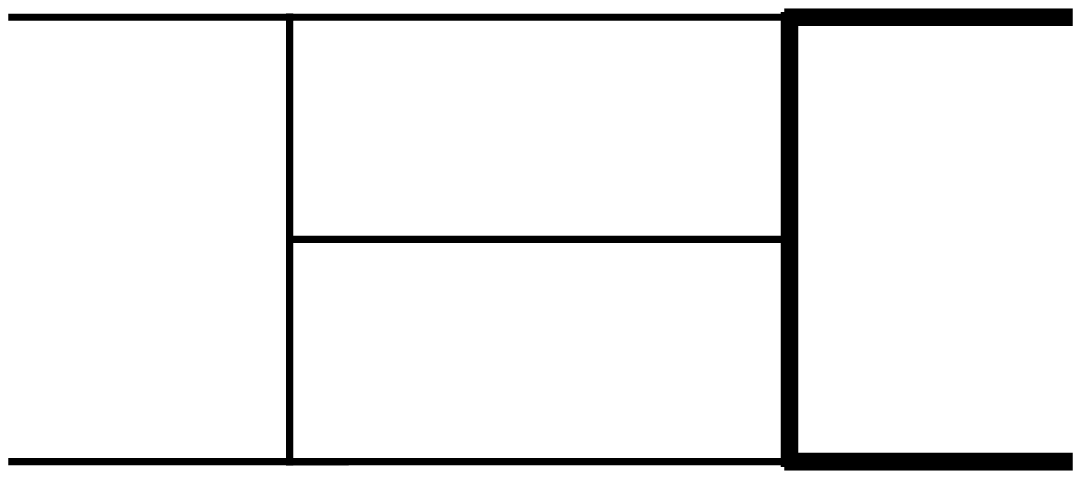}}\qquad
\subfloat[]{\includegraphics[width =3 cm]{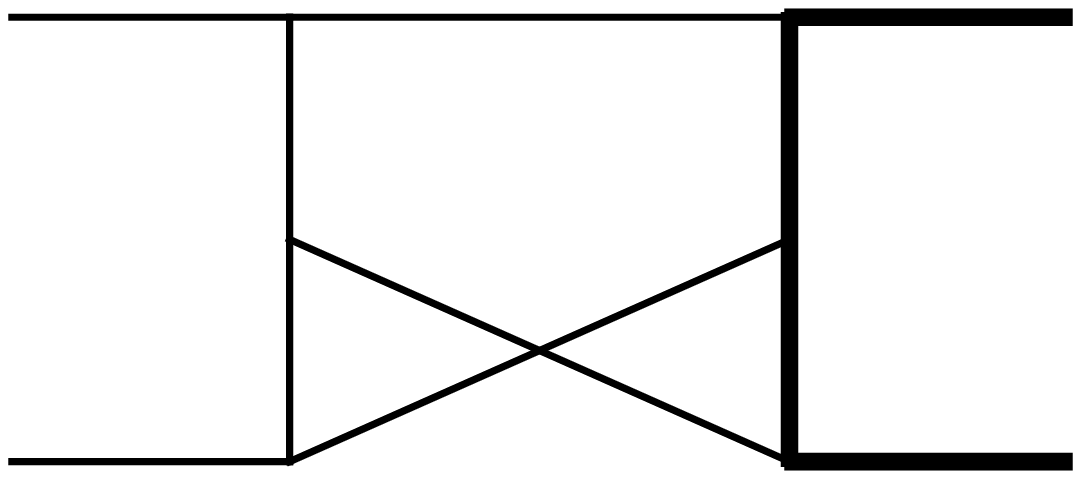}}\qquad
\subfloat[]{\includegraphics[width =2.6 cm]{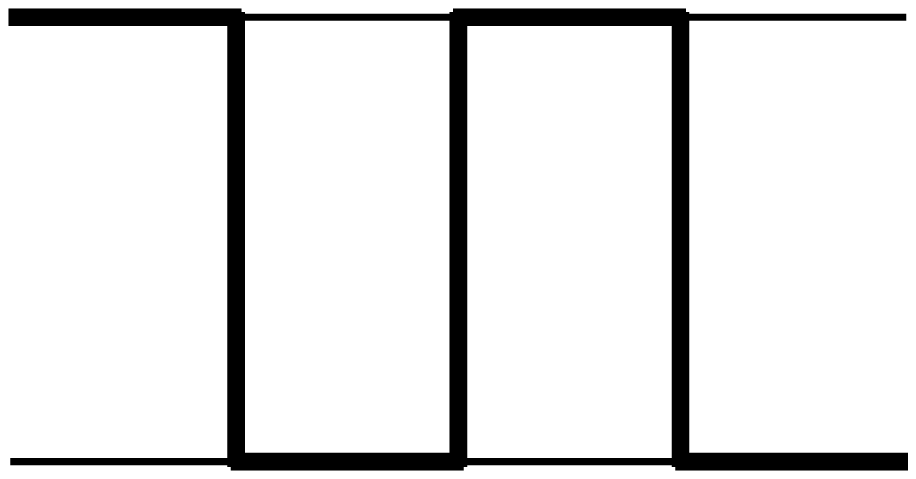}}\qquad
\subfloat[]{\includegraphics[width =3.1 cm]{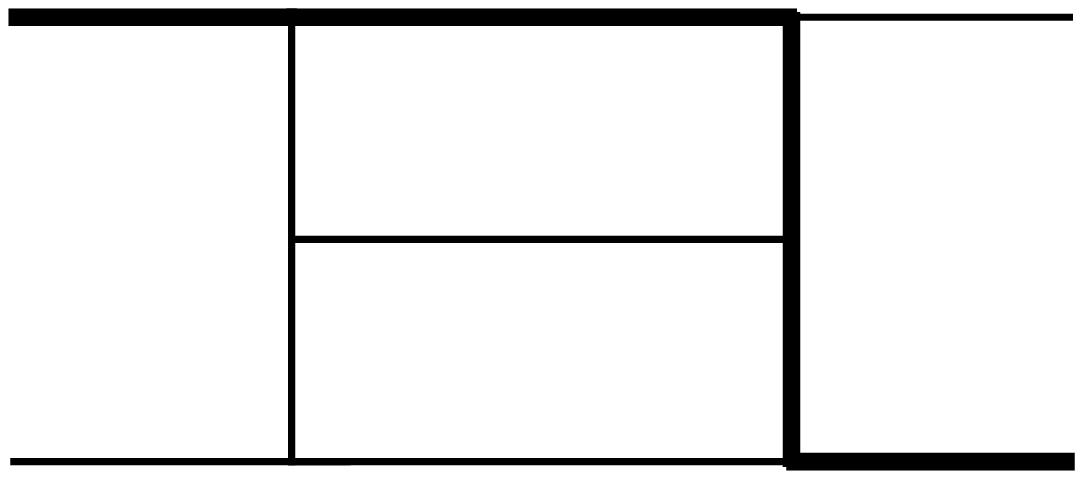}}\qquad
\\[10pt]
\subfloat[]{\includegraphics[width =1.8 cm]{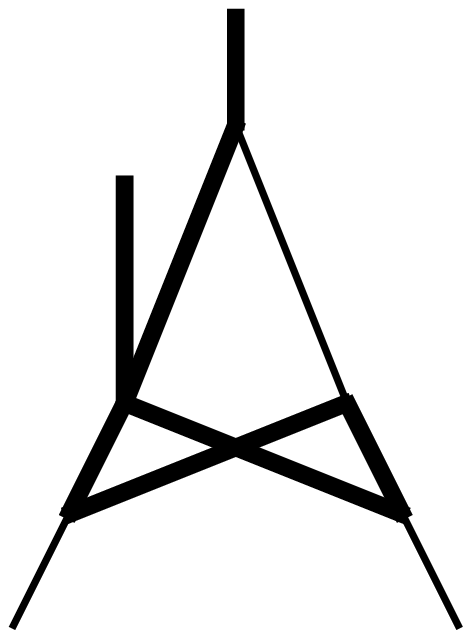}}\quad
\subfloat[]{\includegraphics[width =1.8 cm]{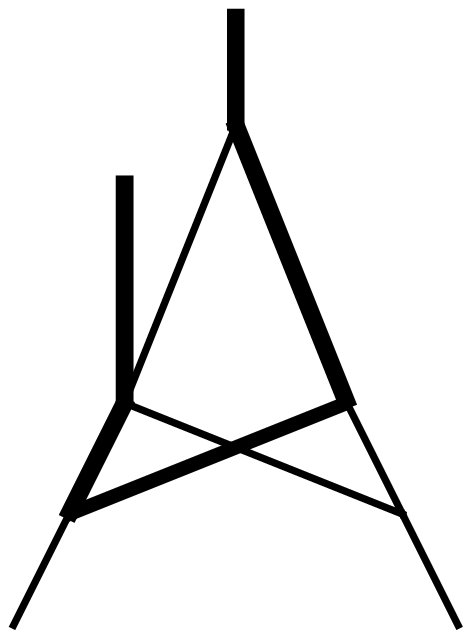}}\quad
\subfloat[]{\includegraphics[width =2.2 cm]{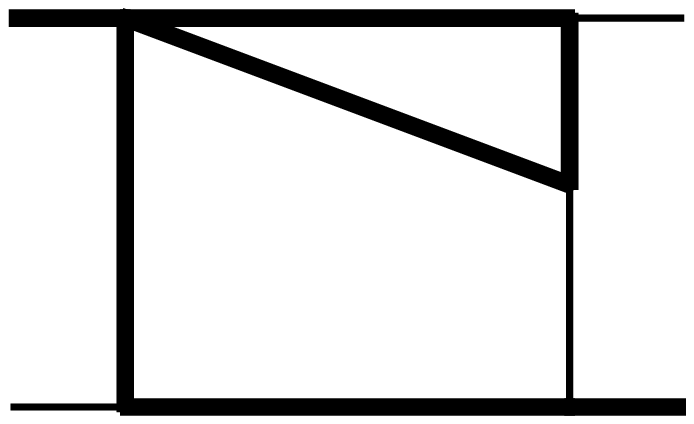}}\quad
\subfloat[]{\includegraphics[width =2.2 cm]{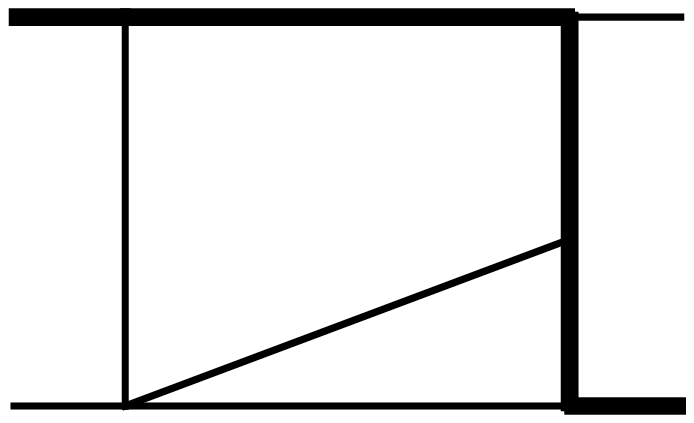}}\quad
\subfloat[]{\includegraphics[width =2.2 cm]{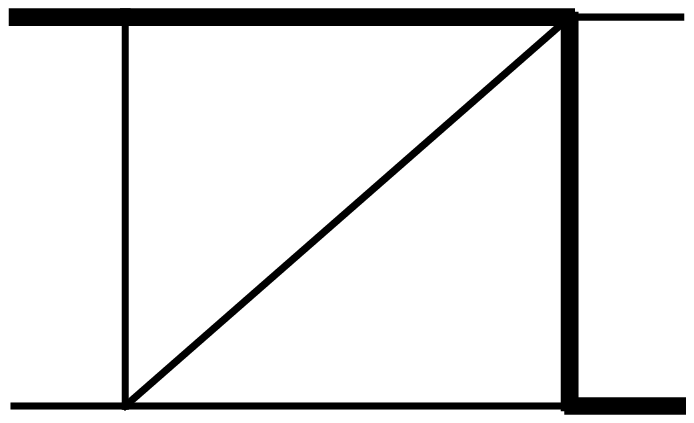}}\quad
\subfloat[]{\includegraphics[width =2.2 cm]{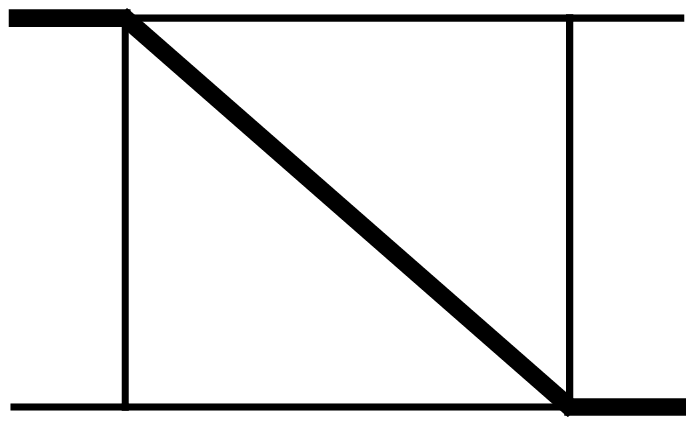}}\quad
\\[10pt]
\subfloat[]{\includegraphics[width =2.2 cm]{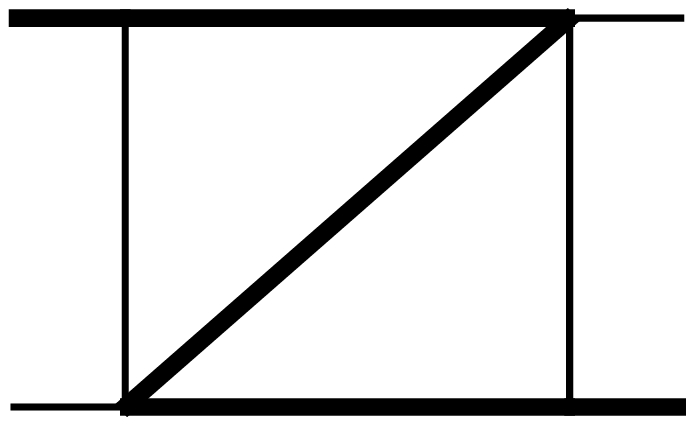}}\quad
\subfloat[]{\includegraphics[width =2.2 cm]{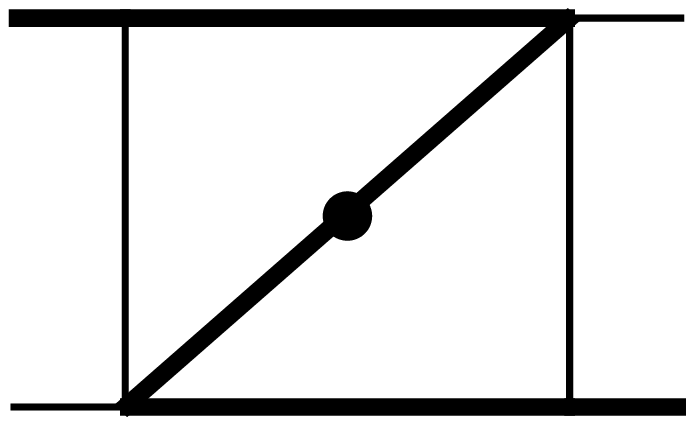}}\quad
\subfloat[]{\includegraphics[width =2.2 cm]{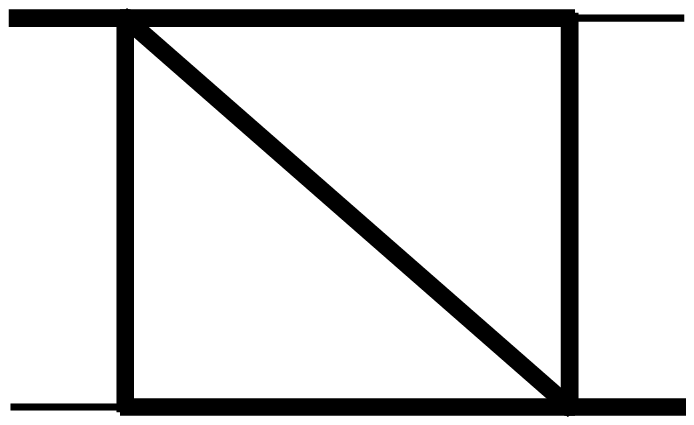}}\quad
\subfloat[]{\includegraphics[width =2.2 cm]{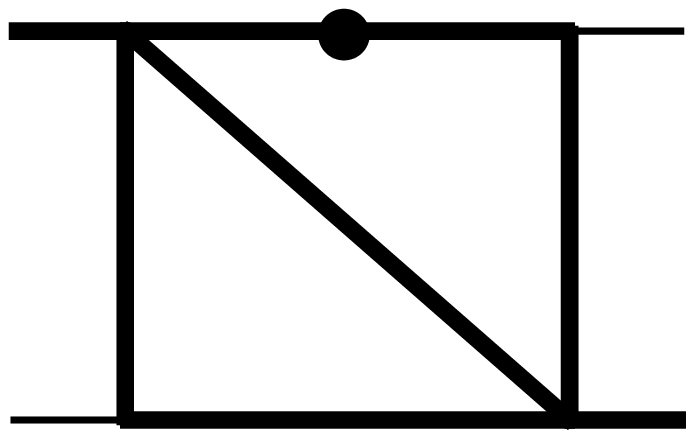}}\quad
\subfloat[]{\includegraphics[width =2.2 cm]{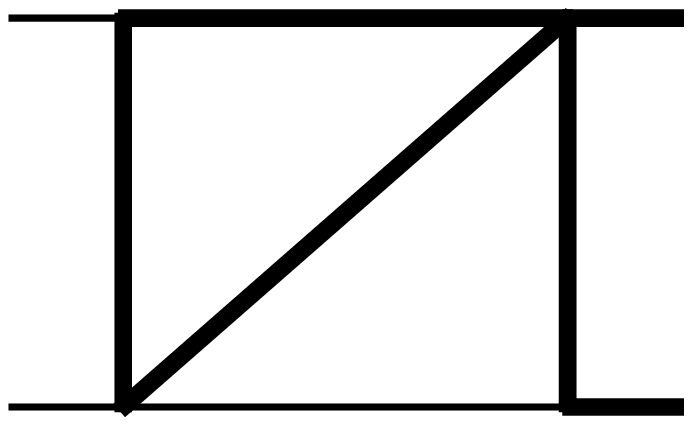}}\quad
\subfloat[]{\includegraphics[width =2.2 cm]{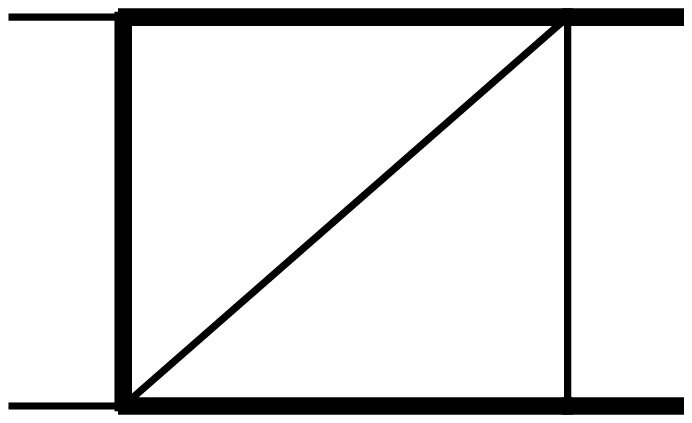}}\quad
\\[10pt]
\subfloat[]{\includegraphics[width =2.2 cm]{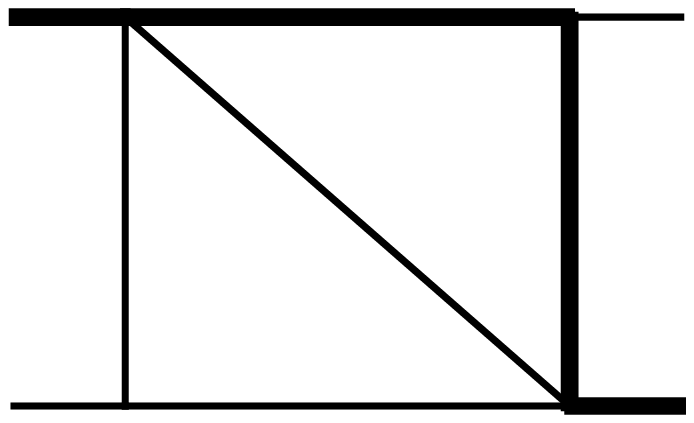}}\quad
\subfloat[]{\includegraphics[width =2.2 cm]{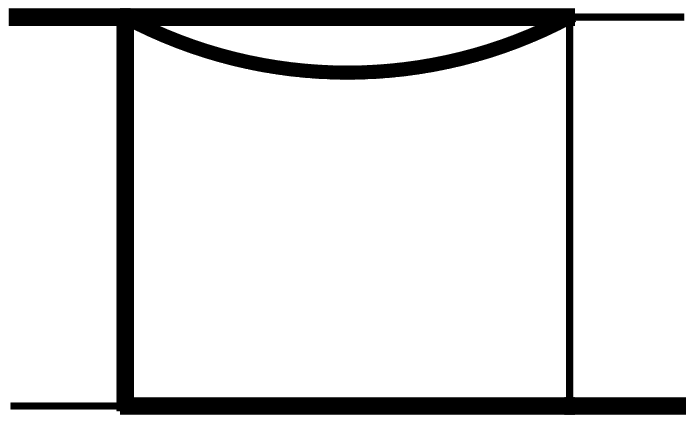}}\quad
\subfloat[]{\includegraphics[width =2.2 cm]{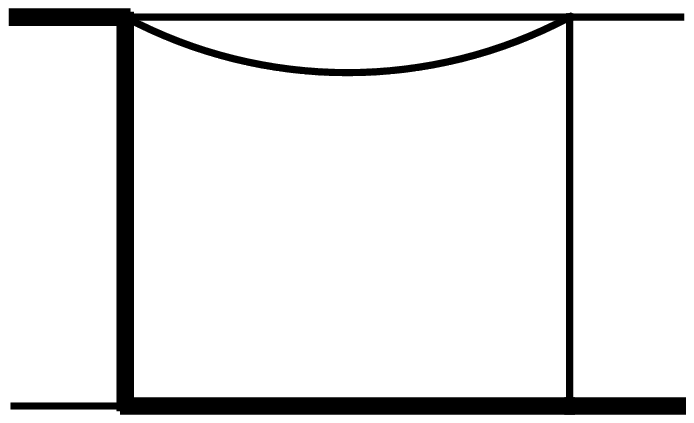}}
\caption{Four-point non-factorisable two-loop integrals. Thin lines are massless,
and thick lines have mass $m_Q$. A dot on a propagator means that
the propagator is squared.} 
\label{fig:four-point}
\end{figure}

\begin{figure}
\setcounter{subfigure}{19}
\captionsetup[subfloat]{labelformat=simple}
\renewcommand*\thesubfigure{$\masterNew{\arabic{subfigure}}$} 
\centering
\vspace*{1.3cm}
\subfloat[]{\includegraphics[width =1.8 cm]{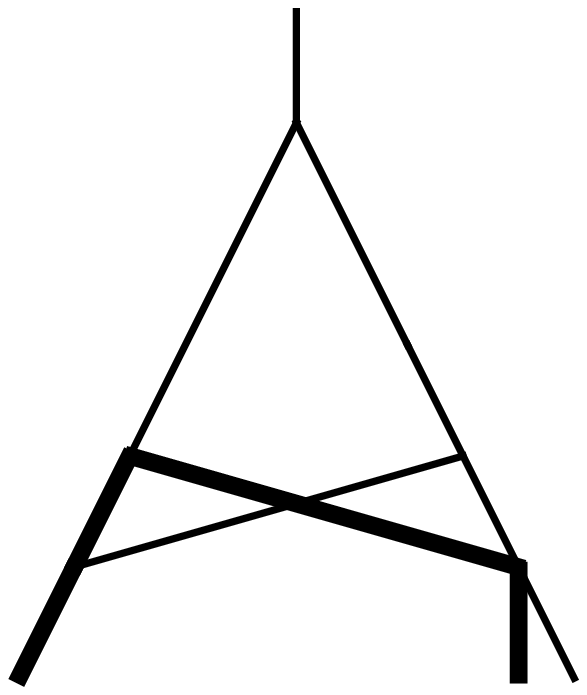}}\quad
\subfloat[]{\includegraphics[width =1.8 cm]{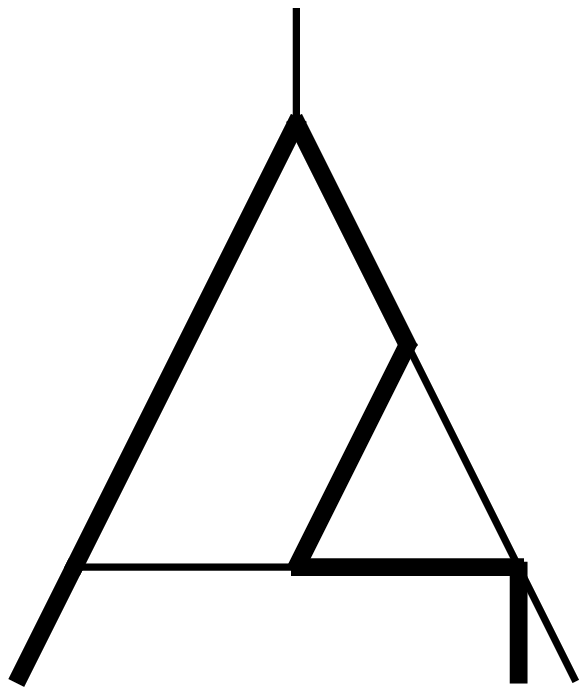}}\quad
\subfloat[]{\includegraphics[width =1.8 cm]{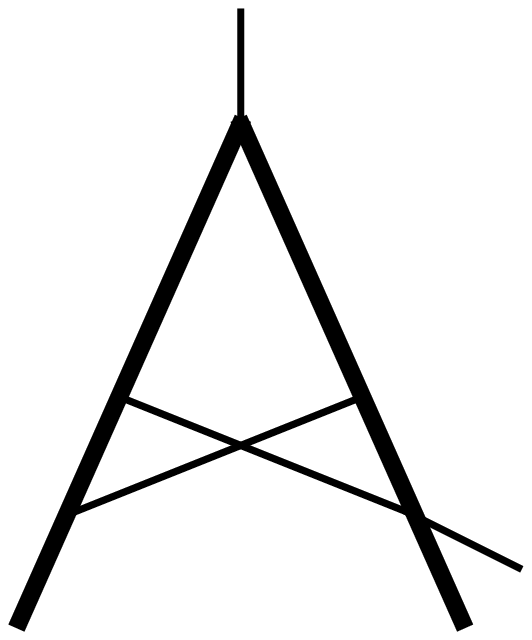}}\quad
\subfloat[]{\includegraphics[width =1.8 cm]{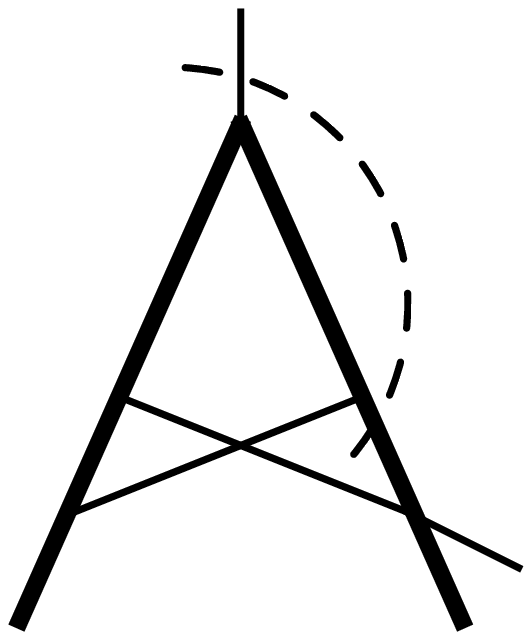}}\quad
\subfloat[]{\includegraphics[width =1.8 cm]{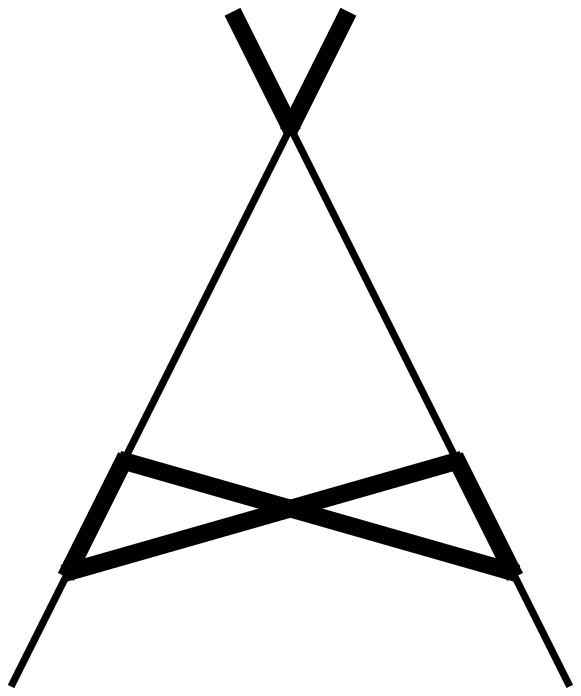}}\quad
\subfloat[]{\includegraphics[width =1.8 cm]{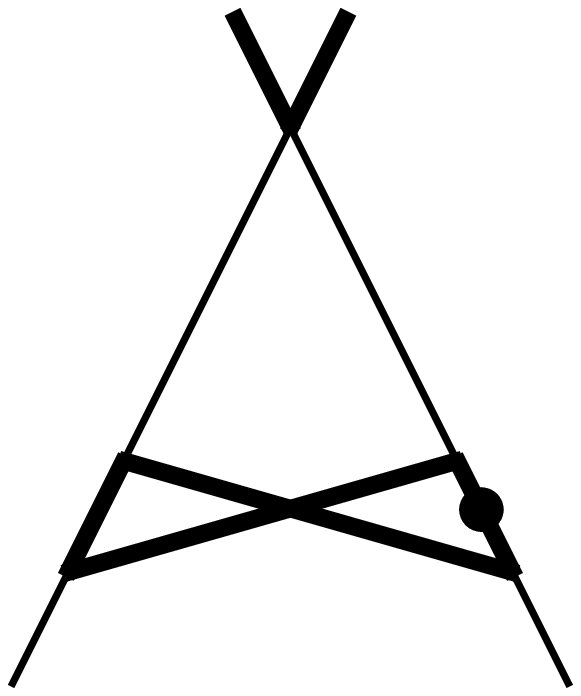}}\quad
\subfloat[]{\includegraphics[width =1.8 cm]{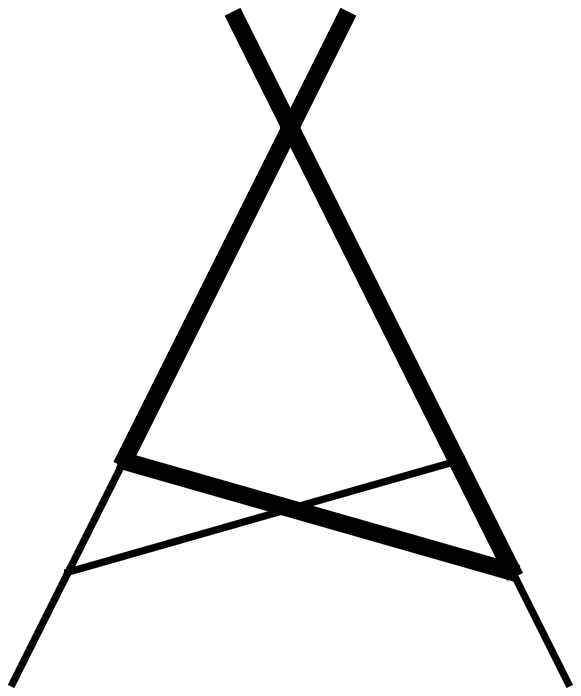}}\quad
\\[12pt]
\subfloat[]{\includegraphics[width =1.8 cm]{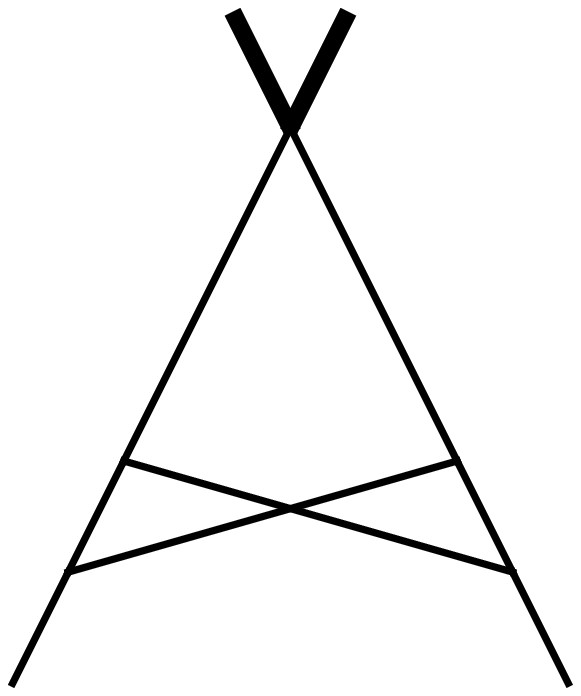}}\quad
\subfloat[]{\includegraphics[width =1.8 cm]{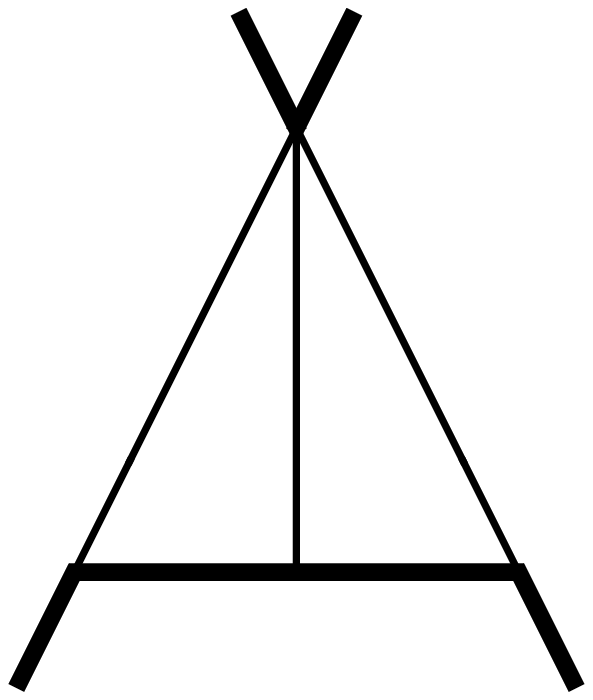}}\quad
\subfloat[]{\includegraphics[width =1.8 cm]{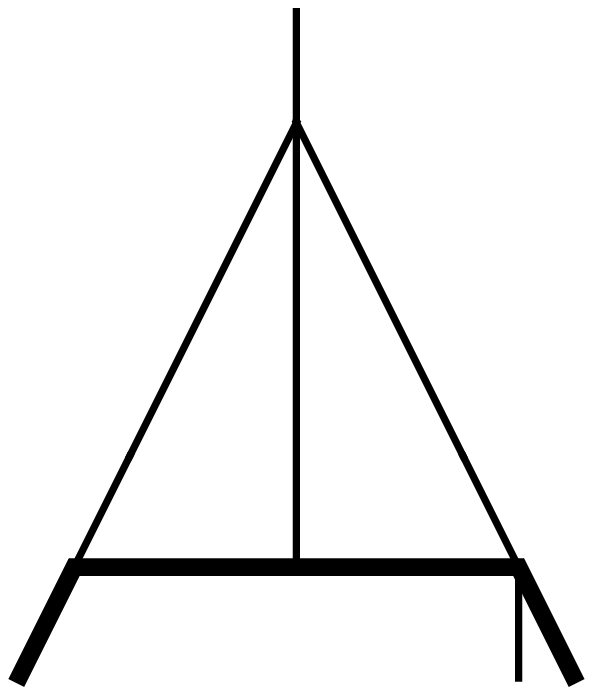}}\quad
\subfloat[]{\includegraphics[width =1.8 cm]{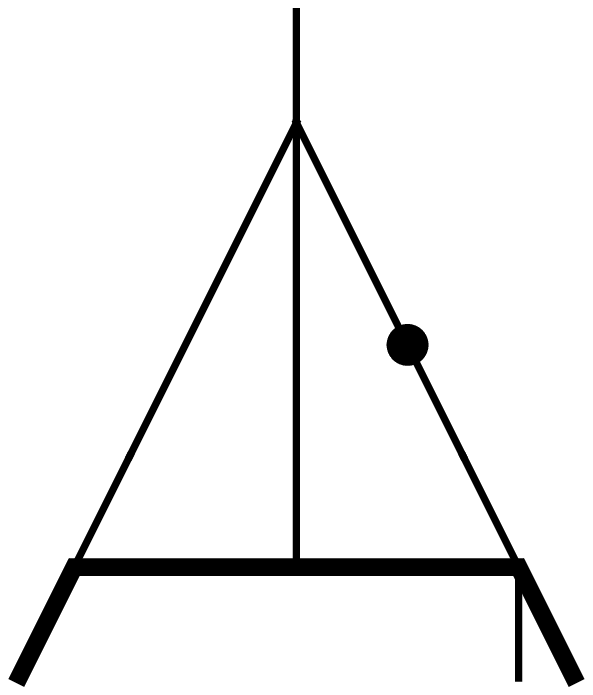}}\quad
\subfloat[]{\includegraphics[width =1.8 cm]{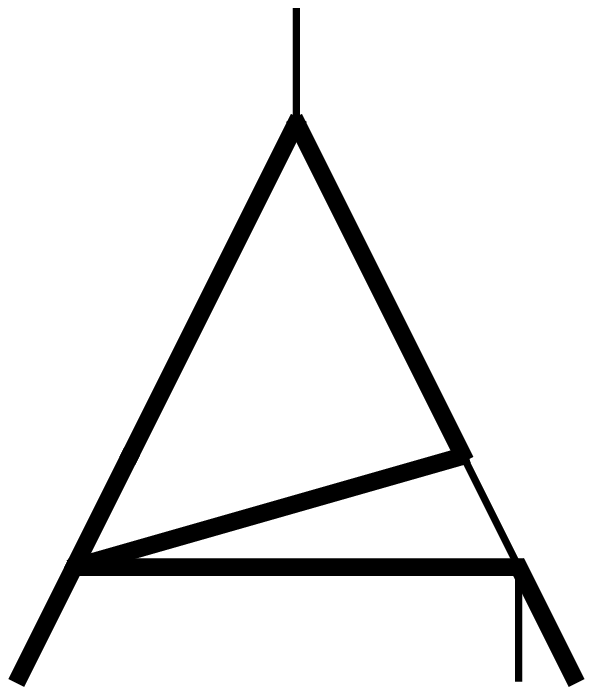}}\quad
\subfloat[]{\includegraphics[width =1.8 cm]{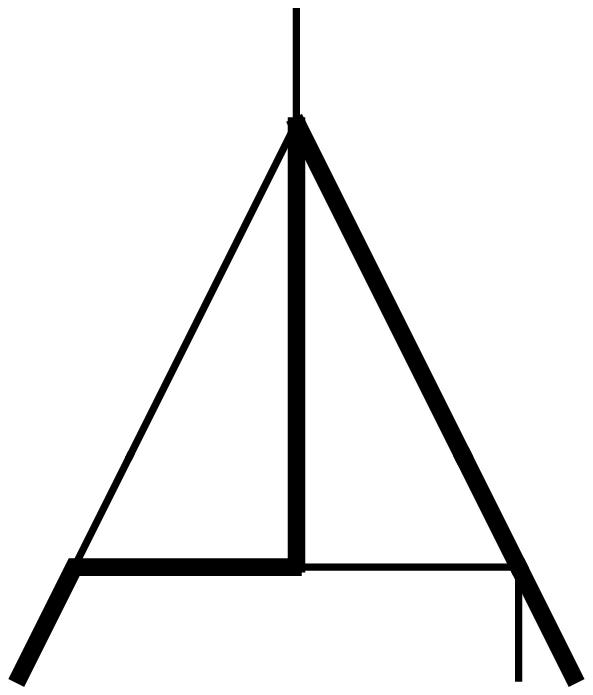}}\quad
\subfloat[]{\includegraphics[width =1.8 cm]{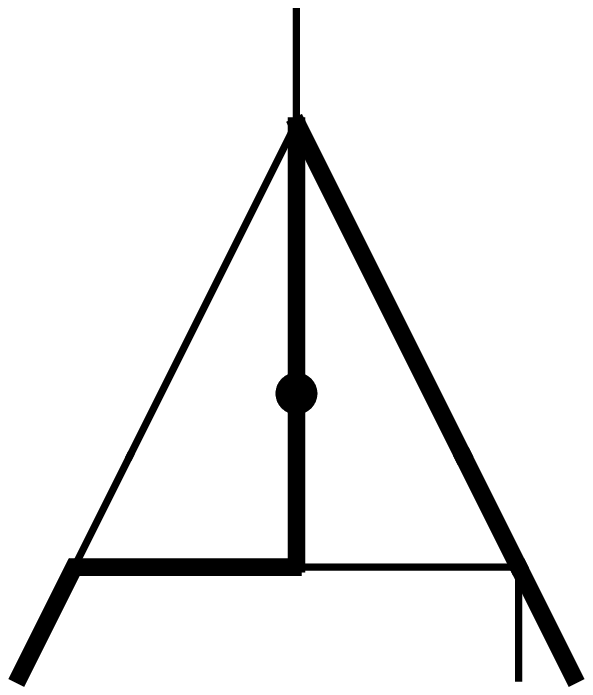}}\quad
\\[12pt]
\subfloat[]{\includegraphics[width =1.8 cm]{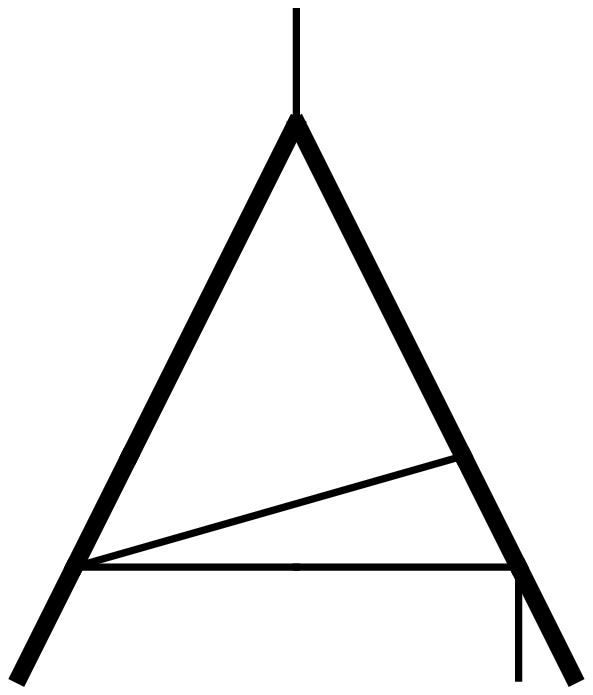}}\quad
\subfloat[]{\includegraphics[width =1.8 cm]{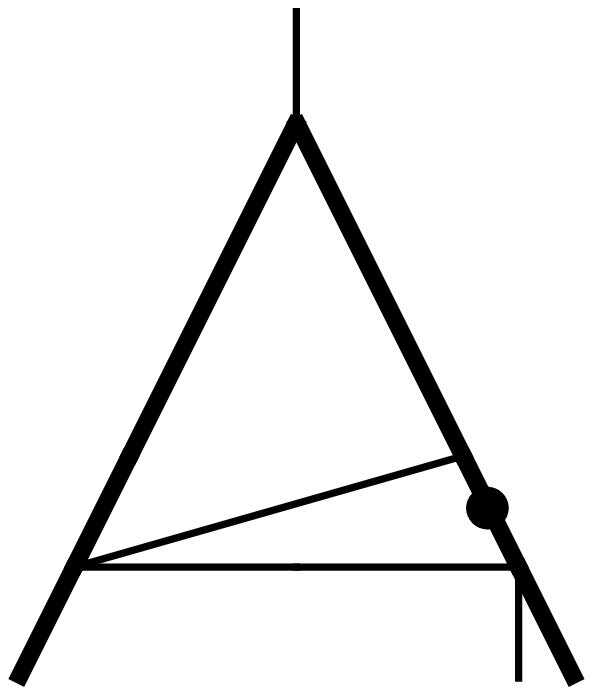}}\quad
\subfloat[]{\includegraphics[width =1.8 cm]{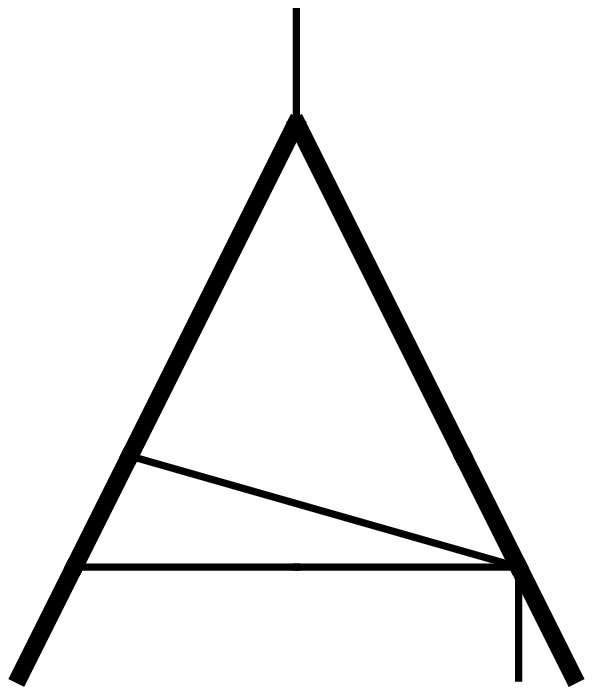}}\quad
\subfloat[]{\includegraphics[width =1.8 cm]{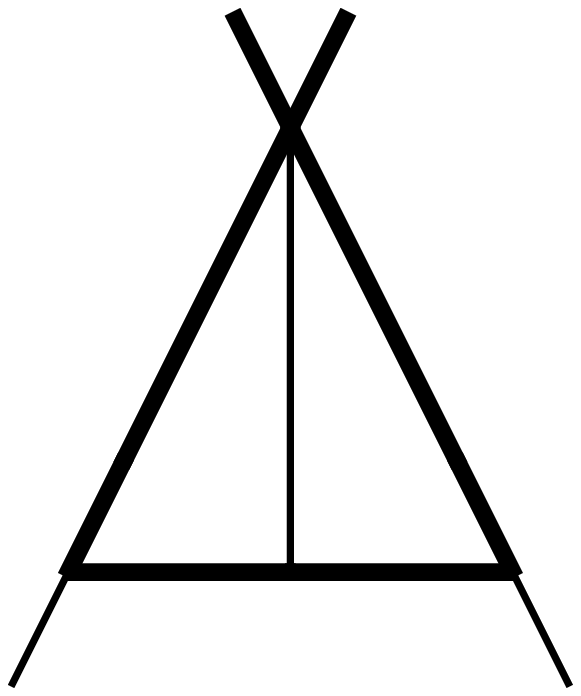}}\quad
\subfloat[]{\includegraphics[width =1.8 cm]{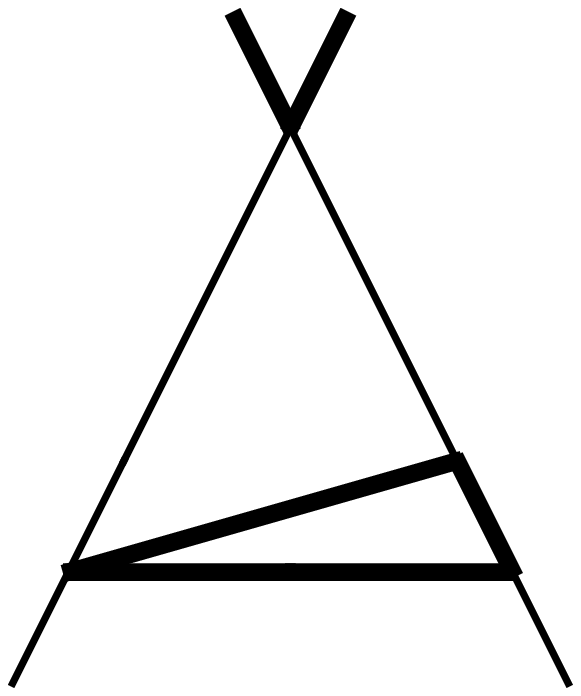}}\quad
\subfloat[]{\includegraphics[width =1.8 cm]{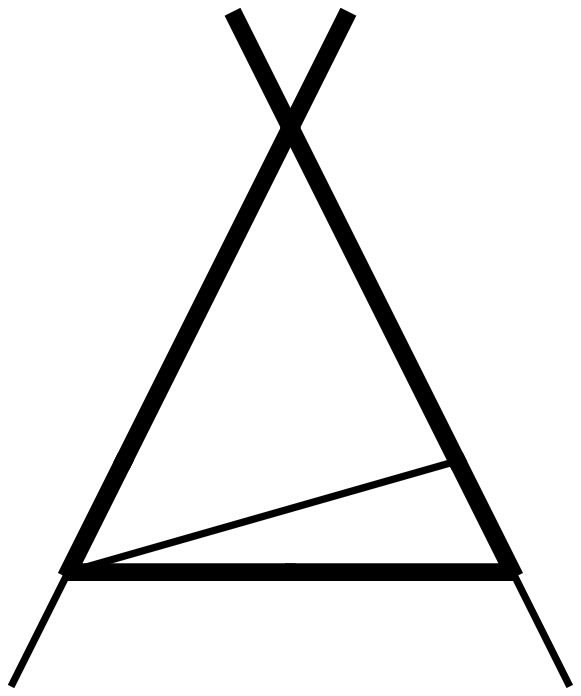}}\quad
\subfloat[]{\includegraphics[width =1.8 cm]{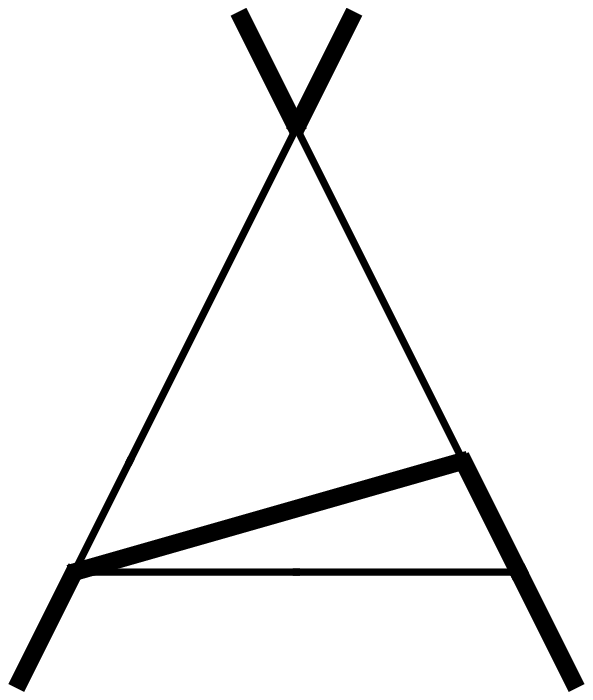}}\quad
\\[12pt]
\subfloat[]{\includegraphics[width =1.8 cm]{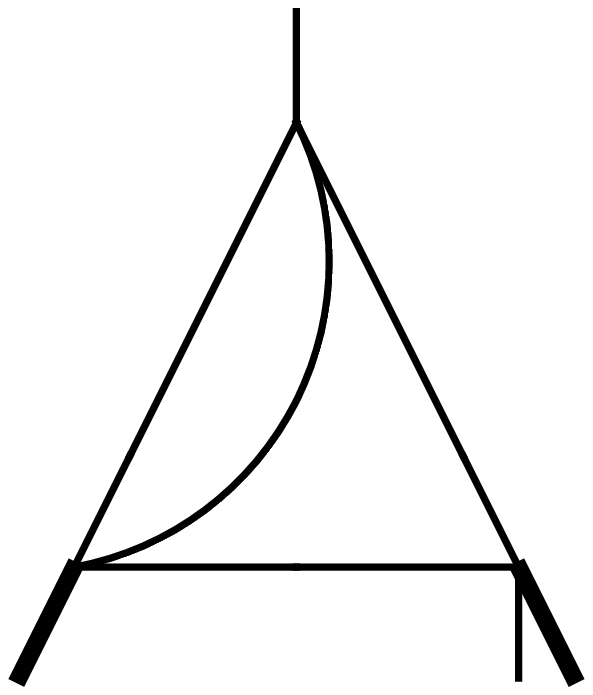}}\quad
\subfloat[]{\includegraphics[width =1.8 cm]{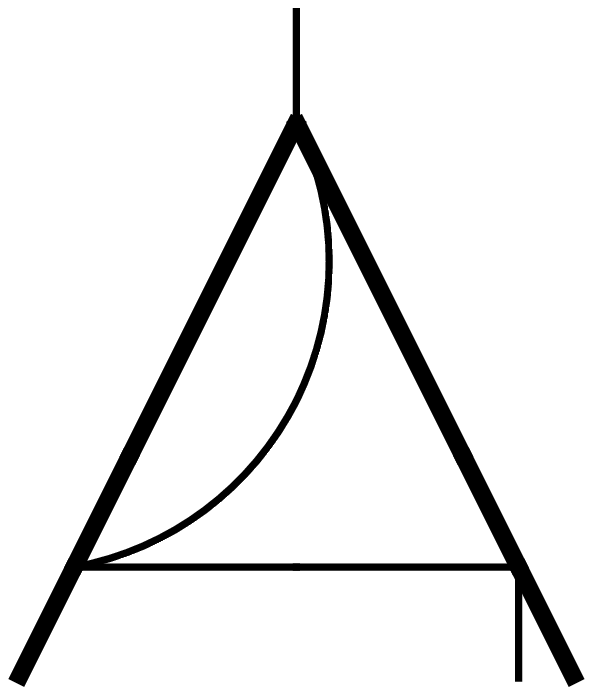}}\quad
\subfloat[]{\includegraphics[width =1.8 cm]{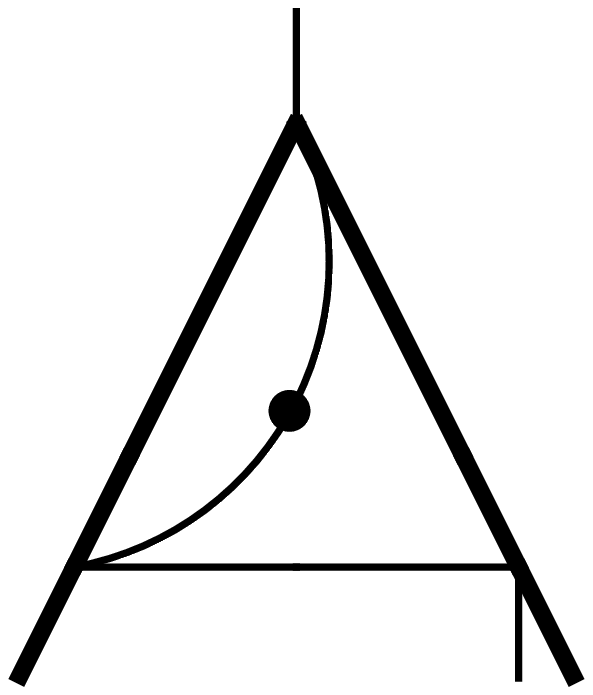}}\quad
\subfloat[]{\includegraphics[width =1.8 cm]{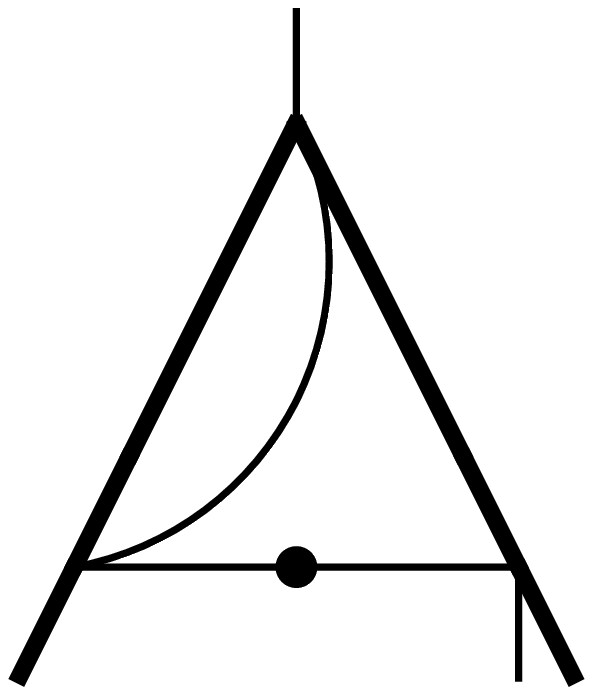}}\quad
\subfloat[]{\includegraphics[width =1.8 cm]{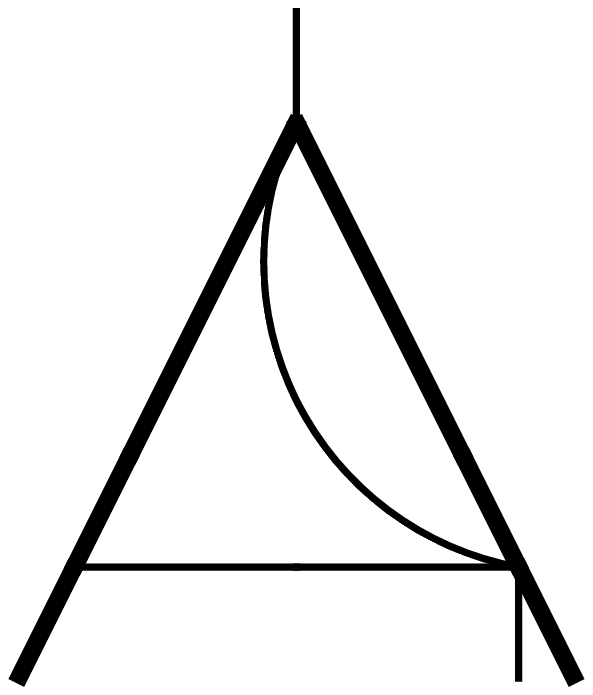}}\quad
\subfloat[]{\includegraphics[width =1.8 cm]{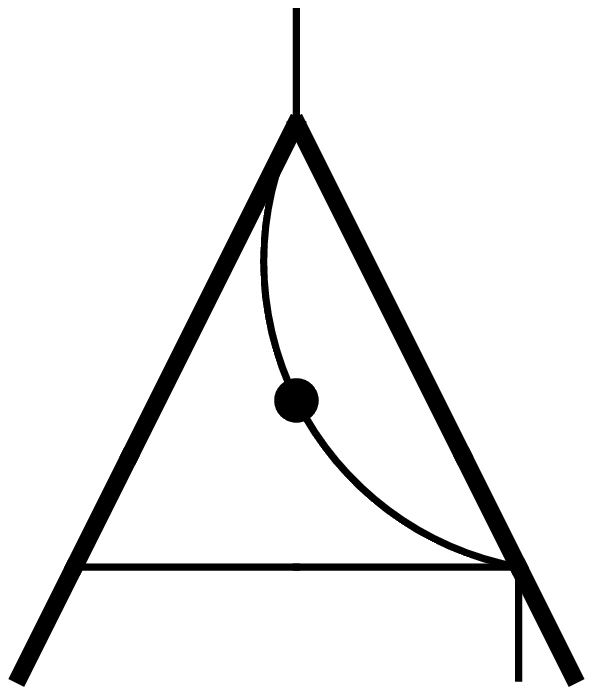}}\quad
\subfloat[]{\includegraphics[width =1.8 cm]{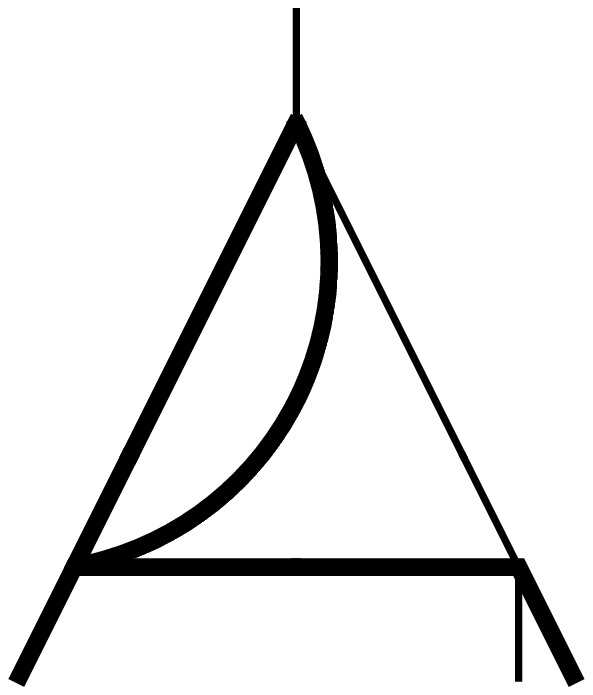}}\quad
\\[12pt]
\subfloat[]{\includegraphics[width =1.8 cm]{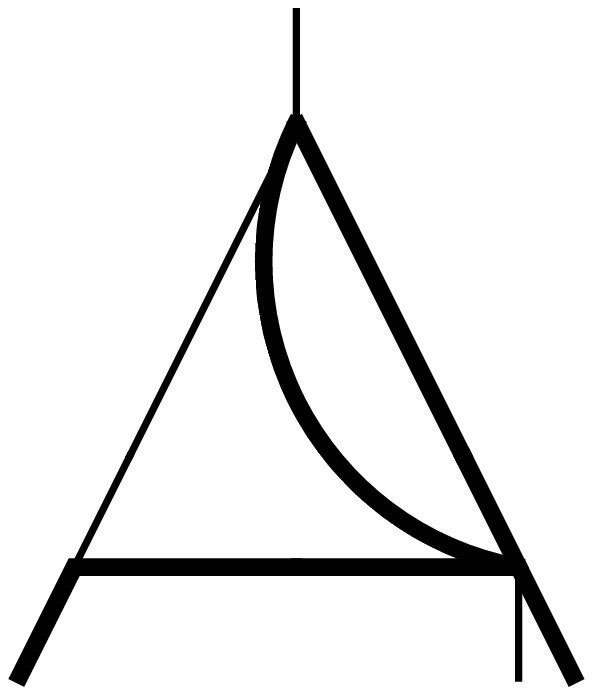}}\quad
\subfloat[]{\includegraphics[width =1.8 cm]{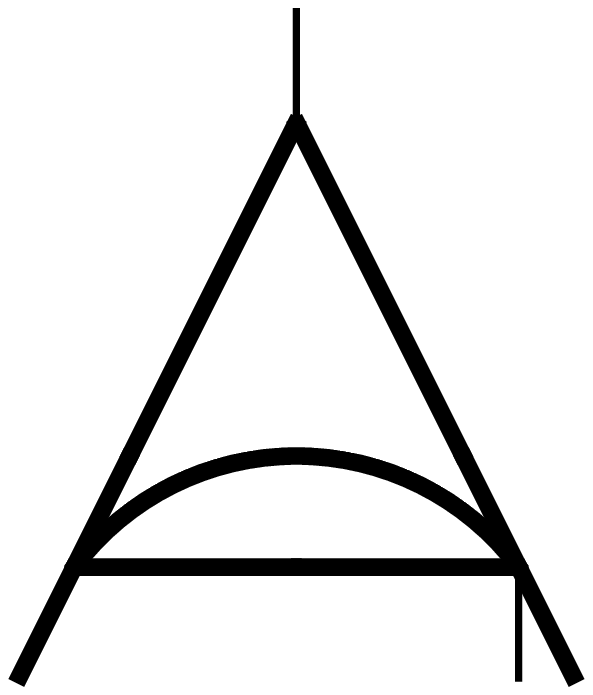}}\quad
\subfloat[]{\includegraphics[width =1.8 cm]{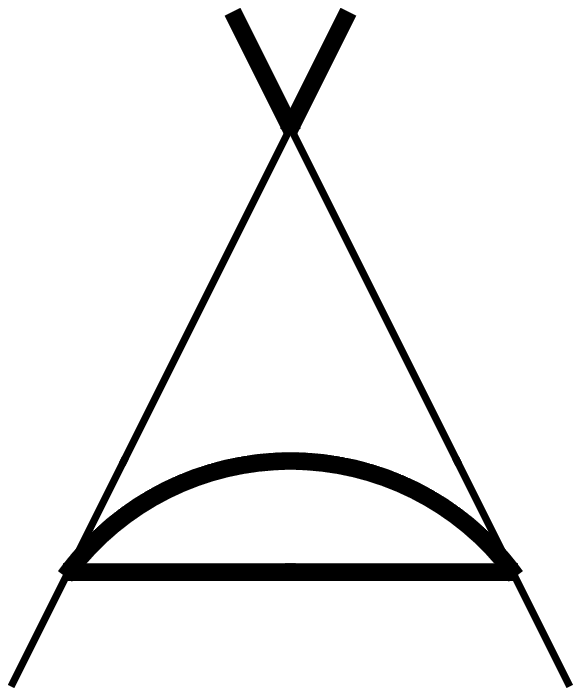}}\quad
\subfloat[]{\includegraphics[width =1.8 cm]{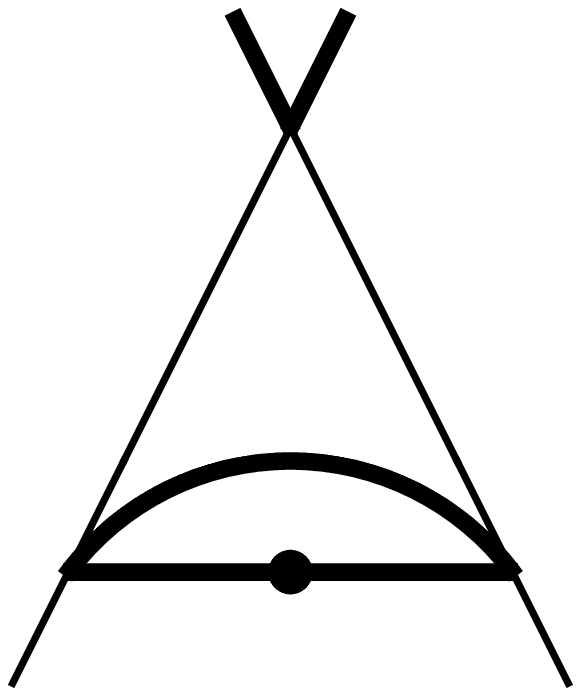}}\quad
\subfloat[]{\includegraphics[width =1.8 cm]{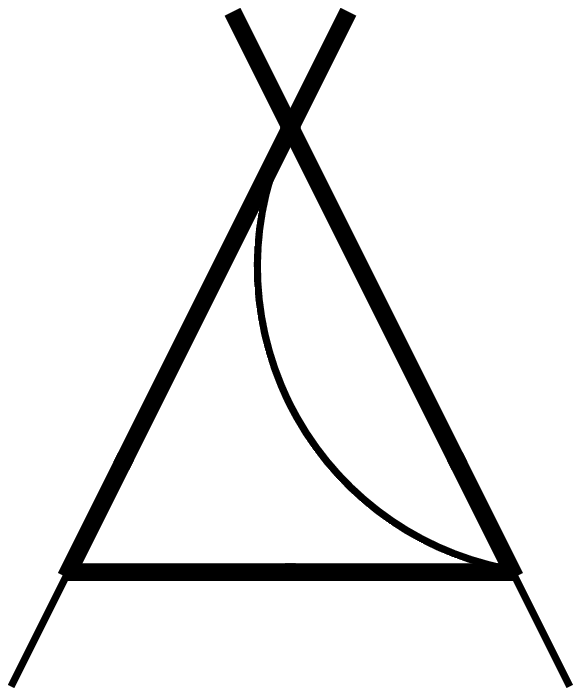}}\quad
\subfloat[]{\includegraphics[width =1.8 cm]{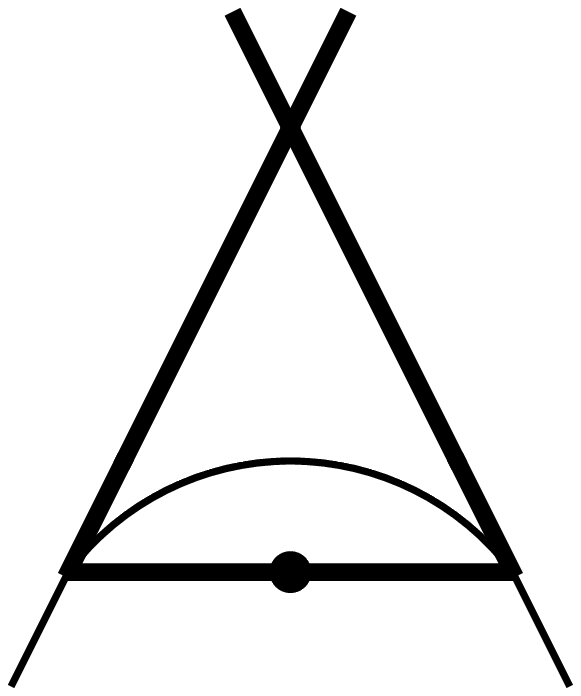}}\quad
\subfloat[]{\includegraphics[width =1.8 cm]{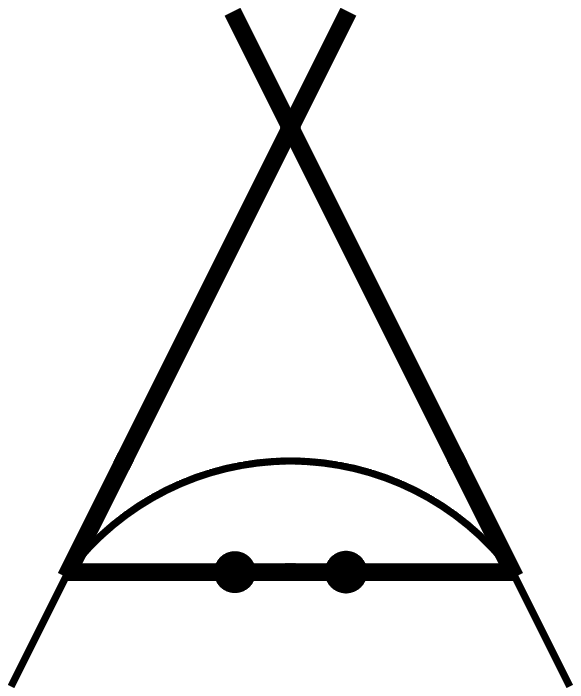}}\quad
\\[12pt]
\subfloat[]{\includegraphics[width =1.8 cm]{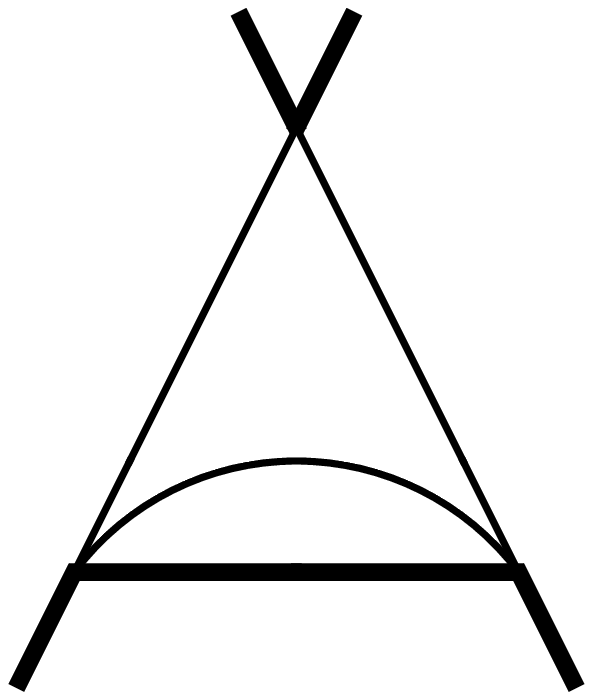}}\quad
\subfloat[]{\includegraphics[width =1.8 cm]{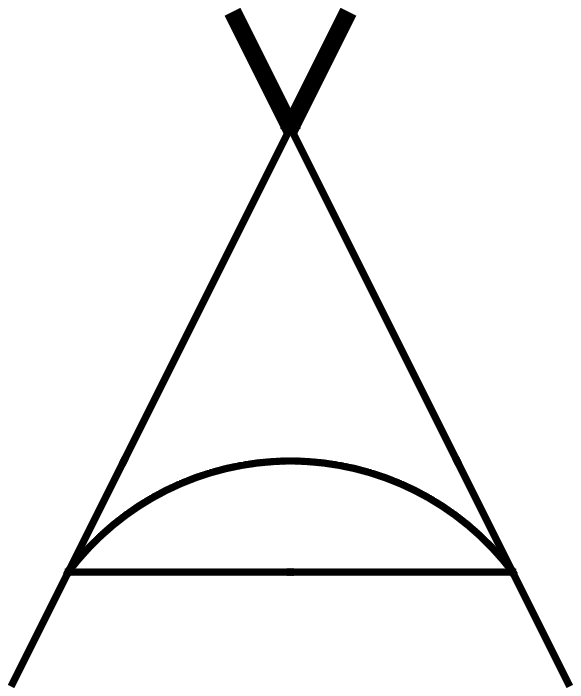}}\quad

\caption{Three-point non-factorisable two-loop integrals. Thin lines are massless,
and thick lines have mass $m_Q$. A dot on a propagator means that
the propagator is squared. The dashed line on $\masterNew{23}$ denotes a numerator,
see \cref{sec:mis} for the explicit definition.} \label{fig:three-point}
\end{figure}

\begin{figure}
\setcounter{subfigure}{56}
\captionsetup[subfloat]{labelformat=simple}
\renewcommand*\thesubfigure{$\masterNew{\arabic{subfigure}}$} 
\centering
\subfloat[]{\includegraphics[width =2.1 cm]{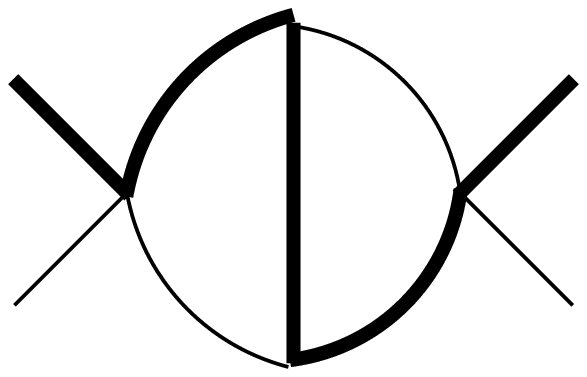}}\quad
\subfloat[]{\includegraphics[width =2.1 cm]{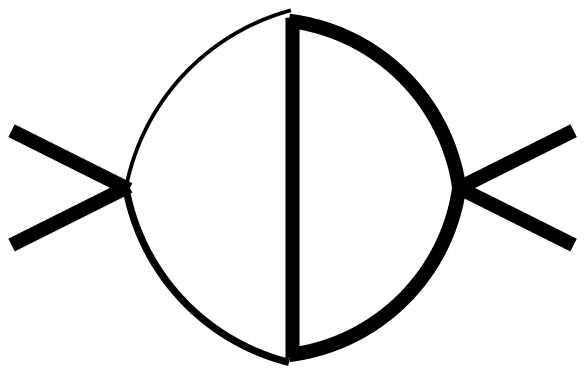}}\quad
\subfloat[]{\includegraphics[width =2.1 cm]{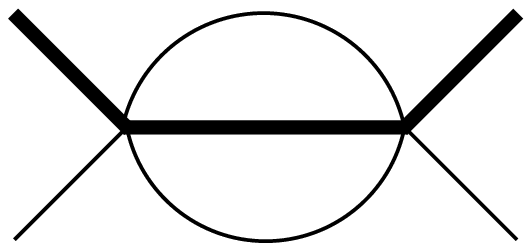}}\quad
\subfloat[]{\includegraphics[width =2.1 cm]{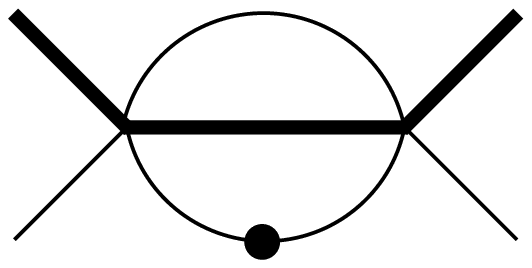}}\quad
\subfloat[]{\includegraphics[width =2.1 cm]{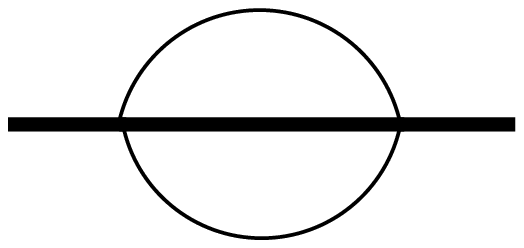}}\quad
\subfloat[]{\includegraphics[width =2.1 cm]{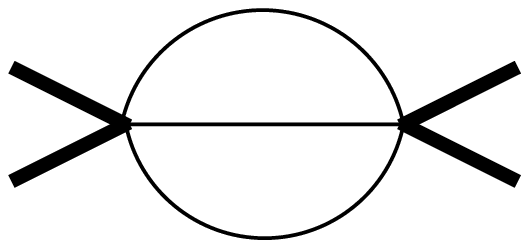}}\quad
\\[10pt]
\subfloat[]{\includegraphics[width =2.1 cm]{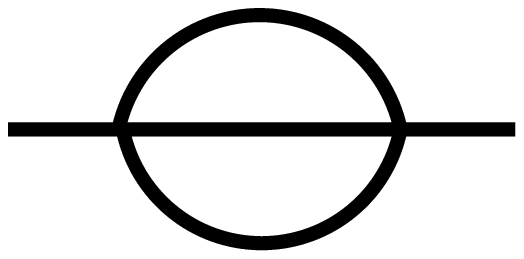}}\quad
\subfloat[]{\includegraphics[width =2.1 cm]{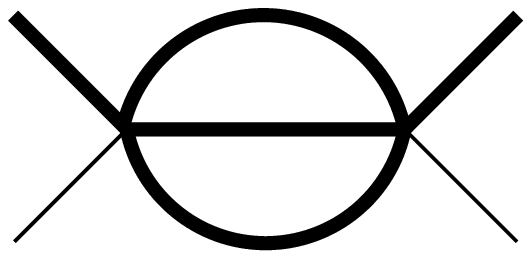}}\quad
\subfloat[]{\includegraphics[width =2.1 cm]{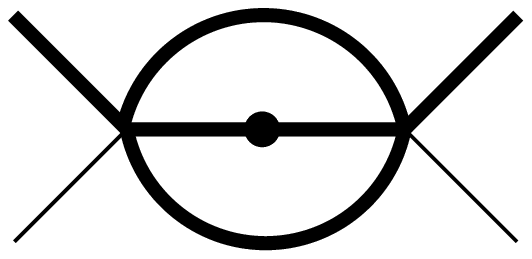}}\quad
\subfloat[]{\includegraphics[width =2.1 cm]{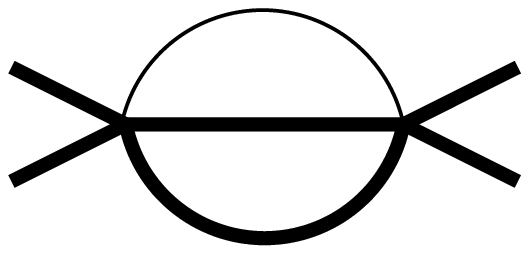}}\quad
\caption{Two-point non-factorisable two-loop integrals. Thin lines are massless,
and thick lines have mass $m_Q$. A dot on a propagator means that
the propagator is squared.} \label{fig:two-point}
\end{figure}

\begin{figure}
\setcounter{subfigure}{66}
\captionsetup[subfloat]{labelformat=simple}
\renewcommand*\thesubfigure{$\masterNew{\arabic{subfigure}}$} 
\centering
\subfloat[]{\includegraphics[width =2 cm]{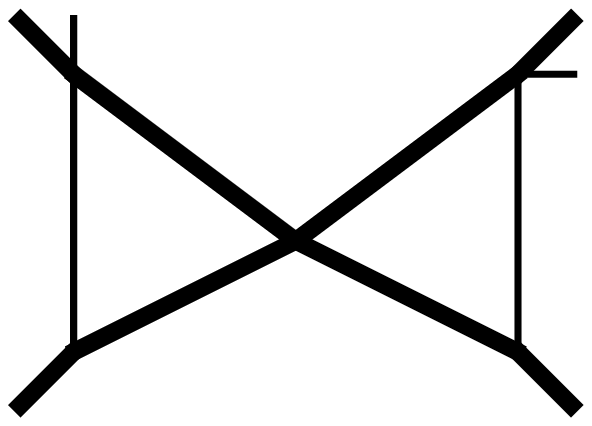}}\quad
\subfloat[]{\includegraphics[width =2 cm]{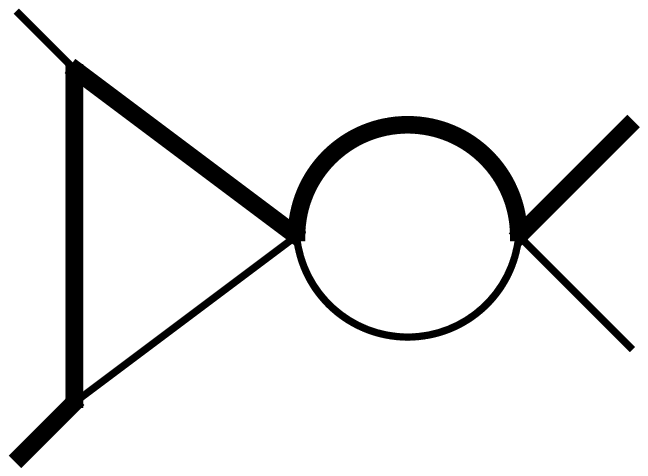}}\quad
\subfloat[]{\includegraphics[width =1.6 cm]{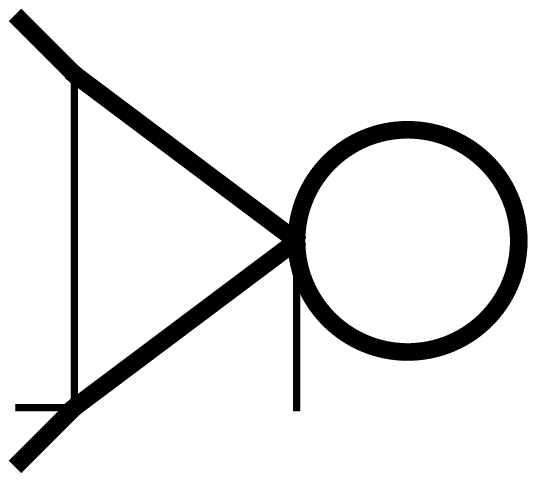}}\quad
\subfloat[]{\includegraphics[width =2 cm]{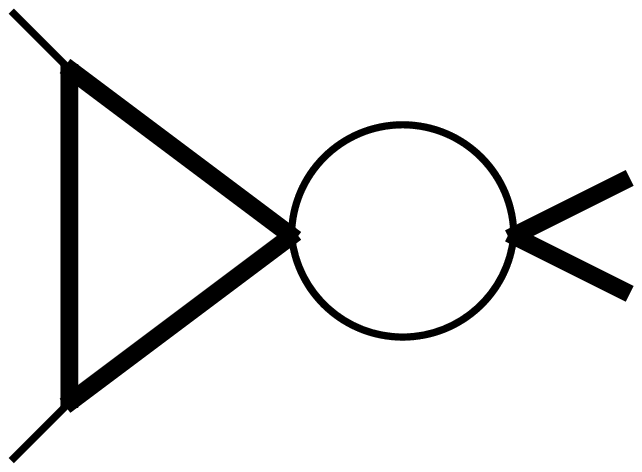}}\quad
\subfloat[]{\includegraphics[width =1.6 cm]{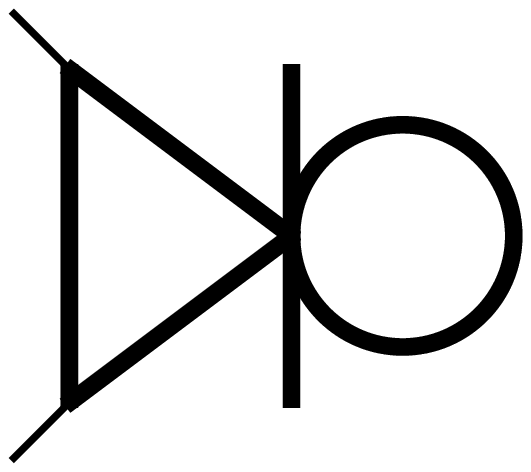}}\quad
\subfloat[]{\raisebox{0.3 cm}{\includegraphics[width =2.6 cm]{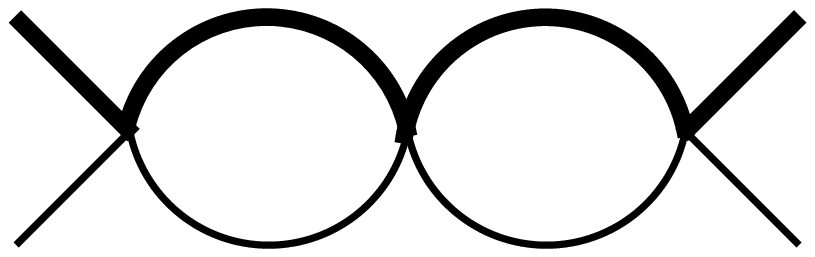}}}\quad
\\[10pt]
\subfloat[]{\raisebox{0.1 cm}{\includegraphics[width =2.4 cm]{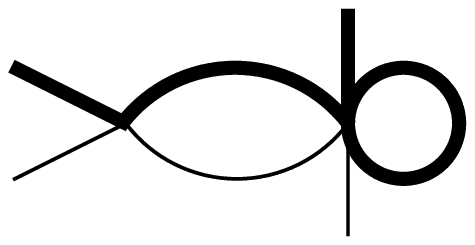}}}\quad
\subfloat[]{\raisebox{0.1 cm}{\includegraphics[width =2.4cm]{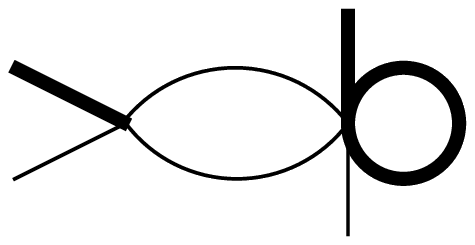}}}\quad
\subfloat[]{\raisebox{0.1cm}{\includegraphics[width =3 cm]{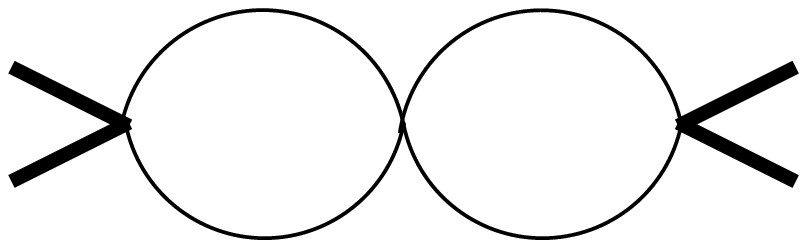}}}\quad
\subfloat[]{\includegraphics[width =2 cm]{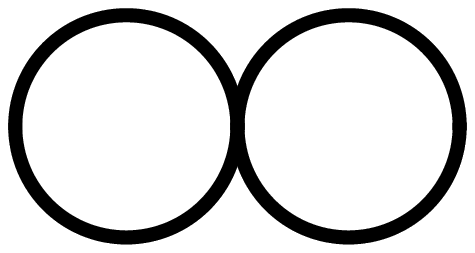}}
\caption{Factorisable two-loop integrals. Thin lines are massless,
and thick lines have mass $m_Q$.} \label{fig:fact}
\end{figure}

After IBP reduction, we find 76 master integrals that contribute to the two-loop
corrections to the process in \cref{eq:proc}. There are 19 four-point integrals
(see \cref{fig:four-point}), 37 three-point integrals
(see \cref{fig:three-point}), 10 two-point integrals
(see \cref{fig:two-point}) and 10
integrals that factorise into a product of one-loop integrals (see \cref{fig:fact}).
The factorised master integrals are trivial to evaluate, and we will not discuss them further.
Several of the genuine two-loop master integrals have been considered previously in the
literature \cite{Gehrmann:1999as,Bonciani:2003hc,Anastasiou:2006hc,Beerli:2008zz,Bonciani:2009nb,
Gehrmann:2010ue,Chen:2017xqd,vonManteuffel:2017hms,Chen:2017soz,DiVita:2018nnh,Chen:2019zoy,
Gerlach:2019kfo,Becchetti:2019tjy,Mandal:2022vju}, often in kinematic configurations more generic than
those of eqs.~\eqref{eq:kin1} and~\eqref{eq:kin2}.

Since there are 7 independent scalar products involving at least one loop momentum, all master integrals can be embedded in topologies involving at most 7 propagators, i.e., they can be written as
\begin{equation}\label{eq:genInt}
m_I(a_1,a_2,a_3,a_4,a_5,a_6,a_7;m_Q^2) = \int \mathcal{D}^{4-2\epsilon} q_1 \mathcal{D}^{4-2\epsilon} q_2 
\frac{1}{D_{1}^{a_1}\cdots D_{7}^{a_7}}\,,
\end{equation}
where the $D_i$ denote inverse propagators and the $a_i$ take integer values.
We refer the reader to \cref{sec:mis} for the explicit
representation of each master integral in the form of \cref{eq:genInt}.
We consider the integrals in dimensional regularisation in $d=4-2\epsilon$ dimensions,
and we normalise the integration measure as
\begin{equation}
\mathcal{D}^{4-2\epsilon} q_k = \frac{\text{d}^{4-2\epsilon} q_k}{i \pi^{2-\epsilon}} e^{\epsilon \gamma_E},
\end{equation}
where $\gamma_E=-\Gamma'(1)$ is the Euler-Mascheroni constant. 
We note that, as a consequence of the degenerate kinematics in \cref{eq:kin2}, 
all master integrals are single-scale integrals whose explicit dependence on $m_Q^2$ can be determined 
from dimensional analysis. Each integral can then be written as a Laurent series in $\epsilon$,
\begin{equation}\label{eq:genLExp}
    m_I(m_Q^2)=(m_Q^2)^{\textrm{dim}(m_I)}
    \sum_{k\ge -4} \epsilon^k F_{I}^{(k)},
\end{equation}
where $\textrm{dim}(m_I)$ is half of the mass dimension of the integral $m_I$, the 
$F_I^{(k)}$ are constants, and we used the fact that two-loop master integrals
have at most quadruple poles in $\epsilon$.
The goal of this paper is to determine these constants for each
of the master integrals, up to the order in $\epsilon$ required to compute the two-loop amplitudes 
for the processes mentioned previously.

\subsection{Partial-fraction and triangle relations}
\label{sec:extraRel}

As we already noted, out of the 76 master integrals obtained after 
IBP reduction, several are available in the literature. To further
reduce the number of integrals we need to compute, we discuss here
a set of special identities beyond IBP or symmetry relations.
The first kind of relations follows from partial-fraction
relations due to the degenerate kinematics in \cref{eq:kin1,eq:kin2}.
The second kind is obtained from a relation between three-point functions with
a special mass configuration.
We stress that we have not tried to find all identities that go beyond IBP relations,
and it would undoubtedly be interesting to find and study such relations in
a more systematic way.

\subsubsection{Partial-fraction relations}
\label{sec:partfrac}

The fact that partial-fraction relations are useful within
the framework of quarkonium physics is well known
(see, e.g., refs.~\cite{Feng:2015uha, Feng:2017hlu}). 
For example, we have already mentioned that partial-fraction relations 
play an important role when sorting integrals into topologies. In addition, 
there are also partial-fraction relations that relate integrals from different topologies. 
We discuss some examples of such relations in this section.

In order to illustrate how partial-fraction relations arise, let us consider the following one-loop integral,
\begin{equation}
    I= \int \text{d}^{d} q_1 \frac{1}{D_1 D_2 D_3 D_4},
\end{equation}
with the propagators
\begin{equation}
\begin{split}
    D_1= \left(q_1-p\right)^2-m_Q^2, \quad D_2=q_1^2,\quad
    D_3= \left(q_1+p\right)^2-m_Q^2, \quad D_4=\left(q_1+p+k_2\right)^2-m_Q^2\,.
\end{split}
\label{eq:defpartfracprop}
\end{equation}
We then note that $D_1+D_3=2D_2$, from which it follows that
\begin{equation}
    I=2 \int \text{d}^{d} q_1 \frac{1}{D_1 D_3^2 D_4}- \int \text{d}^{d} q_1 \frac{1}{D_2 D_3^2 D_4}\,.
\label{eq:partfracrel}
\end{equation}
Diagrammatically,
\begin{equation}
\raisebox{-42pt}{\includegraphics[width=3.3cm]{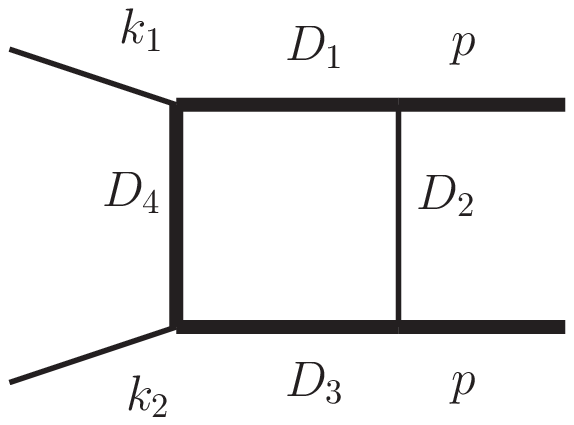}}=
2\raisebox{-40pt}{\includegraphics[width=3.3cm]{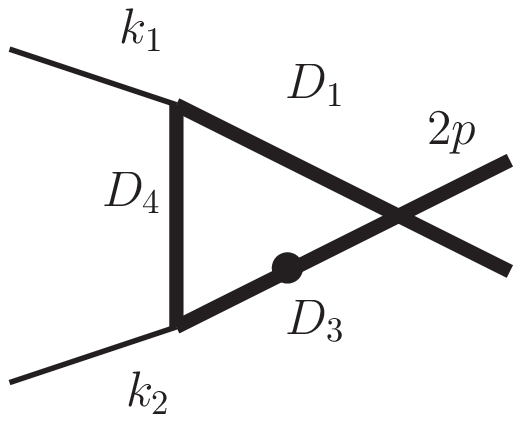}}
-\raisebox{-35pt}{\includegraphics[width=3.3cm]{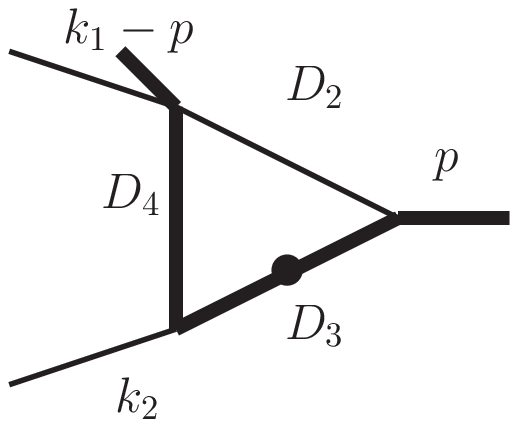}}.
\end{equation}
In summary, this partial-fraction relation relates a four-point function to
simpler three-point functions.

A similar approach can be used to generate relations between two-loop integrals. 
The propagators in \cref{eq:defpartfracprop} satisfy the relation
\begin{equation}
\frac{1}{D_1 D_3}=\frac{1}{2}\frac{1}{D_1 D_2}+\frac{1}{2}\frac{1}{D_2 D_3},
\end{equation}
which can diagrammatically be represented as
\begin{equation}
\begin{split}
\raisebox{-1cm}{\includegraphics[width=2cm]{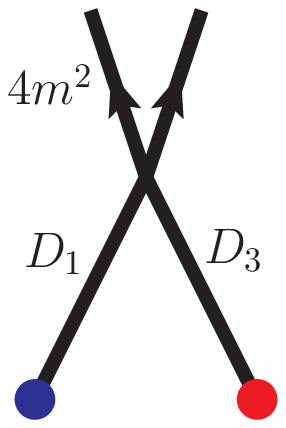}}=
&\frac{1}{2}\raisebox{-1cm}{\includegraphics[width=2cm]{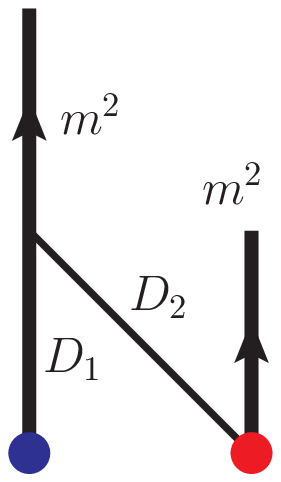}}
+\frac{1}{2}\raisebox{-1cm}{\includegraphics[width=2cm]{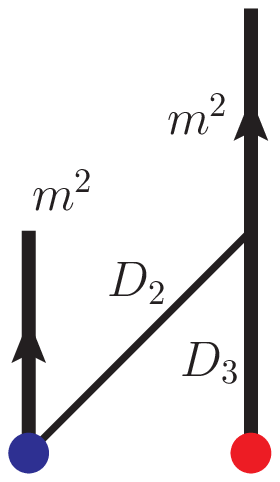}},
\end{split}
\label{eq:specpartrel1}
\end{equation}
where the red and blue dots connect to the rest of the diagram.
It is possible to derive similar partial-fraction relations for
other combinations of propagators. 
We can then use them to find relations among master integrals from different topologies.
Using this type of identities, we found the following relations:
\begin{align}\begin{split}
    \masterNew{6}&=\masterNew{26},\qquad 
    \masterNew{69}=\masterNew{71},\qquad
    \masterNew{36}=2\masterNew{39}-\masterNew{16},\qquad
    \masterNew{40}=\masterNew{58},\\
    \masterNew{45}&=\frac{2\left(3d-11\right)m_Q^2}{\left(d-3\right)\left(3d-10\right)}\masterNew{53}-\frac{8m_Q^4}{\left(d-3\right)\left(3d-10\right)}\masterNew{54}+\frac{\left(d-2\right)^2}{4\left(d-3\right)\left(3d-10\right)m_Q^4}\masterNew{76},\\
   \masterNew{46}&=\masterNew{53}-\frac{4m_Q^2}{\left(d-4\right)}\masterNew{54}.
\end{split}\end{align}
The last two relations have been derived by combining partial fraction 
relations with IBP relations and therefore involve the dimension 
$d=4-2\epsilon$ and the mass scale $m_Q$. The question naturally arises 
whether one could systematically incorporate these partial fraction relations
at intermediate steps in the IBP reduction system to find all possible
relations over the different topologies.

\subsubsection{Triangle relations}

Another special identity follows from a relation between certain triangle integrals.
Consider a one-loop triangle integral with external legs
$p_1$, $p_2$ and $k_1=p_1+p_2$, with $k_1^2=0$ and $p_i^2\neq0$.
We furthermore assume that two of the propagators have the same mass. Explicitly,
we consider
\begin{align}
    C(m_1^2,m_2^2)=&\int d^d q_1 \frac{1}{D_1{\left(m_1^2\right)} D_2{\left(m_1^2\right)} D_3^2{\left(m_2^2\right)}},
\end{align}
with
\begin{equation}
    \begin{split}
        D_1{\left(m^2\right)}=&q_1^2-m^2,
        \\
        D_2{\left(m^2\right)}=&\left(q_1-k_1\right)^2-m^2,
        \\
        D_3{\left(m^2\right)}=&\left(q_1-k_1+p_2\right)^2-m^2.
    \end{split}
\end{equation}
Introducing Feynman parameters and after some manipulations, we find that
\begin{equation}
\begin{split}\label{eq:v1eq}
    C(m_1^2,m_2^2)=\int_0^{\infty}dx \frac{\left(1+x\right)^{2\epsilon}}{\left(p_1^2-p_2^2\right)
    \left(1+\epsilon\right)}&\left[\left(m_1^2\left(1+x\right)+x\left(-p_1^2+m_2^2\left(1+x\right)\right)\right)^{-1-\epsilon}\right.
    \\
    &\left.-\left(m_1^2\left(1+x\right)+x\left(-p_2^2+m_2^2\left(1+x\right)\right)\right)^{-1-\epsilon}\right].
\end{split}
\end{equation}
If we change variables according to $x=1/y$, we get:
\begin{equation}
    \begin{split}\label{eq:v2eq}
    C(m_1^2,m_2^2)=\int_0^{\infty}dy \frac{\left(1+y\right)^{2\epsilon}}{\left(p_1^2-p_2^2\right)\left(1+\epsilon\right)}&\left[\left(m_2^2\left(1+y\right)+y\left(-p_1^2+m_1^2\left(1+y\right)\right)\right)^{-1-\epsilon}\right.
    \\
    &\left.-\left(m_2^2\left(1+y\right)+y\left(-p_2^2+m_1^2\left(1+y\right)\right)\right)^{-1-\epsilon}\right]\,.
\end{split}
\end{equation}
Comparing \cref{eq:v1eq,eq:v2eq}, we see that $C(m_1^2,m_2^2) = C(m_2^2,m_1^2)$ which corresponds diagrammatically to
\begin{equation}
\begin{split}
\raisebox{-38pt}{\includegraphics[width=3.0cm]{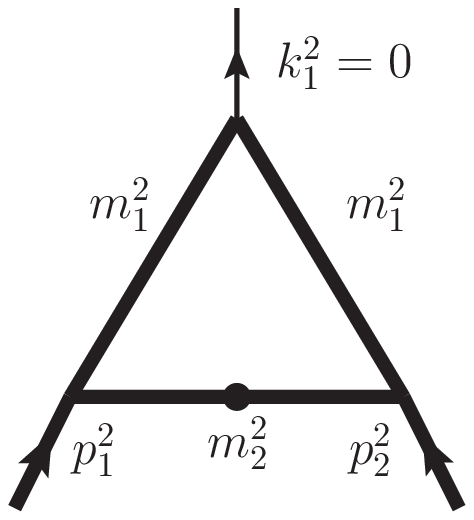}}=\raisebox{-38pt}{\includegraphics[width=3.0cm]{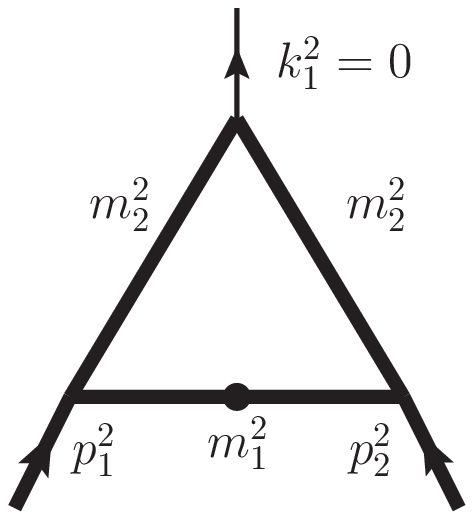}}.
\end{split}
\end{equation}
Since this relation holds for arbitrary $p_1^2$ and $p_2^2$,
it can be used to relate multi-loop integrals having this triangle as a sub-diagram. 
In particular, it follows that
\begin{equation}
\begin{split}
\raisebox{-17pt}{\includegraphics[width=3.3cm]{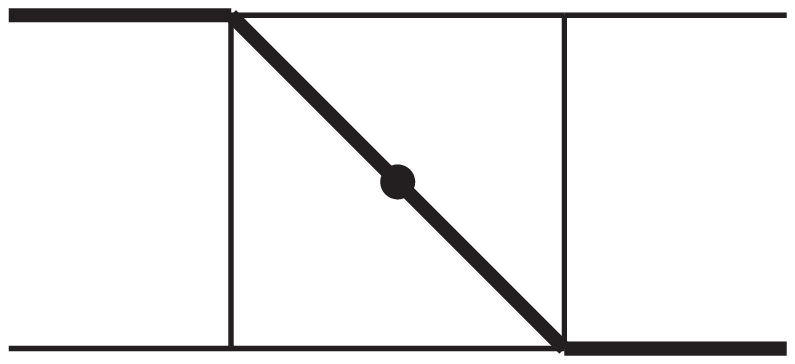}}=
&\raisebox{-17pt}{\includegraphics[width=3.3cm]{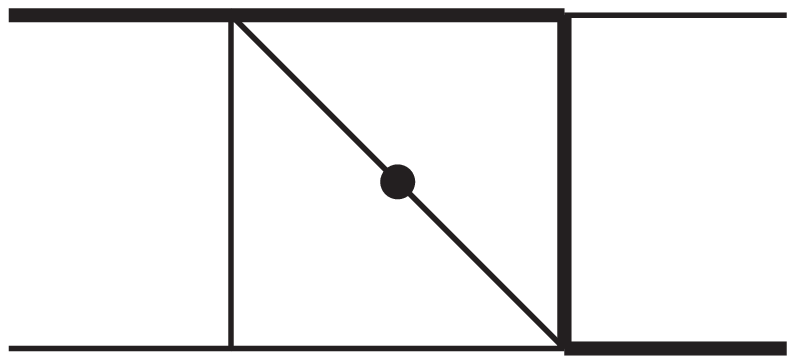}}.
\end{split}
\end{equation}
These integrals are related to $\masterNew{10}$ and $\masterNew{17}$ 
by IBP relations (see \cref{fig:four-point}), and so we can use this
identity to relate $\masterNew{10}$ to $\masterNew{17}$.


\section{Analytic results for the master integrals}
\label{sec:analytic}

In the previous section we have defined our set of master integrals, which is composed of
a total of 76 MIs.
Taking into account results available in the literature~\cite{Gehrmann:1999as,Bonciani:2003hc,Anastasiou:2006hc,Beerli:2008zz,Bonciani:2009nb,
Gehrmann:2010ue,Chen:2017xqd,vonManteuffel:2017hms,Chen:2017soz,DiVita:2018nnh,Chen:2019zoy,
Gerlach:2019kfo,Becchetti:2019tjy,Mandal:2022vju} and the relations
discussed in section~\ref{sec:extraRel}, there are 38 master integrals that we have to consider.
This can be for various reasons: some integrals are unknown, some are known in more general
kinematic configurations and the limit cannot be smoothly taken, some have been computed
but are expressed in different classes of functions, and some are known but not to the required
order in the Laurent expansion around $\epsilon=0$. The integrals we must consider can be classified
as follows:
there are 16 four-point integrals, $\masterNew{1-5}$, $\masterNew{7-15}$ and $\masterNew{18-19}$, 
18 three-point integrals, $\masterNew{20-25}$, $\masterNew{28-33}$, $\masterNew{35}$,
$\masterNew{37-38}$, $\masterNew{48}$, $\masterNew{49}$ and $\masterNew{51}$, 
and 4 two-point integrals, $\masterNew{57}$, $\masterNew{63-65}$.

In this section, we review our approach to the analytic computation of 
the MIs listed above and discuss some of their analytic properties. 
Since we are only interested in their contributions to the NNLO 
corrections to the processes discussed in \cref{sec:setup}  \cite{Abreu:2022cco}, we only
focus on the terms in the Laurent expansion around $\epsilon=0$ 
that contribute at this perturbative order.
In particular three MIs actually are absent from the NNLO
corrections: $\masterNew{5}$, $\masterNew{25}$ and $\masterNew{38}$ contribute 
to the two-loop amplitudes only at $\mathcal{O}{\left(\epsilon\right)}$.
We nevertheless compute the leading order of $\masterNew{5}$ and $\masterNew{38}$ 
both analytically and numerically, while for $\masterNew{25}$ we only present numerical 
results for its leading order.\footnote{We note that the leading order of these integrals is of 
weight/length four, so they
could in principle contribute to other two-loop amplitudes.
}

We recall that for each master integral $m_I$,
our goal is to compute the numbers $F_I^{(k)}$ in eq.~\eqref{eq:genLExp}.
These numbers are special functions evaluated at specific numerical values. 
We will encounter three types of special functions, which are all particular 
instances of iterated integrals~\cite{ChenSymbol}:\footnote{Note that in 
all cases some of these integrals may be divergent and require regularisation.}
\begin{enumerate}
    \item {Multiple polylogarithms} (MPLs)~\cite{GoncharovMixedTate},   
which are iteratively defined by
\begin{equation}\label{eq:mpls}
    G(a_1,\cdots,a_n;x) = \int_0^x \dfrac{dt}{t-a_1}G(a_2,\cdots,a_n;t)\,,
\end{equation}
with $G(;x) = 1$.
\item Elliptic multiple polylogarithms (eMPLs)~\cite{brown2013multiple,Broedel:2014vla,Broedel:2017kkb}:
\begin{equation} \label{eq:gamma_tilde}
    \tilde{\Gamma}\left(\begin{array}{ccc}
    n_1 & \cdots & n_k  \\
    z_1 & \cdots & z_k
    \end{array}; z, \tau\right) = \int_0^z dz' \, g^{(n_1)}(z'-z_1,\tau) \, \tilde{\Gamma}\left(\begin{array}{ccc}
    n_2 & \cdots & n_k  \\
    z_2 & \cdots & z_k
    \end{array}; z', \tau\right)\,,
    \end{equation}
    with $\tilde{\Gamma}\left(; z, \tau\right) = 1$, and with the integration kernels
    $g^{(n)}(z,\tau)$ defined by the Eisenstein-Kronecker series
\begin{equation} \label{eq:EK_series}
    F(z,\tau,\alpha) = \dfrac{1}{\alpha}\sum_{n\geq 0} g^{(n)}(z,\tau) \alpha^n = \dfrac{\theta_1'(0,\tau)\theta_1(z+\alpha,\tau)}{\theta_1(z,\tau)\theta_1(\alpha,\tau)},
\end{equation}
where $\theta_1$ is a Jacobi theta function, and $\theta_1'$ is its derivative with respect to the first argument. The arguments $z_i$, $z$ and $\tau$ are complex numbers, with $\textrm{Im}\,\tau >0$. 
\item {Iterated integrals of Eisenstein series}~\cite{ManinModular,Brown:mmv}:
\begin{equation} \label{eq:IEI_int}
    I(f_1,\cdots,f_k;\tau) = \int_{i\infty}^{\tau} \frac{d\tau'}{2 \pi i} f_1(\tau')I(f_2,\cdots,f_k;\tau')\,,
\end{equation}
with $I(;\tau) = 1$ and the $f_i$ are Eisenstein series of weight $k_i$ for some subgroup $\Gamma\subset \textrm{SL}(2,\mathbb{Z})$. They are special cases of modular forms of weight $k_i$, i.e., holomorphic functions on the upper half-plane such that 
\begin{equation}
    f_i\left(\frac{a\tau+b}{c\tau+d}\right) = (c\tau+d)^{k_i}\,f_i(\tau)\,, \qquad \forall \left(\begin{smallmatrix}a&b\\c&d\end{smallmatrix}\right)\in\Gamma \subseteq \textrm{SL}(2,\mathbb{Z})\,.
\end{equation}
\end{enumerate}
Presenting a detailed account of these functions and their properties would
go beyond the scope of this paper, and we refer instead to the relevant
literature (cf., e.g., refs.~\cite{Duhr:2014woa,Duhr:2019rrs,Broedel:2017kkb,Broedel:2018iwv,Adams:2017ejb,Duhr:2019tlz} 
and references therein, for a discussion in a physics context).

Note that a given integral may be expressible in terms of more than one of these classes of iterated integrals. 
Depending on the representation chosen, the results can be more or less easy to manipulate and evaluate. 
Indeed, the understanding of these different types of iterated integrals is currently not on the same footing.
For example, we know how to systematically simplify expressions involving MPLs
and how to numerically evaluate them very efficiently (see, e.g., refs.~\cite{Duhr:2012fh,
Duhr:2014woa,Panzer:2014caa,Ablinger:2013cf,Vollinga:2004sn,Naterop:2019xaf,Frellesvig:2016ske}, 
and references therein). This is however not (yet) the case for eMPLs and iterated integrals of Eisenstein series. 
While first public codes for the numerical evaluation of eMPLs exist~\cite{Walden:2020odh}, 
these codes are not nearly as efficient as in the MPL case. 
In some cases, it is possible to write eMPLs as iterated integrals of Eisenstein series~\cite{Broedel:2018iwv},
and in this representation the numerical evaluation is much more efficient~\cite{Bogner:2017vim,Duhr:2019rrs}, allowing us to reach
a precision comparable to what can be achieved for MPLs.

\subsection{Direct integration} \label{sec:direct_int}

We have evaluated all the MIs for which no results were available in the literature via direct integration.
Our starting point is the parametric representation of the scalar Feynman integrals
in terms of Feynman parameters:
\begin{equation} \label{eq:feyn_par}
    I = N \int_0^{\infty} dx_1 \cdots \int_0^{\infty}dx_n\, 
    \delta\left(1 - \Sigma_0\right) \prod_{i = 1}^n x_i^{a_i - 1} 
    \frac{\mathcal{U}^{a - \frac{3}{2}d}}{\mathcal{F}^{a - d}},
\end{equation}
where $N$ is some normalisation factor, $d = 4 - 2 \epsilon$, 
$\Sigma_0=\sum_{i=1}^n x_i$,
and $\mathcal{U}$ and $\mathcal{F}$ are the usual Symanzik polynomials. 
The representation in eq.~\eqref{eq:feyn_par} can be integrated order by order
in $\epsilon$ using direct integration.\footnote{The integrals
might have singularities in $\epsilon$ that forbid us to trivially integrate \cref{eq:feyn_par}
order by order in $\epsilon$. For those cases, it is simple to choose a different
master integral belonging to a quasi-finite basis, where poles in $\epsilon$ appear as prefactors (cf., e.g., refs.~\cite{Panzer:2014gra,vonManteuffel:2014qoa}). We thus assume that
we can always expand \cref{eq:feyn_par} under the integration sign and all integrations
are well defined.}
We distinguish the two cases:
\begin{itemize}
    \item \textbf{MPL case}: we are able to perform all the integrations in 
    \cref{eq:feyn_par} in terms of MPLs, and consequently the integral can be expressed
    in terms of MPLs evaluated at algebraic arguments;
    \item \textbf{eMPL case}: we are not able to perform all the integrations in 
    \cref{eq:feyn_par} in terms of MPLs, because the integrand depends on the square root 
    of a polynomial of the fourth order, which describes an elliptic curve. 
    In this case we exploit the strategy outlined in refs.~\cite{Broedel:2019hyg,Hidding:2017jkk} 
    to obtain an analytic expression in terms of eMPLs. (We refer to ref.~\cite{Broedel:2017kkb} for how to relate iterated integrals involving square roots of quartic polynomials to the eMPLs in eq.~\eqref{eq:gamma_tilde}).
\end{itemize}

The MPL case is vastly discussed in the literature~\cite{Brown:2008um,Anastasiou:2013srw,Ablinger:2014yaa,Bogner:2014mha,Panzer:2014caa,Bogner:2015nda}. In a nutshell, one starts from the
Feynman-parameter representation in \cref{eq:feyn_par} and notices that the Feynman-parameter
integrals are projective integrals. This observation leads to what is usually known as the
\emph{Cheng-Wu theorem} in physics~\cite{Cheng:1987ga}, which states that the integral is invariant under some transformations of the argument of the
delta function, such as:
\begin{equation} \label{eq:Cheng}
    \Sigma_0 \rightarrow \Sigma= \sum_{i=1}^n c_i x_i \,\qquad \textrm{for any}\qquad c_i \geq 0\,.
\end{equation}
Using the freedom afforded by this transformation, together with different integration
orderings, one then looks for a way to integrate over all Feynman parameters 
so that each integration can be written in terms of MPLs.
This can be done using public codes such as $\tt{HyperInt}$~\cite{Panzer:2014caa}, $\tt{PolyLogTools}$~\cite{Duhr:2019tlz} or {\tt MPL}~\cite{Bogner:2015nda}.

For the MIs which do not admit a representation in terms of MPLs, 
we use the strategy introduced in refs.~\cite{Broedel:2019hyg,Hidding:2017jkk}. More precisely, we proceed exactly as in the MPL case, but, due to the appearance of a square root defining an elliptic curve, we cannot perform
all integrations in terms of MPLs. Instead, we find a transformation as in \cref{eq:Cheng} and
an integration ordering such that we can perform all but the last integration in terms of MPLs.
The last integration will depend on the final Feynman parameter $x$ and the square root of
a polynomial of degree 4 in $x$ (denoted by $y$) describing the underlying elliptic curve. 
Schematically, the last integration will be of the form:
\begin{equation}
    \int dx \, R(x,y) \, J_{\textrm{MPL}}(x,y),
\end{equation}
where $R(x,y)$ is a rational function in $x$ and $y$, 
and $J_{\textrm{MPL}}(x,y)$ is a linear combination of MPLs with
numerical coefficients, where we note that the square root $y$ can 
also appear in the arguments of the MPLs.
The next step is to rewrite the MPLs that appear in the last integration as eMPLs.
This can be done systematically as described in ref.~\cite{Broedel:2019hyg},
by iteratively differentiating the MPLs and integrating them back in terms
of eMPLs. For our MIs, these rather straightforward steps are complicated
by the appearance of \emph{spurious} square roots at either the second to last or the last
integration. 
By spurious roots, we mean roots which are not related to the square root defining the elliptic curve.
In the remaining part of this subsection we discuss how we overcame this issue (we note that spurious roots can also appear in the calculation of integrals involving only MPLs, and they can be dealt with in the same way; we focus here on the elliptic case since we found it to be by far the most challenging one).

Let us first discuss the MIs for which spurious square roots appear in the second to last integration.
This was the case for $\masterNew{3}$, $\masterNew{5}$, $\masterNew{7}$, $\masterNew{13}$, and $\masterNew{14}$, for which
we could not find a transformation of the type of \cref{eq:Cheng} and
an integration ordering that allowed us to rewrite all but the last Feynman-parameter integration
in terms of MPLs. 
In this scenario, the next-to-last integration is of the form:
\begin{equation} \label{eq:int_roots}
    \int dx_1 \int dx_2 \,S(x_1,x_2)\,I_{\textrm{MPL}}(x_1,x_2),
\end{equation}
where $S$ is a rational function in $x_1$, $x_2$ and the spurious square roots, and 
$I_{\textrm{MPL}}(x_1,x_2)$ is a linear combination of MPLs whose arguments are rational functions 
in $x_1$, $x_2$ and the spurious square roots. If there is a single spurious square root, there is a simple
solution: we find a change of variables that rationalises the spurious root,
for example with the approaches described in refs.~\cite{Becchetti:2017abb,Besier:2018jen,Besier:2019kco}, 
and then proceed with the integration over the new variables with the general strategy outlined 
above.\footnote{To be more precise, this is done at the level of each individual term in the expression. 
Multiple spurious roots may be present in different terms. In this case, one can rationalise 
each spurious root separately, and one must carefully match the integration boundaries.}
This was sufficient for MIs $\masterNew{5}$, $\masterNew{7}$, $\masterNew{13}$, and $\masterNew{14}$.
Integral $\masterNew{3}$ proved to be more challenging. 
With the most convenient choice of transformation of the type of eq.~\eqref{eq:Cheng} 
and most convenient integration order that we could find, we ended up with multiple spurious 
square roots appearing inside the same term. 
We found different terms involving different pairs of spurious roots, 
each still depending on two integration variables.
In principle, one could attempt to
rationalise all these square roots simultaneously, but this typically leads to very complicated
changes of variables, which generate long expressions that are difficult to manipulate.
Instead, we found that the expression could be simplified by inserting
\begin{equation}
    1=\int_{-\infty}^\infty d\tilde x\,\delta(1-\tilde\Sigma)\,,
\end{equation}
where $\tilde\Sigma$ is linear in the Feynman parameters and in $\tilde x$, into \cref{eq:feyn_par}:
\begin{eqnarray}
     I = N \int_0^{\infty} dx_1 \cdots \int_0^{\infty}dx_n \, \delta\left(1 - \Sigma\right)
     \int_{-\infty}^{\infty}d\tilde{x} \,\delta\left(1 - \tilde{\Sigma}\right) 
     \prod_{i = 1}^n x_i^{a_i - 1} \frac{\mathcal{U}^{a - \frac{3}{2}d}}{\mathcal{F}^{a - d}} .
\end{eqnarray}
We then found that there was a choice of integration ordering that allows us to perform
all but the last integral in terms of MPLs (more precisely, there were still
spurious square roots, but all of them could be rationalised), 
bringing us into the general strategy discussed above.

To be more explicit, let us look more closely at the case of $\masterNew{3}$.
Its Feynman parameter representation is
\begin{equation} \label{eq:m23_v1}
    \masterNew{3} = - e^{2\epsilon \gamma_E}\Gamma(3+2\epsilon)\int_0^{\infty} \prod_{i=1}^7 dx_i \, 
    \delta\left(1- \Sigma_{3}\right) \mathcal{U}^{1 + 3\epsilon} \mathcal{F}^{-3 - 2\epsilon},
\end{equation}
where the Symanzik polynomials are:
\begin{eqnarray}
\mathcal{U} & = & x_{1}x_{3} +x_{2}x_{3} +x_{1}x_{4} +x_{2}x_{4} +x_{3}x_{4} +x_{1}x_{5} +x_{2}x_{5} +x_{4}x_{5} +x_{1}x_{6} +x_{2}x_{6} +x_{4}x_{6} \nn \\
&&+x_{3}x_{7} +x_{4}x_{7} +x_{5}x_{7} +x_{6}x_{7}, \nn \\
\mathcal{F} & = & x_1^2\left(x_3+x_4+x_5+x_6\right) + x_2^2\left(x_3+x_4+x_5+x_6\right) + x_1 \left[2 x_3 x_4 + x_4^2 + 4 x_4 x_5 + x_5^2 \right. \nn \\
&& \left. + 2 (x_3 + x_4 + x_5)x_6 + x_6^2 + 2 x_2 (x_3 + x_4 + x_5 + x_6)\right] + x_4 \left[(x_5 + x_6)(x_4 + x_5 + x_6) \right. \nn \\
&& \left. + x_3(x_4 + 2x_6)\right] + \left[2x_3 (x_4 + x_6) + (x_4 + x_5 + x_6)^2\right]x_7 + x_2\left[x_4^2 + 2x_3(x_4 + x_6 + x_7) \right. \nn \\
&& \left. + 2x_4 (x_5 + 2x_6 + x_7) + (x_5 + x_6)(x_5 + x_6 + 2x_7)\right]\,.
\end{eqnarray}
We first note that by choosing $\Sigma=x_1 + x_2 + x_4$ we found the most compact intermediate expressions.
In the most convenient integration order that we could identify, we encounter the following spurious square roots
\begin{align}\begin{split}\label{eq:sqrtM3}
    y_1 & =  \sqrt{1 + 4 x_4^2 -8 x_4^3 + 4 x_4^4 + 8 x_4 x_5 -8 x_4^2 x_5}\,,  \\
    y_2 & =  \sqrt{1 - 4 x_4 + 4 x_4^2 - 8 x_4 x_5 +8 x_4^2 x_5}\,,  \\
    y_3 & =  \sqrt{x_4^2 - 2x_4^3 +x_4^4 - 2x_4^3x_5 + x_5^2 - 2x_4x_5^2 + x_4^2x_5^2}\,,
\end{split}\end{align}
where some of the spurious square roots appear simultaneously in the same terms. However, we found that by rewriting $\masterNew{3}$ as
\begin{equation}
    \masterNew{3} = - e^{2\epsilon \gamma_E}\Gamma(3+2\epsilon)\int_0^{\infty} 
    \prod_{i=1}^7 dx_i \, \delta\left(1- \Sigma\right) \int_{-\infty}^{\infty}d\tilde{x} \,
    \delta\left(x_5 - x_6 + \tilde{x}\right) \mathcal{U}^{1 + 3\epsilon} \mathcal{F}^{-3 - 2\epsilon}\,,
\end{equation}
with $\Sigma$ as given above, and choosing the integration order
\begin{equation}
    \left(x_2,x_5,x_3,x_7,x_1,x_6,\tilde x,x_4\right)\,,
\end{equation}
we obtain an expression with two spurious square roots in the integration 
variables, which do not appear simultaneously within the same term. 
As such, we can proceed and rationalise the square roots in each term 
separately, and perform the next-to-last integration in terms of 
MPLs.\footnote{For the integrals $\masterNew{5}$, $\masterNew{7}$, $\masterNew{13}$, 
and $\masterNew{14}$, we find that the strategy of introducing a second 
delta function allows us to avoid any spurious square roots. 
This provides an alternative to the procedure described previously.}

Let us now discuss how we deal with spurious square roots in the last integration
step, which is a much more common occurrence.
These spurious roots are generated when factorising the denominators in the second-to-last
integration. In most cases, we find only spurious square roots involving a quadratic polynomial
which can always be rationalised. However, for some integrals we encountered different spurious roots 
involving quartic or quintic polynomials, sometimes appearing in the arguments of the MPLs.
This was for instance the case for $\masterNew{3}$, where we encountered more than $20$ different spurious roots.
In order to deal with these roots, we devised an algorithm that allows us to eliminate all spurious roots systematically. 
The key idea is that it is not necessary to bring individual terms into a form that can be integrated in terms 
of MPLs or eMPLs, but only the entire integrand.
Our algorithm is iterative in the transcendental weight and proceeds 
as follows:
\begin{enumerate}
    \item We first consider all terms of transcendental weight one. 
    We then group them according to the $x$-dependence in their prefactor,
    where $x$ denotes the last integration variable. Let us call $g_{1,i}$ 
    one of these combinations of MPLs of weight one with the $x$-dependent 
    prefactor removed. We compute the derivative $\frac{d}{dx}g_{1,i}$ and 
    construct the dlog-kernels. At this stage, all spurious roots in 
    $\frac{d}{dx}g_{1,i}$ must disappear, otherwise they would not be spurious. 
    We then integrate $\frac{d}{dx}g_{1,i}$ back in $x$ and fix the boundary 
    condition either analytically or with \texttt{PSLQ}. We repeat this 
    procedure for all other weight one terms $g_{1,j}$.
    \item We next consider all terms of transcendental weight two. 
    The first step is again to group them according to the $x$-dependence 
    of their prefactor. When computing the derivative of one such term, 
    $\frac{d}{dx}g_{2,i}$, we now have MPLs of weight one which might still 
    involve spurious roots. These are dealt with as in the previous step. 
    Next, we construct the dlog-kernels for the weight two contributions, 
    and again all spurious roots disappear. 
    We then integrate $\frac{d}{dx}g_{2,i}$ back in $x$ and the boundary 
    condition is fixed as at weight one. 
    We repeat this procedure for all other weight two terms $g_{2,j}$.
    \item We proceed with the same steps for weight three and then weight four, 
    where each time the procedure requires dealing with all the lower weight 
    terms that are generated when taking derivatives.
\end{enumerate}
The crucial idea behind this algorithm is that, while individual terms may 
have spurious roots, the combination of terms that have the same weight and 
share the same integration kernel cannot, otherwise the square roots would not 
be spurious. Once all spurious roots have been removed, we can perform the last 
integral in terms of eMPLs for all integrals where spurious square roots 
appeared. 

Following the steps described in this subsection we obtain fully analytic 
results for all the master integrals that we are interested in. 
We also obtained expressions for the integrals that had already
been considered previously in the literature~\cite{Gehrmann:1999as,Bonciani:2003hc,Anastasiou:2006hc,Beerli:2008zz,Bonciani:2009nb,
Gehrmann:2010ue,Chen:2017xqd,vonManteuffel:2017hms,Chen:2017soz,DiVita:2018nnh,Chen:2019zoy,
Gerlach:2019kfo,Becchetti:2019tjy,Mandal:2022vju}, and we provide results for the complete
set of master integrals that contribute to the amplitudes mentioned in \cref{sec:setup}.
The analytic results for the MIs can be very lengthy, so we make them 
available in \texttt{Mathematica}-readable form at 
ref.~\cite{masterIntegralsGit}.
The file \texttt{analytics.m} contains a replacement list of the form
\begin{equation}\label{eq:ancNotation}
    \masterNew{I}\to\sum\epsilon^k F^{(k)}_I\,,
\end{equation}
with the $F^{(k)}_I$ as defined in \cref{eq:genLExp} where we set $m_Q^2=1$,
since the dependence on the single scale $m_Q^2$ can always be reinstated by dimensional analysis.
In the case where the $F^{(k)}_I$ can be written as MPLs, the corresponding 
expressions are explicitly inserted. In the elliptic case, where the expressions
are particularly lengthy, we do not explicitly insert them into 
the expressions for the master integrals. Instead, as discussed in 
\cref{sec:pslq} below, we used the \texttt{PSLQ} algorithm to find 
relations between the different elliptic $F^{(k)}_I$, and
the right-hand side of \cref{eq:ancNotation} incorporates these relations.
An independent set\footnote{This statement should be understood in light of 
what is described in \cref{sec:pslq}: the set is independent up to relations 
that are undetected by \texttt{PSLQ} with the numerical precision at which we 
evaluated the integrals.} of $F^{(k)}_I$ is given in analytic form in the files
\texttt{/elliptics/F*.m} of ref.~\cite{masterIntegralsGit}.

\subsection{Analytic results for the master integrals}

The 54 master integrals
\begin{align}\label{eq:MPL_List}
\begin{split}
\masterNew{1},\masterNew{2},&\masterNew{4},\masterNew{6},\masterNew{8-10},\masterNew{16},\masterNew{17},\masterNew{19},\masterNew{20},\masterNew{26-30},\\
&\masterNew{34-47},\masterNew{50-56},\masterNew{58-63},\masterNew{66-76}
\end{split}\end{align}
can be expressed entirely in terms of MPLs up to weight four evaluated at
algebraic arguments. As already mentioned, the algebraic properties and 
the numerical evaluation of MPLs are well understood. In particular, 
using the public implementation in \texttt{GiNaC}~\cite{Vollinga:2004sn} 
we obtained high-precision evaluations of these MIs to hundreds of digits. 
We then used the {\tt PSLQ} algorithm~\cite{pslq} to fit our results to 
a basis of transcendental numbers.
We observe that all of the MIs in eq.~\eqref{eq:MPL_List},
except
$\masterNew{8}$, $\masterNew{38}$, $\masterNew{50}$ and $\masterNew{51}$,
only involve MPLs evaluated at sixth roots of unity. This space
of transcendental numbers has dimension 88, and a basis is 
known~\cite{Broadhurst:1998rz,Henn:2015sem}.
This observation allows us to obtain very compact expressions for these
integrals. The four remaining master integrals,
$\masterNew{8}$, $\masterNew{38}$, $\masterNew{50}$ and $\masterNew{51}$,
involve fourth roots of unity or algebraic numbers of the form 
$a+b\sqrt{2}$, where $a$ and $b$ are rational numbers. In this cases 
it is possible to construct a (conjectural) basis for these 
transcendental numbers using the results from section 2.5 of 
ref.~\cite{Brown:coaction}.

The remaining 22 MIs not in eq.~\eqref{eq:MPL_List} cannot be expressed 
in terms of MPLs alone, but they involve eMPLs.\footnote{We note, however, 
that MPLs do appear in these expressions. This is to be expected, 
since ordinary MPLs are a subset of eMPLs, cf., e.g., 
ref.~\cite{Broedel:2017kkb}.} 
eMPLs depend on an elliptic curve, 
which is defined either by the square root of a quartic polynomial or,
equivalently, by a value of
$\tau$ as in \cref{eq:gamma_tilde}.
We find two different elliptic curves in our computation,
defined by:
\begin{equation}\begin{split}\label{eq:tau_values}
    \tau^{(a)} &\,=  i\frac{\K\left(\frac{1}{2}-\frac{1}{2\sqrt{5}}\right)}{\K\left(\frac{1}{2}+\frac{1}{2\sqrt{5}}\right)} = i\,0.805192\ldots\,,\\
        \tau^{(b)} &\,= i \frac{\K\left(\frac{7}{2}-\frac{3 \sqrt{5}}{2}\right)}{\K\left(\frac{3 \sqrt{5}}{2}-\frac{5}{2}\right)} = i\,0.680035\ldots\,,
    \end{split}
\end{equation}
where $\K(\lambda)$ denotes the complete elliptic integral of the first kind:
\begin{equation}\label{eq:K_def}
    \K(\lambda) = \int_0^1\frac{d t}{\sqrt{(1-t^2)(1-\lambda t^2)}}\,.
\end{equation}
More specifically, every master integral depends on (at most) one 
elliptic curve, and the only MIs that depend on the elliptic curve 
defined by $\tau^{(b)}$ are $\masterNew{24}$ and $\masterNew{25}$. 
At this point we have to address an important question: while every $\tau$ 
in the complex upper half-plane defines an elliptic curve, 
the same elliptic curve may arise from different values of $\tau$. 
It is therefore natural to ask if the two elliptic curves defined by 
eq.~\eqref{eq:tau_values} are indeed different (i.e., non-isomorphic) 
elliptic curves. This can easily be checked by computing the $j$-invariant 
of the elliptic curve,\footnote{The $j$-invariant can be evaluated in Mathematica using $j(\tau) = 1728\,${\tt KleinInvariantJ[}$\tau${\tt]}.}
\begin{equation}
    j(\tau) = \frac{1}{q}+744 +196884\, q +21493760\, q^2 +864299970\, q^3 + \begin{cal}O\end{cal}(q^4)\,,\qquad q = e^{2\pi i\tau}\,,
\end{equation}
and the two elliptic curves defined by $\tau_1$ and $\tau_2$ are the same if and only if $j(\tau_1) = j(\tau_2)$. We find:
\begin{equation}
j(\tau^{(a)})  = \frac{16384}{5} \textrm{~~~and~~~} 
j(\tau^{(b)})  =\frac{55296}{5}\,.
\end{equation}
This shows that our two elliptic curves are indeed distinct. 
They are in fact particular members of the families of elliptic 
curves associated to the sunrise integral, $\masterNew{64}$, \cite{Sabry,Broadhurst:1987ei,Bogner:2017vim,Adams:2015gva,Adams:2015ydq,Bogner:2019lfa,Broedel:2017siw,Campert:2020yur} 
and the two-loop non-planar three-point function $\masterNew{24}$ 
considered in refs.~\cite{vonManteuffel:2017hms,Broedel:2019hyg}.

The arguments $z_i$ of the eMPLs defined in
\cref{eq:gamma_tilde} which appear in our computation have the form
\begin{equation}\label{eq:z_i_def}
    z_i = \frac{m_i}{12} + \frac{n_i}{12}\,\tau^{(x)} + \xi_i\,, 
    \qquad x = a,b
\end{equation}
where $m_i$ and $n_i$ are integers and the $\xi_i$ are 
irrational numbers that are not rational multiples of~$\tau^{(x)}$. 
Instead, they can be expressed in terms of the complete and incomplete 
elliptic integrals of the first kind evaluated at special arguments. 
The complete elliptic integral of the first kind was already defined in eq.~\eqref{eq:K_def}, while its incomplete version is defined by
\begin{equation}\label{eq:F_def}
    \F(\phi|\lambda) = \int_0^\phi\frac{d\theta}{\sqrt{1-\lambda\,\sin^2\theta}}\,.
\end{equation}
All the $\xi_i$ can be written as rational linear combination
of the $\tilde{z}_i$ listed in appendix~\ref{app:elliptic_args}.

If all arguments of an eMPL have the form $a \tau + b$, with $a,b\in\mathbb{Q}$ 
(i.e., whenever $\xi_i=0$ in eq.~\eqref{eq:z_i_def}), the eMPL can 
be expressed in terms of iterated integrals of Eisenstein series, defined 
in \cref{eq:IEI_int}, in an algorithmic fashion~\cite{Broedel:2018iwv}. 
Writing integrals in terms of iterated integrals of Eisenstein series presents
the advantage that we know a basis for this type of iterated integrals
and also efficient techniques for their numerical evaluation. 
We identified seven MIs for which their analytic expressions involve iterated Eisenstein series for two distinct subgroups 
$\Gamma\subseteq\textrm{SL}(2,\mathbb{Z})$: $\masterNew{15}$, $\masterNew{23}$, $\masterNew{48}$, $\masterNew{49}$, 
$\masterNew{64}$ and $\masterNew{65}$ involve Eisenstein series of 
the congruence subgroup $\Gamma_1(6)$,\footnote{The MIs $\masterNew{64}$ and $\masterNew{65}$ 
correspond to the two master integrals for the equal-mass sunrise, 
for which a representation in terms of iterated integrals of Eisenstein 
series for $\Gamma_1(6)$ is known in the literature, 
cf.~refs.~\cite{Adams:2017ejb,Broedel:2018iwv}. In this paper we explicitly compute them up to the order $\epsilon^2$ in dimensional regularisation.} while $\masterNew{24}$ and $\masterNew{25}$ involves 
Eisenstein series of the congruence subgroup 
$\Gamma_1(4)$, with
\begin{equation}
    \Gamma_1(N) = \left\{\left(\begin{smallmatrix}a&b\\c&d\end{smallmatrix}\right)\in\textrm{SL}(2,\mathbb{Z}) : a,d = 1\!\!\!\!\mod N\textrm{~and~} c=0\!\!\!\!\mod N\right\}\,.
\end{equation}

To summarise, we can classify the 76 MIs into four categories, depending on the classes of transcendental numbers that appear in the analytic results:
\begin{itemize}
    \item 50 MIs can be expressed in terms of MPLs evaluated at sixth
    roots of unity,
    \item 4 MIs can be expressed in terms of MPLs evaluated at fourth roots of unity and arguments of the form $a+b\sqrt{2}$, $a,b\in\mathbb{Q}$,
    \item 20 MIs can be expressed in terms of eMPLs associated to the elliptic curve defined by $\tau^{(a)}$, 
    and iterated integrals of Eisenstein series for $\Gamma_1(6)$.
    \item 2 MI can be expressed in terms of eMPLs associated to the elliptic curve defined by $\tau^{(b)}$, 
    and iterated integrals of Eisenstein series for the congruence subgroup $\Gamma_1(4)$. We note once more that we only provide numerical results for $\masterNew{25}$ since it is not strictly speaking
    needed for the processes we are concerned with. Nevertheless, given that it is coupled
    to $\masterNew{24}$, we know that it can be written in terms of the same set of functions. 
\end{itemize}


\section{Numerical results for the master integrals}
\label{sec:numerics}

In the previous section we discussed how we could obtain fully 
analytic results for the master integrals and we briefly reviewed the class 
of special functions that appear in the answer. We also noted that
the numerical evaluation of MIs integrating to MPLs was a solved problem, 
and even used high-precision evaluations to write them in bases of 
transcendental numbers.

For the numerical evaluation of eMPLs and iterated integrals of Eisenstein
series, we follow the strategy presented in 
refs.~\cite{Duhr:2019rrs,eMPL_num_toappear,marzucca_elliptics20}.
In a nutshell, the integration kernels $g^{(n)}(z,\tau)$ and $f_j(\tau)$ 
in \cref{eq:gamma_tilde,eq:IEI_int} are expanded in a so-called 
$q$-expansion (a Fourier expansion with $q = e^{2\pi i \tau}$).
This series is truncated at a finite order, which, together with the convergence
properties of the series representation, determines the precision of 
the numerical evaluation. The coefficients of this series expansion are 
MPLs, which can be numerically evaluated with standard tools. In the case 
of iterated integrals of Eisenstein series, it is known how to transform 
the integrals into a representation where the $q$-expansion converges very 
rapidly, and we can evaluate the integrals to hundreds 
of digits in an acceptable amount of time. For eMPLs, however, the convergence 
is in general very slow which in practice means we can only get a very 
small number of digits. For all integrals that involve eMPLs we obtained
numerical results valid to $\mathcal{O}(10)$ digits.
We have validated all our results by comparing them to completely
independent numerical evaluations obtained with \texttt{pySecDec} 
\cite{Borowka:2017idc}, with which one can achieve a similar level of 
precision.

While the efficiency of the numerical evaluation of eMPLs will surely improve in 
the future, alternative approaches can already be used to obtain high-precision
numerical evaluations for MIs involving eMPLs. These 
evaluations can then be used in phenomenological applications.
In this section we explain how we bypassed the 
evaluation of eMPLs to obtain numerical results for all master integrals with a precision 
of 1000 digits.
This precision is more than sufficient for any phenomenological application, 
and also allows us to use the {\tt PSLQ} algorithm to simplify and find 
relations between the various $F_{I}^{(k)}$, as will be discussed in 
\cref{sec:pslq}. We discuss two independent calculations, which have in 
common the fact that they are based on numerically solving systems of 
differential equations.

\subsection{High-precision numerical results from differential equations}
\label{sec:diffEq}

To obtain high-precision numerical evaluations for the elliptic MIs that 
involve eMPLs, we construct systems of differential equations for 
them \cite{Kotikov:1990kg,Kotikov:1991hm,Kotikov:1991pm,Gehrmann:1999as}, 
which are then solved numerically.
These calculations are performed with two alternative approaches.
In the first one, we build the differential equations and provide the required
boundary conditions ourselves. The system is then solved
in terms of generalised power series~\cite{Moriello:2019yhu,Hidding:2020ytt}
using \texttt{diffexp} \cite{Hidding:2020ytt}.
In the second approach, we use \texttt{AMFlow}~\cite{Liu:2022chg}, 
which automatises the construction and solution of the system of differential 
equations~\cite{Liu:2017jxz,Liu:2021wks,Liu:2022chg}.
The integrals involving eMPLs can be collected in four topologies 
(topologies 3, 4, 5 and 6 in the notation of \cref{sec:mis}). 
In topology 6, however, the only elliptic integrals are 
$\masterNew{5}$, which only contributes at order $\epsilon$ to the two-loop
amplitudes of ref.~\cite{Abreu:2022cco}, and $\masterNew{15}$, which can be written
in terms of iterated integrals of Eisenstein series, and can thus be evaluated
efficiently to hundreds of digits from its analytic representation. We have thus
only evaluated the elliptic integrals of topology 6 with \texttt{AMFlow},
since the evaluation with \texttt{diffexp} is only used as a check in the other
topologies. 

\begin{figure} 
\centering
\subfloat[]{\raisebox{17pt}{\includegraphics[width =3.6 cm]{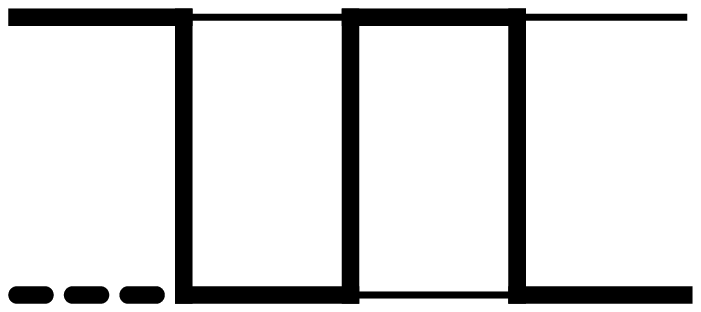}}\label{fig:evalDE1}}\qquad \qquad
\subfloat[]{\includegraphics[width =2.3 cm]{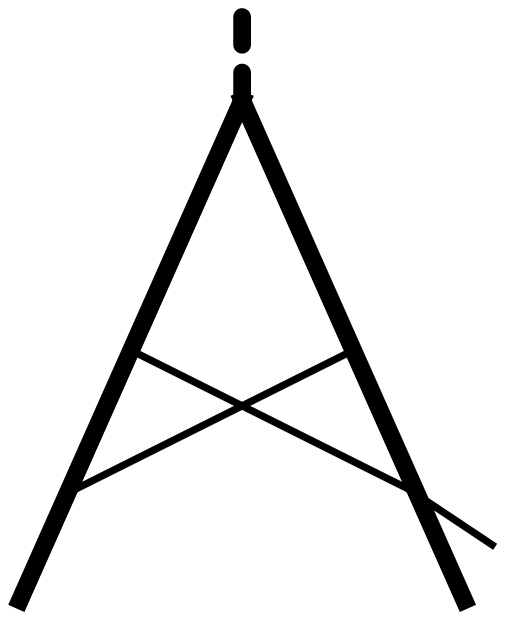}\label{fig:evalDE2}}\qquad\qquad 
\subfloat[]{\includegraphics[width =2.2 cm]{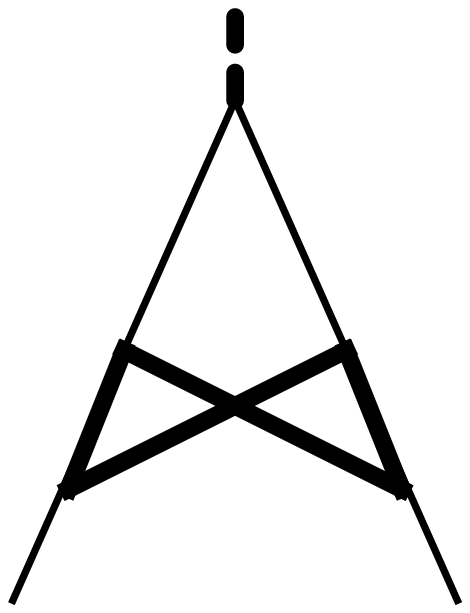}\label{fig:evalDE3}} 
\caption{Diagrams defining the topologies for the differential equations used in the numerical
evaluation of the integrals involving eMPLs. The thick dashed line has mass $\lambda^2$, while the solid
thick lines have mass $m_Q^2$. Thin lines are massless.} \label{fig:evalDE}
\end{figure}

Let us briefly discuss the calculation with \texttt{diffexp}, since
\texttt{AMFlow} implements similar steps in an automated way.
In a nutshell, if $\vec I$ is a basis of master integrals for a given
topology, we compute the system of differential equations
\begin{equation}\label{eq:diffeqGen}
    \partial_{x_i}\vec I=M(\vec x;\epsilon)\vec I\,,
\end{equation}
where $\vec x$ denotes all the kinematic variables on which the integrals $\vec I$
depend. We then numerically solve this differential equation order by order in $\epsilon$.
We note, however, that we cannot straightforwardly apply the 
differential-equation approach to the master integrals 
in \cref{fig:four-point,fig:three-point,fig:two-point,fig:fact}.
Indeed, as noted in \cref{eq:genLExp}, 
they depend on a single variable and as such they
satisfy a trivial differential equation. 
Instead, we will consider two-scale versions of
the master integrals and evaluate them at the point corresponding to the
kinematics of \cref{eq:kin2}. We are assisted in this task by the fact that 
a generalisation of topologies 3 and 4 was considered in 
refs.~\cite{Chen:2017xqd,Chen:2017soz} and a generalisation of topology 5 
in ref.~\cite{vonManteuffel:2017hms}, precisely from the perspective 
of their differential equations. For each topology we setup 
the calculation in the following way:
\begin{itemize}
    \item {\bf For topology 3}: we consider the topology defined by the diagram in \cref{fig:evalDE1}, evaluated at $\lambda^2=0$.
    Our basis is similar to the one chosen in 
    refs.~\cite{Chen:2017xqd,Chen:2017soz}.\footnote{We thank the authors of refs.~\cite{Chen:2017xqd,Chen:2017soz} for
    discussions in identifying and correcting some typos in their publications.} For completeness, all the information required to completely define the basis we chose can be found in the folder \texttt{/diffexp/} 
    of ref.~\cite{masterIntegralsGit}.
    \item {\bf For topology 4}: we consider the topology defined by the diagram in \cref{fig:evalDE2}, evaluated at $\lambda^2=0$.
    The same comments as for topology 3 apply. The complete information to define our basis can be found
    in the folder \texttt{/diffexp/} 
    of ref.~\cite{masterIntegralsGit}.
    \item {\bf For topology 5}: we consider the topology defined by the diagram in \cref{fig:evalDE3}, evaluated at $\lambda^2=4 m_Q^2$.
    The basis we choose is that of ref.~\cite{vonManteuffel:2017hms}, up to some trivial normalisation factors.
    The complete information to define our basis can be found 
    in the folder \texttt{/diffexp/} 
    of ref.~\cite{masterIntegralsGit}.
\end{itemize}
Constructing the differential equations for these three sets of master integrals is straightforward with
standard approaches \cite{Gehrmann:1999as}.
In all cases, our bases are such that the differential equation matrices $M(\vec x;\epsilon)$ 
in \cref{eq:diffeqGen} are polynomials in $\epsilon$ (with our choice of bases, they are polynomials
of degree~2). In the sub-sectors that only couple integrals which evaluate to MPLs, the differential
equation is in canonical (dlog) form, that is, the corresponding entries of $M(\vec x;\epsilon)$
consist of dlog-forms and are proportional to $\epsilon$.
To solve the differential equations we use the initial conditions given in 
refs.~\cite{Chen:2017xqd,Chen:2017soz,vonManteuffel:2017hms} within
\texttt{diffexp}. This allows us
to numerically solve the three systems of differential equations to hundreds
of digits at the relevant kinematic points (if the integrals are logarithmically divergent
at this point, we instead compute a generalised series expansion). 
We then use IBP relations to relate the bases chosen to solve the differential equations to the master integrals
in \cref{fig:four-point,fig:three-point,fig:two-point,fig:fact}. 

Up to some differences in the way the boundary conditions are determined, the evaluation
within \texttt{AMFlow} is based on the same procedure. The different steps 
are automated so that one can simply require the evaluation of the integral
at a phase-space point. Using \texttt{AMFlow}, we were able to evaluate the integrals
to very high precision ($\mathcal{O}(1500)$ digits), but to be conservative
we keep only the first 1000 digits. We found that we could achieve
a higher precision with \texttt{AMFlow} than with \texttt{diffexp}, so we quote
the former as our final results. While the difference is irrelevant for phenomenological
studies, the highest precision of the \texttt{AMFlow} evaluations were useful in identifying
extra relations between different integrals using {\tt PSLQ}.

We find complete agreement between the numerical evaluations within the two approaches,
and also with the {\tt pySecDec} evaluations.
Furthermore, in this process we also reevaluate master integrals that do not involve eMPLs,
which we find to completely agree with the high-precision numerical evaluation of their analytic
expressions.
Our results, correct to 1000 digits, can be found in 
the file \texttt{numerics.m} at ref.~\cite{masterIntegralsGit}.

\subsection{Relations among elliptic master integrals}
\label{sec:pslq}

The fact that the algebraic properties and the numerical evaluation of 
MPLs is well understood allowed us to use the {\tt PSLQ} algorithm to 
express the MIs that evaluate to MPLs in a basis of MPLs with specific arguments.
This results  in rather compact analytic representations for those MIs.
This is in contrast to the case of MIs that involve eMPLs and/or iterated 
integrals of Eisenstein series where, as already noted in \cref{sec:direct_int},
analytic expressions can be extremely long. Indeed, it is currently very
complicated to simplify expressions involving eMPLs, because not much 
is known about how to manipulate such functions.

In order to simplify our analytic expressions, we were nevertheless able to 
find relations between the coefficients in the Laurent expansion of certain 
elliptic master integrals. We were motivated by the fact that the poles of the 
two-loop amplitudes for quarkonium or leptonium production or decay considered 
in ref.~\cite{Abreu:2022cco} are free of elliptic contributions.\footnote{This is a 
consequence of the fact that the pole-structure of the two-loop amplitudes 
is determined by one-loop amplitudes~\cite{Catani:1998bh}, which do not involve 
elliptic integrals.}
Using the $\tt{PSLQ}$ algorithm we found 
the following relations:
\begin{align}\label{eq:ellipticRel}
    F^{(0)}_{7}&=\frac{1}{12}F^{(0)}_{15}-\frac{1}{4}F^{(0)}_{31}+\frac{247}{768}\zeta_4-\frac{1}{4}\zeta_2 \log^2{2}+\frac{7}{24}\zeta_3\log{2} \nonumber
    \\
    &-\frac{3}{16}\text{Re}{\left[G{\left(0,0,e^{-i\pi/3},-1;1\right)}\right]}-\frac{9}{64}\text{Re}{\left[G{\left(0,0,e^{-2i\pi/3},1;1\right)}\right]}\,, \nonumber
    \\
    F^{(0)}_{11}&=2 F^{(0)}_{18}-5 F^{(0)}_{57}-\frac{25}{2}\zeta_3+3\zeta_2+3 \zeta_2 \log{2}
    \,, \nonumber
    \\
    F^{(1)}_{11}&=2 F^{(1)}_{12}-\frac{28}{25} F^{(0)}_{15}+12 F^{(0)}_{18}-\frac{284}{25} F^{(0)}_{31}-\frac{112}{5} F^{(0)}_{57}-\frac{19}{5} F^{(1)}_{57}+18\zeta_2 \nonumber
    \\
    &+\frac{2391}{800}\zeta_4+18\zeta_2\log{(2)}+\frac{24}{25}\zeta_2 \log^2{2}-75\zeta_3-\frac{77}{10}\zeta_3\log{2} \nonumber
    \\
    &-\frac{2187}{50}\text{Re}{\left[G{\left(0,0,e^{-i\pi/3},-1;1\right)}\right]}-\frac{6561}{200}\text{Re}{\left[G{\left(0,0,e^{-2i\pi/3},1;1\right)}\right]}\,, \nonumber
    \\
    F^{(0)}_{12}&=F^{(0)}_{18}-\frac{3 }{5}F^{(0)}_{57}-\frac{29}{40} \zeta_3
    +\frac{3}{10}\zeta_2 \log (2)+\frac{3}{2}\zeta_2\,, \nonumber
    \\
    F^{(0)}_{13}&=\frac{2}{15} F^{(0)}_{15}+\frac{2}{5}F^{(0)}_{31}-\frac{157}{240}\zeta_4-\frac{2}{5}\zeta_2\log^2{2}+\frac{7}{6}\zeta_3 \log{2} \nonumber
    \\
    &+\frac{3}{5}\text{Re}{\left[G{\left(0,0,e^{-i\pi/3},-1;1\right)}\right]}+\frac{9}{20}\text{Re}{\left[G{\left(0,0,e^{-2i\pi/3},1;1\right)}\right]}\,, \nonumber
    \\
    F^{(0)}_{21}&=-\frac{2}{5}F^{(0)}_{15}-\frac{6}{5}F^{(0)}_{31}+\frac{23}{320}\zeta_4+\frac{6}{5}\zeta_2\log^2{2}-\frac{7}{2}\zeta_3 \log{2} \nonumber
    \\
    &-\frac{63}{10}\text{Re}{\left[G{\left(0,0,e^{-i\pi/3},-1;1\right)}\right]}-\frac{189}{40}\text{Re}{\left[G{\left(0,0,e^{-2i\pi/3},1;1\right)}\right]}\,, \nonumber
    \\
    F^{(0)}_{23}&=-\frac{1}{5}F^{(0)}_{15}-3F^{(0)}_{22}-\frac{13}{5}F^{(0)}_{31}+\frac{51479}{11520}\zeta_4-\frac{5}{12}\pi \textrm{Cl}_2{\left(\frac{\pi}{3}\right)} \log{2} \nonumber
    \\
    &+\frac{57}{20} \zeta_2 \log^2{2}-\frac{1}{8}\pi \textrm{Cl}_2{\left(\frac{\pi}{3}\right)} \log{3} +\frac{9}{2}\zeta_2\, \text{Li}_2{\left(-\frac{1}{2}\right)} - \frac{187}{24} \zeta_3 \log{2} \nonumber
    \\
    &-\frac{3}{8}\pi \text{Im}{\left[G{\left(0,1,e^{-2i\pi/3};1\right)}\right]}-\frac{1}{2}\pi \text{Im}{\left[G{\left(0,e^{-i\pi/3},-1;1\right)}\right]} \nonumber
    \\
    &+\frac{33}{80}\text{Re}{\left[G{\left(0,0,e^{-i\pi/3},-1;1\right)}\right]}+\frac{99}{320}\text{Re}{\left[G{\left(0,0,e^{-2i\pi/3},1;1\right)}\right]} \nonumber
    \\
    &+3\text{Re}{\left[G{\left(e^{-i\pi/3},1,1,-1;1\right)}\right]}+\frac{3}{2}\text{Re}{\left[G{\left(e^{-i\pi/3},1,-1;1\right)}\right]}\log{2}\,, \nonumber
    \\
    F^{(0)}_{32}&=-F^{(0)}_{15}-F^{(0)}_{31}-\frac{36259}{6912}\zeta_4+\frac{5}{36}\pi\textrm{Cl}_2{\left(\frac{\pi}{3}\right)}\log{2}+\frac{9}{4}\zeta_2\log^2{2} \nonumber
    \\
    &+\frac{1}{24}\pi \textrm{Cl}_2{\left(\frac{\pi}{3}\right)} \log{3}-\frac{3}{2}\zeta_2\, \text{Li}_2{\left(-\frac{1}{2}\right)}-\frac{317}{72} \zeta_3 \log{2}
    \\
    &+\frac{1}{8}\pi \text{Im}{\left[G{\left(0,1,e^{-2i\pi/3};1\right)}\right]}+\frac{1}{6}\pi \text{Im}{\left[G{\left(0,e^{-i\pi/3},-1;1\right)}\right]} \nonumber
    \\
    &-\frac{103}{16}\text{Re}{\left[G{\left(0,0,e^{-i\pi/3},-1;1\right)}\right]}-\frac{309}{64}\text{Re}{\left[G{\left(0,0,e^{-2i\pi/3},1;1\right)}\right]} \nonumber
    \\
    &-\text{Re}{\left[G{\left(e^{-i\pi/3},1,1,-1;1\right)}\right]}-\frac{1}{2}\text{Re}{\left[G{\left(e^{-i\pi/3},1,-1;1\right)}\right]}\log{2}\,, \nonumber
    \\
    F^{(0)}_{33}&=\frac{1}{2}F^{(0)}_{64}-\frac{1}{2}F^{(0)}_{65}+\frac{5}{4}\zeta_2-\frac{89}{16}\,, \nonumber
    \\
    F^{(1)}_{33}&=-\frac{23}{5}F^{(0)}_{57}-\frac{7}{4}F^{(0)}_{64}+F^{(0)}_{65}+\frac{1}{2}F^{(1)}_{64}-\frac{1}{2}F^{(1)}_{65}-\frac{5}{12} \pi  \,\text{Cl}_2\left(\frac{\pi}{3}\right) \nonumber
    \\
    &-\frac{3}{4} \text{Li}_2\left(-\frac{1}{2}\right) \log{2}
    -\frac{847}{120}\zeta_3+\frac{3}{4} \log^2{2}\, \log{3}-
    \frac{3}{8}\log^3{2} \nonumber
    \\
    &+\frac{9}{2} \zeta_2 \log{3}-\frac{39}{20} \zeta_2 \log{2}+\frac{57}{8} - \frac{3}{2} \text{Re}\left[G\left(e^{-i\pi/3},1,-1;1\right)\right]\,, \nonumber
    \\
    F^{(2)}_{33}&=\frac{4}{25}F^{(0)}_{15}-\frac{288}{25}F^{(0)}_{31}+\frac{46}{5}F^{(0)}_{57}+\frac{3}{2}F^{(0)}_{64}-\frac{23}{5}F^{(1)}_{57}-\frac{7}{4}F^{(1)}_{64}+F^{(1)}_{65}+\frac{1}{2}F^{(2)}_{64} \nonumber
    \\
    &-\frac{1}{2}F^{(2)}_{65}+\frac{1543657}{28800}\zeta_4+\frac{261}{50}\zeta_2\log^2{2}+\frac{3}{4}\log^4{2}-\frac{1}{4}\pi \textrm{Cl}_2{\left(\frac{\pi}{3}\right)} \log{3} \nonumber
    \\
    &-6 \zeta_2 \log{2}\, \log{3} -\frac{3}{2} \log^3{2}\, \log{3}+ 6 \zeta_2\, \text{Li}_2{\left(-\frac{1}{2}\right)} +\frac{3}{2}\log^2{2}\, \text{Li}_2{\left(-\frac{1}{2}\right)} \nonumber
    \\
    &-\frac{313}{20}\zeta_3\log{2}-\frac{3}{4}\pi\text{Im}{\left[G{\left(0,1,e^{-2i\pi/3};1\right)}\right]} - \pi \text{Im}{\left[G{\left(0,e^{-i\pi/3},-1;1\right)}\right]} \nonumber
    \\
    &-\frac{12561}{200}\text{Re}{\left[G{\left(0,0,e^{-i\pi/3},-1;1\right)}\right]} - \frac{26883}{800}\text{Re}{\left[G{\left(0,0,e^{-2i\pi/3},1;1\right)}\right]} \nonumber
    \\
    &+6\text{Re}{\left[G{\left(e^{-i\pi/3},1,1,-1;1\right)}\right]}+6\text{Re}{\left[G{\left(e^{-i\pi/3},1,-1;1\right)}\right]}\log{2}\,, \nonumber
    \\
    F^{(0)}_{48}&=-\frac{12 }{5}F^{(0)}_{57}+F^{(0)}_{64}-F^{(0)}_{65}-\frac{32}{5} \zeta_3
    +\frac{6}{5}\zeta_2 \log{2}-\frac{1}{2}\zeta_2-\frac{13}{8}\,, \nonumber
    \\
    F^{(1)}_{48}&=\frac{16}{25}F^{(0)}_{15}-\frac{152}{25}F^{(0)}_{31}-\frac{12}{5}F^{(0)}_{57}+\frac{3}{2}F^{(0)}_{64}-3 F^{(0)}_{65}-\frac{12}{5}F^{(1)}_{57}+F^{(1)}_{64}-F^{(1)}_{65} \nonumber
    \\
    &-\frac{71}{8}-\frac{5}{2}\zeta_2+\frac{149}{400}\zeta_4+\frac{18}{5}\zeta_2\log{2}-\frac{78}{25}\zeta_2 \log^2{2}-\frac{283}{15}\zeta_3 +\frac{7}{5}\zeta_3\log{2} \nonumber
    \\
    &-\frac{693}{25}\text{Re}{\left[G{\left(0,0,e^{-i\pi/3},-1;1\right)}\right]}-\frac{2079}{100}\text{Re}{\left[G{\left(0,0,e^{-2i\pi/3},1;1\right)}\right]}\,, \nonumber
    \\
    F^{(0)}_{49}&=-\frac{6 }{5}F^{(0)}_{57}-F^{(0)}_{64}+2 F^{(0)}_{65}
    +\frac{3}{5}\zeta_2  \log{2}-\frac{16 }{5}\zeta_3+\zeta_2+\frac{73}{8}\,, \nonumber
    \\
    F^{(1)}_{49}&=-\frac{2}{25} F^{(0)}_{15}-\frac{56}{25} F^{(0)}_{31}+\frac{6}{5}F^{(0)}_{57}+\frac{1}{2}F^{(0)}_{64}+2F^{(0)}_{65}-\frac{6}{5}F^{(1)}_{57}-F^{(1)}_{64}+2F^{(1)}_{65} \nonumber
    \\
    &+\frac{61}{8}+\frac{13}{2}\zeta_2+\frac{1719}{800}\zeta_4+\frac{3}{5}\zeta_2 \log{2}-\frac{9}{25} \zeta_2 \log^2{2}-\frac{58}{15}\zeta_3 +\frac{7}{10}\zeta_3\log{2} \nonumber
    \\
    &-\frac{783}{50}\text{Re}{\left[G{\left(0,0,e^{-i\pi/3},-1;1\right)}\right]}-\frac{2349}{200}\text{Re}{\left[G{\left(0,0,e^{-2i\pi/3},1;1\right)}\right]}\,, \nonumber
\end{align}
where $\zeta_n$ denotes the Riemann zeta function evaluated at $n$ (in particular, $\zeta_2=\pi^2/6$ and $\zeta_4=\pi^4/90$) and
$\textrm{Cl}_2(x)$ denotes the Clausen function, $\text{Cl}_2\left(x\right) = \textrm{Im}\left[\textrm{Li}_2(e^{ix})\right]$.
It would be very interesting to find a systematic way to derive such identities,
and some might follow from the type of relations 
discussed in \cref{sec:extraRel}. In practice, using these relations we are able to write
some of the lengthiest elliptic integrals in terms of much more compact 
expressions.

\section{Analytic results for the para-positronium decay up to NNLO}
\label{sec:parap}

As mentioned in \cref{sec:setup}, the master integrals we have
discussed in this paper allow us to compute the NNLO contributions to the
production and decay of several quarkonium and leptonium 
pseudo-scalar bound states.
While we leave a more extensive discussion of such calculations to a 
separate publication~\cite{Abreu:2022cco}, we finish this paper
by discussing the NNLO corrections to the para-positronium ($e^+ e^-$) 
decay to two photons, presenting for the first time complete
analytic results for this contribution.

The para-positronium state is a pseudo-scalar particle with 
spectroscopic notation $0^{+-}$ ($J^{PC}$).\footnote{The vector 
particle case would be the ortho-positronium with notation $1^{--}$.}
The most precise experimental measurement for its decay 
width includes the decay to any even 
number of photons, giving~\cite{Al-Ramadhan:1994num}
\begin{equation}\label{eq:expRes}
    \Gamma^{\text{exp.}}_{\text{p-Ps decay}}=\left(7990.9 \pm 1.7\right) \left(\mu s\right)^{-1}.
\end{equation}
The leading-order of the decay to four photons is of the
same order in perturbation theory as the NNLO corrections
to the decay to two photons. We write
\begin{equation}
    \Gamma_{\text{p-Ps decay}}=\Gamma_{\text{p-Ps}\rightarrow \gamma \gamma}
    +\Gamma_{\text{p-Ps}\rightarrow 4\gamma},
\end{equation}
where the leading-order contribution to  
$\Gamma_{\text{p-Ps}\rightarrow 4\gamma}$ has been computed analytically 
in ref.~\cite{Lee:2017ftw},
\begin{equation}\label{eq:dec4y}
    \Gamma_{\text{p-Ps}\rightarrow 4\gamma}=
    \frac{m_e\,\alpha_{em}^5}{2} \left(\frac{\alpha_{em}}{\pi}\right)^2 
    \left(
    \frac{112}{5}-\frac{3\pi^2}{2}+\frac{3}{10}\zeta_3+\frac{\pi^4}{24}+\frac{697}{15}\pi^2\log{2}-\frac{152}{5}\pi^2 \log{3}
    \right)\,.
\end{equation}
The NNLO corrections to $\Gamma_{\text{p-Ps}\rightarrow \gamma \gamma}$
were first computed in purely numerical form more than 20 years 
ago~\cite{Czarnecki:1999ci,Czarnecki:1999gv} (see also ref.~\cite{Adkins:2003eh}), 
and in this section we present them for the first time
in analytic form.

We can express the decay of the para-positronium to two photons up to NNLO accuracy in QED as \cite{Czarnecki:1999ci,Melnikov:2000fi,Adkins:2003eh},
\begin{equation}
\begin{split}
\Gamma_{\text{p-Ps}\rightarrow \gamma \gamma}=&\frac{m_e\,\alpha_{em}^5}{2} 
\left[1+\left(\frac{\alpha_{em}}{\pi}\right)\left(\frac{\pi^2}{4}-5\right)
-2\alpha_{em}^2\log{\alpha_{em}}
-\frac{3\alpha_{em}^3}{2\pi}\log^2{\alpha_{em}}
\right.
\\
&\left.
+\frac{\alpha_{em}^3}{\pi} \log{\alpha_{em}} 
\left(\frac{533}{90}+10\log{2}-\frac{\pi^2}{2}\right)
+\left(\frac{\alpha_{em}}{\pi}\right)^2\left(K_2+K_{2,\text{soft}}\right)
+\mathcal{O}\left(\alpha_{em}^3\right)\right],
\end{split}
\label{eq:DecNNLOexpression}
\end{equation}
where $\alpha_{em}$ is the electromagnetic 
fine structure constant and $m_e$ is the electron mass.
The terms of the form $\alpha_{em}^n \log^k{\alpha_{em}}$ and
$K_{2,\text{soft}}$ are corrections related to the leptonium bound 
state \cite{Khriplovich:1990eh,Karshenboim:1993,Melnikov:2000fi,Kniehl:2000dh}.
These and the LO and NLO corrections were already known 
in analytic form in the literature. We note that $K_2$ and $K_{2,\text{soft}}$
are independently divergent (each has a so-called Coulomb singularity)
but their sum is finite. The master integrals that we
have computed in this paper allow us to obtain for the 
first time the analytic results for the two-loop
contribution $K_2$. 

\begin{figure}
\centering
    \subfloat[]{\includegraphics[width=4.2cm,angle=180]{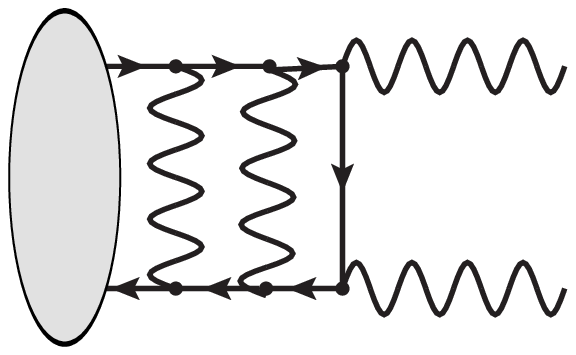}\label{fig:positroniumabelian}}\, \subfloat[]{\includegraphics[width=4.6cm,angle=180]{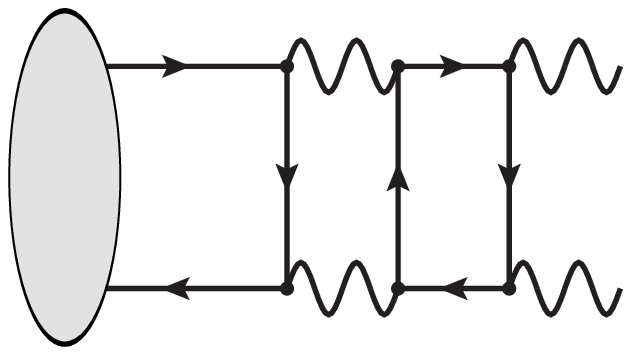} \label{fig:positroniumlbl}}\, \subfloat[]{\includegraphics[width=4.2cm,angle=180]{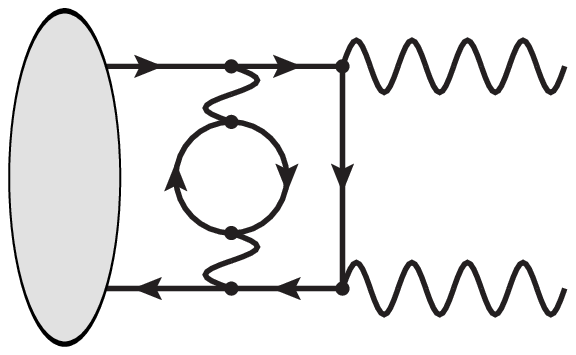} \label{fig:positroniumvac}}
  \caption{Two-loop diagrams for the decay of para-positronium into two photons
    with (a) regular corrections, (b) light-by-light contributions and (c) vacuum polarisation corrections.}
  \label{fig:positroniumdiagrams}
\end{figure}

We follow the presentation of refs.~\cite{Czarnecki:1999ci,Czarnecki:1999gv}
and express the two-loop virtual amplitudes contributing
to $K_2$ in terms of three different types of terms:
(a) regular diagrams without fermion loops as shown in 
fig.~\ref{fig:positroniumabelian}, 
(b) contributions with fermion loops in diagrams of type light-by-light scattering as in fig.~\ref{fig:positroniumlbl} and (c) 
contributions with vacuum polarisations as in fig.~\ref{fig:positroniumvac}.
These different contributions have ultraviolet poles
which can be removed by renormalisation.  After this process,
a single pole remains which is related to a Coulomb singularity
(see, e.g.,~ref.~\cite{Czarnecki:1999ci}).
We then write:
\begin{equation}\label{eq:paraDecay}
    \mathcal{A}^{(2-\text{loop})}_{\text{ren.}}=
    \mathcal{A}^{(2-\text{loop})}_{\text{reg ren.}}+
    \mathcal{B}^{(2-\text{loop})}_{\text{lbl ren.}}+
    \mathcal{B}^{(2-\text{loop})}_{\text{vac ren.}}+
    \mathcal{A}^{(2-\text{loop})}_{\text{Coul.}},
\end{equation}
where $\mathcal{A}^{(2-\text{loop})}_{\text{ren.}}$ denotes
the two-loop amplitude obtained after subtracting the ultraviolet
poles, which can be done contribution by contribution
in the decomposition of \cref{fig:positroniumdiagrams}, and 
\begin{equation}
\mathcal{A}^{(2-\text{loop})}_{\text{Coul.}}=-\frac{\pi^2}{4 \epsilon}.
\label{eq:Coulombsing}
\end{equation}
The analytic expressions for the remaining terms in \cref{eq:paraDecay} are:

\begin{align}
    \mathcal{A}^{(2-\text{loop})}_{\text{reg, ren.}}=&\frac{1261}{96}-\frac{2579}{1728}\pi^2-\frac{57743}{4147200}\pi^4+\frac{35}{288}\pi\Cl_2{\left(\frac{\pi}{3}\right)}+\frac{3}{32}\pi\Ime{\left[G{\left(0,1,e^{-\frac{2i\pi}{3}};1\right)}\right]}\nonumber
    \\
    & \hspace{-1cm} +\frac{1}{8}\pi \Ime{\left[G{\left(0,e^{-\frac{i\pi}{3}},-1;1\right)}\right]}+\frac{7}{16}\Ree{\left[G{\left(e^{-\frac{i\pi}{3}},1,-1;1\right)}\right]}-\frac{681}{320}\Ree{\left[G{\left(0,0,e^{-\frac{i\pi}{3}},-1;1\right)}\right]}\nonumber
    \\
    & \hspace{-1cm} -\frac{2043}{1280}\Ree{\left[G{\left(0,0,e^{-\frac{2i\pi}{3}},1;1\right)}\right]}-\frac{3}{4}\Ree{\left[G{\left(e^{-\frac{i\pi}{3}},1,1,-1;1\right)}\right]}+\frac{11}{6}\log{2}-\frac{1253}{960}\pi^2 \log{2}
    \nonumber\\
    & \hspace{-1cm} +\frac{5}{48}\pi\Cl_2{\left(\frac{\pi}{3}\right)}\log{2}-\frac{3}{8}\Ree{\left[G{\left(e^{-\frac{i\pi}{3}},1,-1;1\right)}\right]}\log{2}-\frac{37}{24}\log^2{2}-\frac{43}{160}\pi^2\log^2{2}+\frac{7}{64}\log^3{2}
    \nonumber\\
    & \hspace{-1cm} -\frac{7}{32}\pi^2 \log{3}+\frac{1}{32}\pi \Cl_2{\left(\frac{\pi}{3}\right)}\log{3}-\frac{7}{32}\log^2{2}\,\log{3}-\frac{3}{16}\pi^2\Li_2{\left(-\frac{1}{2}\right)}+\frac{7}{32}\log{2}\,\Li_2{\left(-\frac{1}{2}\right)}
    \nonumber\\
    & \hspace{-1cm} +\frac{3923}{2880}\zeta_3+\frac{355}{96}\zeta_3\log{2}-F_3^{(0)}-3 F_{14}^{(0)}+\frac{7}{20} F_{15}^{(0)} -\frac{11}{6} F_{18}^{(0)}-F_{22}^{(0)}+\frac{31}{20} F_{31}^{(0)}+\frac{59}{15} F_{57}^{(0)}
    \nonumber\\
    & \hspace{-1cm} -\frac{29}{36} F_{64}^{(0)}+\frac{37}{36} F_{65}^{(0)},
    \\[12pt]
    \mathcal{B}^{(2-\text{loop})}_{\text{lbl}}=&\left[-\frac{631}{24}+\frac{7}{12}\pi^2+\frac{2}{3}\log{2}+\frac{113}{60}\pi^2\log{2}+\frac{2}{3}\pi^2\log{\left(-1+\sqrt{2}\right)}-\frac{4}{3}\log^3{\left(-1+\sqrt{2}\right)}\right.
    \nonumber\\
    &\left. -2\Li_3{\left(3-2\sqrt{2}\right)}-\frac{647}{120}\zeta_3\right]+i \pi\left[-\frac{1}{3}+\frac{1}{4}\pi^2-2\log^2{\left(-1+\sqrt{2}\right)}\right]
    \nonumber\\
    & -6 F_{14}^{(0)}+\frac{2}{3} F_{24}^{(0)}+\frac{2}{5}F_{57}^{(0)}+\frac{7}{3} F_{64}^{(0)}-3 F_{65}^{(0)},
    \\[12pt]
    \mathcal{B}^{(2-\text{loop})}_{\text{vac, ren.}}=&-\frac{15061}{1440}-\frac{107}{2160}\pi^2-\frac{1}{60}\pi^2 \log{2}+\frac{8}{15}\zeta_3+\frac{1}{5} F_{57}^{(0)}+\frac{217}{180} F_{64}^{(0)}-\frac{19}{9} F_{65}^{(0)},
\end{align}
where we have kept the elliptic contributions in the symbolic form $F_{I}^{(k)}$,
whose analytic representation can be found at ref.~\cite{masterIntegralsGit}. We note that the 
two-loop amplitude exhibits functions of maximal weight four (for MPLs) and 
maximal length four (for elliptic functions), 
which is in full agreement with the conjectural property 
that scattering amplitudes at $n$-loops should exhibit functions of maximal 
weight/length $2n$.
Using our high-precision numerical evaluations, these
different contributions evaluate to
\begin{align}
\mathcal{A}^{(2-\text{loop})}_{\text{reg, ren.}}=&-21.10789796731067145661\ldots,
\label{eq:ampregren}
\\
\mathcal{B}^{(2-\text{loop})}_{\text{lbl}}=&0.64696557211233073992\ldots+i\,2.07357555846158085167 \ldots,
\label{eq:amplblren}
\\
\mathcal{B}^{(2-\text{loop})}_{\text{vac, ren.}}=&0.22367201327357266787\ldots,
\label{eq:ampvacren}
\end{align}
where the numbers are truncated to 20 digits after the decimal. 

The NNLO contribution $K_2$ in eq.~\eqref{eq:DecNNLOexpression} can then be 
decomposed as
\begin{equation}
    K_2=\frac{1}{4}K_1^2+K_{2, \text{reg}}+K_{2, \text{lbl}}+K_{2, \text{vac}}+K_{2, \text{Coul.}}\,,
\end{equation}
where $K_1=\frac{\pi^2}{4}-5$ is the NLO contribution, and
the remaining terms are twice the real part of eqs.~\eqref{eq:ampregren}, \eqref{eq:amplblren}, 
\eqref{eq:ampvacren} and \eqref{eq:Coulombsing} respectively. 
The soft contribution takes the form \cite{Czarnecki:1999ci},
\begin{align}
    K_{2, \text{soft}}=\frac{107\pi^2}{24}-K_{2, \text{Coul.}}.
\end{align}
Therefore, combining all these expressions, we obtain the high-precision numerical result
\begin{equation}
    K_2+K_{2, \text{soft}}=5.1309798210659600230\ldots,
\end{equation}
where, as before, all number are truncated to 20 digits after the decimal. 
We note that the finite part of $K_{2}$ is dominated by the (negative) 
contribution of $K_{2, \text{reg}}$. The finite piece of $K_{2, \text{soft}}$ 
has a similarly sized positive contribution, leading to a NNLO perturbative 
correction of $\mathcal{O}(1)$.

Since the precision of our numerical evaluation of $K_2$ depends on the numerical evaluation of our master integrals, which are correct up to over 1000-digits accuracy, they do not contribute to the theory error associated with the NNLO decay width. The theory error is determined by the precision to which the QED coupling $\alpha_{em}$ and the electron mass $m_e$ are known.
We use for the coupling
\begin{equation}
    \alpha_{em}=\left(7.2973525693 \pm 0.0000000011\right)\times 10^{-3},
\end{equation}
and for the electron mass
\begin{equation}
    m_e= \left(0.5109989500 \pm 0.00000000015\right)~\text{MeV},
\end{equation}
where the errors are correlated with $r=-0.99998$~\cite{Zyla:2020zbs,221451}.
The propagation of these correlated errors to the decay width can be determined with:
\begin{equation}
\Delta{\Gamma}=\sqrt{\left(\Delta{m_e}\right)^2 \left(\frac{\partial \Gamma}{\partial m_e}\right)^2 + \left(\Delta{\alpha_{em}}\right)^2 \left(\frac{\partial \Gamma}{\partial \alpha_{em}}\right)^2+2 r\; \Delta{m_e} \Delta{\alpha_{em}} \frac{\partial \Gamma}{\partial \alpha_{em}}\frac{\partial \Gamma}{\partial m_e}}.
\end{equation}
We find that the decay width of para-Positronium to two photons is given by
\begin{align}
\begin{split}
    \Gamma^{\text{theory, LO}}_{\text{p-Ps}\rightarrow \gamma \gamma}=&\left(8032.502933 \pm 0.000004\right) \left(\mu s\right)^{-1},
\end{split}
\\
\begin{split}
    \Gamma^{\text{theory, NLO}}_{\text{p-Ps}\rightarrow \gamma \gamma}=&\left(7989.458752 \pm 0.000004\right) \left(\mu s\right)^{-1},
\end{split}
\\
\begin{split}
    \Gamma^{\text{theory, NNLO}}_{\text{p-Ps}\rightarrow \gamma \gamma}=&\left(7989.606333 \pm 0.000004\right) \left(\mu s\right)^{-1},
\end{split}
\end{align}
where for comparison we also quote the LO and NLO results. After a sizeable change in going
from LO to NLO, we find that the NNLO result is very close to the NLO. Our results
agree with those of refs.~\cite{Czarnecki:1999ci,Adkins:2003eh}, but we find
that the authors of ref.~\cite{Adkins:2003eh} have slightly underestimated their error.
Combining our results with those for the decay into four photons quoted in \cref{eq:dec4y},
we obtain
\begin{equation}
\begin{split}
    \Gamma^{\text{theory, NNLO}}_{\text{p-Ps decay}}=&\left(7989.618221 \pm 0.000004\right) \left(\mu s\right)^{-1},
\end{split}
\label{eq:ultraprecision}
\end{equation}
which is in full agreement with the experimental measurement quoted in \cref{eq:expRes}.


\section{Conclusions}\label{sec:conclusion}

In this paper we have computed the complete set of two-loop master integrals
contributing to the NNLO corrections to the production or decay of a 
pseudo-scalar bound state of two massive fermions of the same flavour.
We have presented both analytic results as well as high precision numerical
evaluations. The analytic results involve both MPLs and eMPLs (and
iterated integrals of Eisenstein series). The presence of elliptic
integrals presented the main challenge to the evaluation of this set of 
integrals. The high-precision numerical evaluations are comparatively
simple to obtain with modern tools such as 
\texttt{diffexp} \cite{Hidding:2020ytt}
and \texttt{AMFlow} \cite{Liu:2022chg}.
Besides being important for phenomenological studies,
these high-precision numerical evaluations also
allowed us to considerably simplify the analytic expressions
using the \texttt{PSLQ} algorithm. All our results can be obtained from 
the repository at ref.~\cite{masterIntegralsGit}. 
As an example of the type of processes these integrals can be used for,
we recomputed the two-loop corrections to the decay of para-positronium 
into two photons \cite{Czarnecki:1999ci,Czarnecki:1999gv,Adkins:2003eh},
presenting both analytic results and high-precision
numerical evaluations.

Computing Feynman integrals involving elliptic functions is still a very 
challenging task. Moreover, once analytic expressions are obtained,
it is not yet known how to simplify them or even how to efficiently
evaluate them. We expect this situation to improve substantially in
the coming years as more and more processes involving elliptic
functions are computed, and the analytic results we obtained in this paper
will be very useful in this process. For instance, understanding how
to systematically find the relations in \cref{sec:pslq} will be of
great importance: in our calculation, it allowed us to rewrite the lengthiest
elliptic integrals in terms of simpler ones. We also expect that there will
be important developments in the numerical evaluation of eMPLs, 
and reproducing the high-precision numerical evaluations of the master 
integrals from their analytic representation will provide an important test.

The set of master integrals we have computed will open the door
to new NNLO predictions for processes involving bound states of 
heavy fermions. Besides the para-positronium decay to two photons
discussed in this paper, for which experimental precision is expected
to increase at future experiments, see e.g.~\cite{Bass:2019ibo,Dulski:2020pqi}, and the equivalent process for (true) para-muonium and (true) para-tauonium,
our results also allow us to study the production and decay of
quarkonium states at an unprecedented level of precision. 
An interesting prospect is to use charmonium production to study the gluon 
parton distribution function of the proton, as these are rather unconstrained at 
scales close to the mass of the charm quark. 
Furthermore, it is interesting to study the convergence of the perturbative
expansion as the strong coupling $\alpha_s$ is not so small at these scales,
see refs.~\cite{Lansberg:2020ejc,Ozcelik:2021zqt} and references therein.
We will discuss several of these processes in a companion paper \cite{Abreu:2022cco}.

Regarding the evaluation of two-loop master integrals for processes with
quarkonium or leptonium states, the next steps are clear. The set of integrals
we considered here correspond to very simple kinematics, only depending
on the mass of the heavy fermion. Adding an extra particle in the final
state would lead to richer kinematic configurations and allow us to study other processes, for instance, hadro-production or photo-production of vector bound states with spectroscopic notation~${^3S_1}$, commonly called $J/\psi$ ($c\overline{c}$) and $\Upsilon$ ($b\overline{b}$). Due to the Landau-Yang theorem, the LO of these production processes involves an additional gluon in the final state making it effectively a 2-to-2 process ($gg\rightarrow J/\psi+g$). Similarly, we could study the $p_T$-distribution of $\eta_Q$ at NNLO accuracy by computing the two-loop corrections to the process $gg\rightarrow \eta_Q+g$. Tackling such calculations will pose many challenges. In particular, the numerical evaluation will be much more involved as the integrals will depend on several variables. As such, a single high-precision
evaluation of master integrals will no longer be sufficient.

\acknowledgments{We thank Stephen Jones for  assistance in evaluating
the most challenging integrals with \texttt{pySecDec}, and 
Martijn Hidding for his help in using \texttt{diffexp} for numerical
evaluations. We acknowledge useful discussions with Jean-Philippe Lansberg, Kirill Melnikov and Hua-Sheng Shao.
M.B.~acknowledges the financial support from the European Union Horizon 
2020 research and innovation programme: High precision multi-jet dynamics 
at the LHC (grant agreement no. 772009). 
The research of C.D.~and M.A.O.~was supported by the ERC Starting Grant 637019 ``MathAm''. 
M.A.O.~thanks the TH Department at CERN for hospitality while part of this work was carried out.
M.A.O.~also acknowledges the financial support from the following sources while at IJCLab, Universit\'e Paris-Saclay: the European Union’s Horizon 2020 research and innovation programme under grant agreement STRONG–2020 No 824093 in order to contribute to the EU Virtual Access NLOAccess (VA1-WG10), the funding from the Agence Nationale de la Recherche (ANR) via the grant ANR-20-CE31-0015 (“PrecisOnium”) and via the IDEX Paris-Saclay “Investissements d’Avenir” (ANR-11-IDEX-0003-01) through the GLUODYNAMICS project funded by the “P2IO LabEx (ANR-10-LABX-0038)”, and partially via the IN2P3 project GLUE@NLO funded by the French CNRS.
}

\appendix


\section{Master integrals}
\label{sec:mis}

In this appendix we give the definition of the 76 MIs studied in the present paper in terms of 12 different sets of denominators, which we list below:
\begin{itemize}
    \item \textbf{Set $t_1$}:
    \begin{align}
\begin{split}
D_{t_1} = & \left\{q_1^2, \left(q_1-k_1\right)^2, q_2^2, \left(q_2+k_2\right)^2, \left(q_1+q_2-k_1\right)^2, \left(q_2-\frac{k_1}{2}+\frac{k_2}{2}\right)^2-m_Q^2,\right.
\\
&\left. \left(q_1-\frac{k_1}{2}-\frac{k_2}{2}\right)^2-m_Q^2\right\},
\end{split}
\end{align}
\item \textbf{Set $t_2$}:
\begin{align}
\begin{split}
D_{t_2} = & \left\{q_1^2,\left(q_1-k_1\right)^2,q_2^2,\left(q_2+k_2\right)^2,\left(q_1+q_2-k_1\right)^2,\left(q_1-\frac{k_1}{2}-\frac{k_2}{2}\right)^2-m_Q^2,\right.
\\
&\left.\left(q_1+q_2-\frac{k_1}{2}+\frac{k_2}{2}\right)^2-m_Q^2\right\},
\end{split}
\end{align}
\item \textbf{Set $t_3$}:
\begin{align}
\begin{split}
D_{t_3} = &\left\{q_1^2-m^2_{Q}, \left(q_1-k_1\right)^2-m^2_{Q}, q_2^2, \left(q_1+q_2-k_1\right)^2-m^2_{Q}, \left(q_2-\frac{k_1}{2}-\frac{k_2}{2}\right)^2-m^2_{Q}, \right.
\\
& \left. \left(q_2-\frac{k_1}{2}+\frac{k_2}{2}\right)^2-m^2_{Q}, \left(q_1-\frac{k_1}{2}-\frac{k_2}{2}\right)^2\right\};
\end{split}
\end{align}
\item \textbf{Set $t_4$}:
\begin{align}
\begin{split}
D_{t_4} = &\left\{q_1^2-m^2_{Q}, \left(q_1+k_1\right)^2-m^2_{Q}, q_2^2-m^2_{Q}, \left(q_1+q_2\right)^2, \left(q_2-\frac{k_1}{2}-\frac{k_2}{2}\right)^2, \right.
\\
& \left. \left(q_1+q_2+\frac{k_1}{2}-\frac{k_2}{2}\right)^2-m^2_{Q}, \left(q_1+k_2\right)^2\right\};
\end{split}
\end{align}
\item \textbf{Set $t_5$}:
\begin{align}
\begin{split}
D_{t_5} = &\left\{q_1^2-m^2_{Q}, \left(q_1+k_1\right)^2-m^2_{Q}, \left(q_1+q_2\right)^2, \left(q_1+q_2+k_1+k_2\right)^2, q_2^2-m^2_{Q}, \right.
\\
& \left. \left(q_2+k_2\right)^2-m^2_{Q}, \left(q_1+k_2\right)^2\right\};
\end{split}
\end{align}
\item \textbf{Set $t_6$}:
\begin{align}
\begin{split}
D_{t_6} = &\left\{q_1^2-m^2_{Q}, \left(q_1+k_1\right)^2-m^2_{Q}, \left(q_1+q_2+k_1+k_2\right)^2, q_2^2-m^2_{Q}, \left(q_2+k_2\right)^2-m^2_{Q}, \right. 
\\
& \left. \left(q_1+q_2+\frac{k_1}{2}+\frac{k_2}{2}\right)^2-m^2_{Q}, \left(q_1+k_2\right)^2\right\};
\end{split}
\end{align}
\item \textbf{Set $t_7$}:
\begin{align}
\begin{split}
D_{t_7} = &\left\{q_1^2,\left(q_1+k_1\right)^2,q_2^2-m_Q^2,\left(q_2+k_2\right)^2-m_Q^2,\left(q_1+q_2\right)^2-m_Q^2, \left(q_1+q_2+k_1+k_2\right)^2-m_Q^2, \right.
\\
&\left.\left(q_1+k_2\right)^2\right\};
\end{split}
\end{align}
\item \textbf{Set $t_8$}:
\begin{align}
\begin{split}
D_{t_8} = &\left\{q_1^2,\left(q_1-\frac{k_1}{2}-\frac{k_2}{2}\right)^2-m_Q^2,\left(q_1-\frac{k_1}{2}+\frac{k_2}{2}\right)^2-m_Q^2,q_2^2,\left(q_1+q_2-k_1\right)^2,\right.
\\
&\left.\left(q_2-\frac{k_1}{2}-\frac{k_2}{2}\right)^2-m_Q^2, \left(q_1+q_2\right)^2\right\};
\end{split}
\end{align}
\item \textbf{Set $t_9$}:
\begin{align}
\begin{split}
D_{t_9} = &\left\{q_1^2-m_Q^2,\left(q_1-k_1\right)^2-m_Q^2,\left(q_1+k_2\right)^2-m_Q^2,q_2^2,\left(q_2-k_1-k_2\right)^2,\right.
\\
&\left.\left(q_1+q_2-k_1\right)^2-m_Q^2,\left(q_1+q_2\right)^2\right\};
\end{split}
\end{align}
\item \textbf{Set $t_{10}$}:
\begin{align}
\begin{split}
D_{t_{10}} = &\left\{q_1^2,\left(q_1-k_1\right)^2,\left(q_1+k_2\right)^2,\left(q_2-\frac{k_1}{2}-\frac{k_2}{2}\right)^2,q_2^2-m_Q^2,\left(q_1-q_2+k_2\right)^2-m_Q^2,\right.
\\
&\left.\left(q_1+q_2\right)^2\right\};
\end{split}
\end{align}
\item \textbf{Set $t_{11}$}:
\begin{align}
\begin{split}
D_{t_{11}} = &\left\{q_1^2,\left(q_1-k_1\right)^2,\left(q_1+k_2\right)^2,q_2^2,\left(q_2-k_1-k_2\right)^2,\left(q_1+q_2-k_1\right)^2,\left(q_1+q_2\right)^2\right\};
\end{split}
\end{align}
\item \textbf{Set $t_{12}$}:
\begin{align}
\begin{split}
D_{t_{12}} = &\left\{q_1^2, (q_1-k_1)^2, q_2^2, (q_1+q_2-k_1)^2, \left(q_2-\frac{k_1}{2}-\frac{k_2}{2}\right)^2-m_Q^2,\right.
\\
&\left.\left(q_2-\frac{k_1}{2}+\frac{k_2}{2}\right)^2-m_Q^2,\left(q_1-\frac{k_1}{2}-\frac{k_2}{2}\right)^2-m_Q^2\right\}.
\end{split}
\end{align}
\end{itemize}

In terms of the previous 12 sets, the four-point non-factorisable two-loop integrals in fig.~\ref{fig:four-point} can be written as:
\begin{eqnarray}
\masterNew{1} & = & m_{I_{(t_1)}}(1,1,1,1,1,1,1;m_Q^2), \quad
\masterNew{2}  =  m_{I_{(t_2)}}(1,1,1,1,1,1,1;m_Q^2), \nn \\
\masterNew{3} & = & m_{I_{(t_3)}}(1,1,1,1,1,1,1;m_Q^2), \quad
\masterNew{4}  =  m_{I_{(t_{12})}}(1,1,1,1,1,1,1;m_Q^2), \nn \\
\masterNew{5} & = & m_{I_{(t_6)}}(1,1,1,1,1,1,0;m_Q^2), \quad
\masterNew{6}  =  m_{I_{(t_8)}}(0,1,1,1,1,1,1;m_Q^2), \nn \\
\masterNew{7} & = & m_{I_{(t_3)}}(1,1,0,1,1,1,1;m_Q^2), \quad
\masterNew{8}  =  m_{I_{(t_8)}}(1,1,1,1,1,1,0;m_Q^2), \nn \\
\masterNew{9} & = & m_{I_{(t_2)}}(1,0,1,0,1,1,1;m_Q^2), \quad
\masterNew{10}  =  m_{I_{(t_2)}}(1,1,1,1,0,0,1;m_Q^2), \nn \\
\masterNew{11} & = & m_{I_{(t_3)}}(1,0,1,1,1,0,1;m_Q^2), \quad
\masterNew{12}  =  m_{I_{(t_3)}}(1,0,1,2,1,0,1;m_Q^2), \nn \\
\masterNew{13} & = & m_{I_{(t_3)}}(1,1,0,1,1,1,0;m_Q^2), \quad
\masterNew{14}  =  m_{I_{(t_3)}}(2,1,0,1,1,1,0;m_Q^2), \nn \\
\masterNew{15} & = & m_{I_{(t_6)}}(1,0,1,1,1,1,0;m_Q^2), \quad
\masterNew{16}  =  m_{I_{(t_8)}}(0,1,1,1,0,1,1;m_Q^2), \nn \\
\masterNew{17} & = & m_{I_{(t_8)}}(0,1,1,1,1,0,1;m_Q^2), \quad
\masterNew{18}  =  m_{I_{(t_3)}}(1,0,1,1,1,1,0;m_Q^2), \nn \\
\masterNew{19} & = & m_{I_{(t_8)}}(1,1,1,1,1,0,0;m_Q^2).
\end{eqnarray}
Similarly, the three-point integrals in fig.~\ref{fig:three-point} admit the expressions:
\begin{eqnarray}
\masterNew{20} & = & m_{I_{(t_2)}}(0,1,1,1,1,1,1;m_Q^2), \quad
\masterNew{21}  =  m_{I_{(t_3)}}(0,1,1,1,1,1,1;m_Q^2), \nn \\
\masterNew{22} & = & m_{I_{(t_4)}}(1,1,1,1,1,1,0;m_Q^2), \quad
\masterNew{23}  =  m_{I_{(t_4)}}(1,1,1,1,1,1,-1;m_Q^2), \nn \\
\masterNew{24} & = & m_{I_{(t_5)}}(1,1,1,1,1,1,0;m_Q^2), \quad
\masterNew{25}  =  m_{I_{(t_5)}}(2,1,1,1,1,1,0;m_Q^2), \nn \\
\masterNew{26} & = & m_{I_{(t_7)}}(1,1,1,1,1,1,0;m_Q^2), \quad
\masterNew{27}  =  m_{I_{(t_{11})}}(1,0,1,1,1,1,1;m_Q^2), \nn \\
\masterNew{28} & = & m_{I_{(t_1)}}(1,0,0,1,1,1,1;m_Q^2), \quad
\masterNew{29}  =  m_{I_{(t_1)}}(0,1,0,1,1,1,1;m_Q^2), \nn \\
\masterNew{30} & = & m_{I_{(t_1)}}(0,2,0,1,1,1,1;m_Q^2), \quad
\masterNew{31}  =  m_{I_{(t_3)}}(0,1,0,1,1,1,1;m_Q^2), \nn \\
\masterNew{32} & = & m_{I_{(t_4)}}(1,0,1,1,1,1,0;m_Q^2), \quad
\masterNew{33}  =  m_{I_{(t_4)}}(2,0,1,1,1,1,0;m_Q^2), \nn \\
\masterNew{34} & = & m_{I_{(t_4)}}(1,1,0,1,1,1,0;m_Q^2), \quad
\masterNew{35}  =  m_{I_{(t_4)}}(1,1,0,1,1,2,0;m_Q^2), \nn \\
\masterNew{36} & = & m_{I_{(t_4)}}(1,1,1,1,1,0,0;m_Q^2), \quad
\masterNew{37}  =  m_{I_{(t_5)}}(1,1,0,1,1,1,0;m_Q^2), \nn \\
\masterNew{38} & = & m_{I_{(t_5)}}(0,1,1,1,1,1,0;m_Q^2), \quad
\masterNew{39}  =  m_{I_{(t_7)}}(0,1,1,1,1,1,0;m_Q^2), \nn \\
\masterNew{40} & = & m_{I_{(t_{10})}}(0,1,1,1,1,1,0;m_Q^2), \quad
\masterNew{41}  =  m_{I_{(t_1)}}(0,1,0,1,1,0,1;m_Q^2), \nn \\
\masterNew{42} & = & m_{I_{(t_2)}}(0,1,1,0,0,1,1;m_Q^2), \quad
\masterNew{43}  =  m_{I_{(t_2)}}(0,1,2,0,0,1,1;m_Q^2), \nn \\
\masterNew{44} & = & m_{I_{(t_2)}}(0,2,1,0,0,1,1;m_Q^2), \quad
\masterNew{45}  =  m_{I_{(t_2)}}(0,0,1,0,1,1,1;m_Q^2), \nn \\
\masterNew{46} & = & m_{I_{(t_2)}}(0,0,2,0,1,1,1;m_Q^2), \quad
\masterNew{47}  =  m_{I_{(t_3)}}(0,1,0,1,1,0,1;m_Q^2), \nn \\
\masterNew{48} & = & m_{I_{(t_3)}}(1,0,0,1,1,0,1;m_Q^2), \quad
\masterNew{49}  =  m_{I_{(t_3)}}(0,1,0,1,1,1,0;m_Q^2), \nn \\
\masterNew{50} & = & m_{I_{(t_5)}}(0,1,1,1,1,0,0;m_Q^2), \quad
\masterNew{51}  =  m_{I_{(t_5)}}(0,2,1,1,1,0,0;m_Q^2), \nn \\
\masterNew{52} & = & m_{I_{(t_5)}}(0,1,1,0,1,1,0;m_Q^2), \quad
\masterNew{53}  =  m_{I_{(t_7)}}(0,0,2,0,1,1,1;m_Q^2), \nn \\
\masterNew{54} & = & m_{I_{(t_7)}}(0,0,3,0,1,1,1;m_Q^2), \quad
\masterNew{55}  =  m_{I_{(t_{10})}}(0,1,1,1,0,1,0;m_Q^2), \nn \\
\masterNew{56} & = & m_{I_{(t_{11})}}(0,0,1,1,1,0,1;m_Q^2),
\end{eqnarray}
and the two-point integrals in fig.~\ref{fig:two-point}:
\begin{eqnarray}
\masterNew{57} & = & m_{I_{(t_3)}}(0, 1, 1, 1, 0, 1, 1;m_Q^2), \quad
\masterNew{58}  =  m_{I_{(t_9)}}(0,1,1,1,1,1,0;m_Q^2), \nn \\
\masterNew{59} & = & m_{I_{(t_1)}}(0,0,1,0,1,0,1;m_Q^2), \quad
\masterNew{60}  =  m_{I_{(t_1)}}(0,0,2,0,1,0,1;m_Q^2), \nn \\
\masterNew{61} & = & m_{I_{(t_1)}}(0,0,0,1,1,0,1;m_Q^2), \quad
\masterNew{62}  =  m_{I_{(t_1)}}(1,0,0,1,1,0,0;m_Q^2), \nn \\
\masterNew{63} & = & m_{I_{(t_3)}}(0,1,0,1,1,0,0;m_Q^2), \quad
\masterNew{64}  =  m_{I_{(t_3)}}(0, 1, 0, 1, 0, 1, 0;m_Q^2), \nn \\
\masterNew{65} & = & m_{I_{(t_3)}}(0, 2, 0, 1, 0, 1, 0;m_Q^2), \quad
\masterNew{66}  =  m_{I_{(t_5)}}(0,1,1,0,0,1,0;m_Q^2).
\end{eqnarray}
Finally, the two-loop factorisable integral in fig.~\ref{fig:fact} are written as:
\begin{eqnarray}
\masterNew{67} & = & m_{I_{(t_3)}}(1, 1, 1, 0, 1, 1, 1;m_Q^2), \quad
\masterNew{68}  =  m_{I_{(t_3)}}(0, 1, 1, 0, 1, 1, 1;m_Q^2), \nn \\
\masterNew{69} & = & m_{I_{(t_3)}}(0, 0, 1, 1, 1, 1, 0;m_Q^2), \quad
\masterNew{70}  =  m_{I_{(t_9)}}(1,1,1,1,1,0,0;m_Q^2), \nn \\
\masterNew{71} & = & m_{I_{(t_9)}}(1,1,1,0,0,1,0;m_Q^2), \quad
\masterNew{72}  =  m_{I_{(t_1)}}(0, 1, 1, 0, 0, 1, 1;m_Q^2), \nn \\
\masterNew{73} & = & m_{I_{(t_1)}}(0,0,1,0,0,1,1;m_Q^2), \quad
\masterNew{74}  =  m_{I_{(t_5)}}(0,0,1,1,0,1,0;m_Q^2), \nn \\
\masterNew{75} & = & m_{I_{(t_{11})}}(0,1,1,1,1,0,0;m_Q^2), \quad
\masterNew{76}  =  m_{I_{(t_1)}}(0,0,0,0,0,1,1;m_Q^2).
\end{eqnarray}


\section{Arguments of the eMPLs}
\label{app:elliptic_args}

In this appendix we present the analytic expressions for the arguments of the eMPLs as defined in eq.~\eqref{eq:z_i_def}. 
We find that we need six different values of $\tilde{z}_i$ for the eMPLs 
associated to the elliptic curve $\tau^{(a)}$, and just one, which we denote 
$\tilde{z}_b$, for the eMPLs described by $\tau^{(b)}$. Furthermore, we checked
using {\tt PSLQ} that there is no rational linear combination of the form
\begin{equation}
    \sum_{i=1}^6c_i\,\tilde{z}_i = c_7+c_8\,\tau^{(a)}\,,\qquad c_i \in \mathbb{Q}\,.
\end{equation}
The numbers $\tilde{z}_i$ are special values of  the elliptic integral:
\begin{equation}\label{eq:Abel_Def}
    \textrm{\bf A}^{(a)}(x) = \frac{\sqrt[4]{5}}{2\,\K\left(\lambda_a\right)}\,\int_{(1-\sqrt{1+2i})/2}^x\frac{dt}{\sqrt{4t^4-8t^3+4t^2+1}}\,,
\end{equation}
with 
\begin{equation}
    \lambda_a = \frac{1}{2}+\frac{1}{2\sqrt{5}}\,.
\end{equation}
We then have\footnote{We note that when evaluating
the quantities $\tilde{z}_a$ with \texttt{Mathematica} using its native implementation of the incomplete elliptic integral of the first kind, {\tt EllipticF},
we obtained different numerical values than those obtained from the numerical evaluation of eq.~\eqref{eq:Abel_Def} with \texttt{NIntegrate}. 
We take the \texttt{NIntegrate} value as the correct one.}
\begin{equation}\begin{split} \label{eq:z_tilde}
\tilde{z}_1 & =  \operatorname{Im}\,\textrm{\bf A}\left({\frac{1}{2}+\frac{i}{2}}\right) = \frac{i}{4}-i \frac{\F(\phi_1|\lambda_a)}{2 \K(\lambda_a)}=0.3486\ldots\,,  \\
\tilde{z}_2 & = \operatorname{Im}\,\textrm{\bf A}\left({\frac{1}{2}-\frac{1}{2} i \sqrt{\frac{5}{3}}}\right) = \frac{i}{4} - i\frac{\F(\phi_2|\lambda_a)}{2 \K(\lambda_a)}=0.0204\ldots \,, \\
\tilde{z}_3 & = \operatorname{Re}\,\textrm{\bf A}\left({-i}\right) = -\frac{\tau^{(a)}}{4} + \frac{\F(\phi_3|\lambda_a)+ \F(\phi_4|\lambda_a)}{4 \K(\lambda_a)}=0.1904\ldots\,,  \\
\tilde{z}_4 & =  \operatorname{Im}\,\textrm{\bf A}\left({-i}\right) = -\frac{i\tau^{(a)}}4 - i\frac{\F(\phi_3|\lambda_a)-\F(\phi_4|\lambda_a)}{4 \K(\lambda_a)}=-0.0615\ldots \,, \\
\tilde{z}_5 & =  \operatorname{Re}\,\textrm{\bf A}(2) = -\frac{\tau^{(a)}}{4} + \frac{\F(\phi_5|\lambda_a)}{2 \K(\lambda_a)}=0.6280\ldots \,, \\
\tilde{z}_6 & = \operatorname{Re}\,\textrm{\bf A}\left({\frac{3}{2}}\right) = -\frac{\tau^{(a)}}{4} + \frac{\F(\phi_6|\lambda_a)}{2 \K(\lambda_a)}=0.5625...\,,
\end{split}\end{equation}
where $\F(\phi|m)$ was defined in eq.~\eqref{eq:F_def} and  we set $\phi_i = \sin^{-1}\alpha_i$, with
\begin{equation}\begin{split}
\alpha_1 & =  \sqrt{\frac{2 \sqrt{\frac{(2+i)+\sqrt{-5+10 i}}{\sqrt{5}+1}}}{i+\sqrt{1-2 i}}}\,,  \\
\alpha_2 & =  \sqrt{\frac{2 \sqrt{30 \left((-1-i)+\sqrt{-3+6 i}\right)}}{(6+3 i)+3 i \sqrt{5}+\sqrt{15+30 i}+\sqrt{15-30 i}}}\,,  \\
\alpha_3 & =  \frac{2^{3/4} \sqrt[4]{(5+5 i)-5 \sqrt{1+2 i}}}{\sqrt{(1+2 i)^{3/2}+(-1+2 i)-\sqrt{5}+\sqrt{5+10 i}}}\,,  \\
\alpha_4 & =  \sqrt{\frac{\left((-1+2 i)+\sqrt{1+2 i}\right) \sqrt{\frac{2-4 i}{\sqrt{5}+1}}}{(-1+2 i)+\sqrt{1-2 i}}}\,,  \\
\alpha_5 & =  \sqrt{\frac{\left(3+\sqrt{1+2 i}\right) \sqrt{\frac{2-4 i}{\sqrt{5}+1}}}{3+\sqrt{1-2 i}}}\,,  \\
\alpha_6 & =  \sqrt{\frac{\left(2+\sqrt{1+2 i}\right) \sqrt{\frac{2-4 i}{\sqrt{5}+1}}}{2+\sqrt{1-2 i}}}\,.
\end{split}\end{equation}

Similarly, the number $\tilde{z}_b$ is defined by the elliptic integral:
\begin{equation}\label{eq:Abel_Def_b}
    \textrm{\bf A}^{(b)} = \frac{1+\sqrt{5}}{4\,\K\left(\lambda_b\right)}\,\int_{(1-\sqrt{5})}^{(1+\sqrt{2})}\frac{dt}{\sqrt{t^4-4 t^3+8 t}}\,,
\end{equation}
with
\begin{equation}
    \lambda_b = \frac{1}{2} \left(3 \sqrt{5}-5\right).
\end{equation}
We therefore have
\begin{equation}
\tilde{z}_b = \textrm{\bf A}^{(b)} -\frac{1}{2} = (3+\sqrt{5})\frac{F(\phi_b|-5-\lambda_b) - K(-5-\lambda_b)}{4 K(\lambda_b)}=i\, 0.1853\ldots,
\end{equation}
where in this case $\phi_b$ is defined as $\phi_b = \sin^{-1}\alpha_b$, with
\begin{equation}
    \alpha_b = \sqrt{\frac{4 \sqrt{2}+\sqrt{5}+3}{2\sqrt{5}}}\, .
\end{equation}

\bibliographystyle{JHEP}
\bibliography{biblio}

\end{document}